\documentclass[usenatbib]{mn2e}
\usepackage{epsfig}
\usepackage{amsmath}
\usepackage{amssymb}
\usepackage{graphicx,graphics,subfigure}
\def\gtrsim{\mathrel{\raise0.35ex\hbox{$\scriptstyle >$}\kern-0.6em
\lower0.40ex\hbox{{$\scriptstyle \sim$}}}}
\def\lesssim{\mathrel{\raise0.35ex\hbox{$\scriptstyle <$}\kern-0.6em
\lower0.40ex\hbox{{$\scriptstyle \sim$}}}}

\def\h50{h_{50}}

\title[IFU Observations of BCGs]{The Diverse Nature of Optical Emission Lines in Brightest Cluster Galaxies: IFU Observations of the Central Kiloparsecs}

\author [Edwards et al.]{Louise~O.V.~Edwards\thanks{Moved to: Infrared Processing and Analysis Center, California Institute of Technology, Pasadena, CA, USA, 91125; louise@ipac.caltech.edu}$^{1,2}$, Carmelle Robert$^1$, Mercedes Moll\'{a}$^3$, Sean~L.~McGee$^4$\\
$^1$D\'{e}partement de physique, g\'{e}nie physique, et optique, Universit\'{e} Laval, and Centre de recherche en astrophysique du Qu\'{e}bec,\\ ~Qu\'{e}bec, QC G1K 7P4, Canada\\
$^2$Department of Physics and Astronomy, Trent University, Peterborough, ON, Canada, K9J 7B8\\
$^3$CIEMAT, Avda. Complutense 22, 28040 Madrid, Spain\\
$^4$Department of Physics and Astronomy, University of Waterloo, Waterloo, ON, Canada, N2L 3G1
}

\date{\today}

\begin{document}
\maketitle

\begin{abstract}

We present integral field spectroscopy of the nebular line emission in a sample of 9 brightest cluster galaxies (BCGs). The sample was chosen to probe both cooling flow and non-cooling flow clusters, as well as a range of cluster X-ray luminosities. The line emission morphology and velocity gradients suggest a great diversity in the properties of the line emitting gas. While some BGCs show evidence for filamentary or patchy emission (Abell 1060, Abell 1668 and MKW3s), others have extended emission (Abell 1204, Abell 2199), while still others have centrally concentrated emission (Abell 2052). We examine diagnostic line ratios to determine the dominant ionization mechanisms in each galaxy. Most of the galaxies show regions with AGN-like spectra, however for two BCGs, Abell 1060 and Abell 1204, the emission line diagnostics suggest regions which can be described by the emission from young stellar populations. The diversity of emission line properties in our sample of BCGs suggests that the emission mechanism is not universal, with different ionization processes dominating different systems. Given this diversity, there is no evidence for a clear distinction of the emission line properties between cooling flow and non-cooling flow BCGs. It is not always cooling flow BCGs which show emission (or young stellar populations), and non-cooling flow BCGs which do not.

\end{abstract}

\begin{keywords}
  galaxies: clusters, cooling flows:general -- galaxies: evolution --
  stars: formation -- galaxies: stellar content
\end{keywords}

\section{Introduction}

The Brightest Cluster Galaxy (BCG) is a giant elliptical, often a cD, and is typically located at the center of the cluster's gravitational potential.  The formation and evolutionary history of these large galaxies has remained an area of active research. Close pairs and multiple nuclei are common in cD galaxies \citep{lai03}, and hence it has been hypothesized that mergers and galactic cannibalism \citep{ost77} drive their formation. 

However, cDs are often found at the peak of the cluster's X-ray surface brightness, where the hot intracluster gas can potentially  cool to form molecular clouds, therefore an additional stellar or gaseous component, arising via the cooling intracluster medium, may be expected. Often BCGs in cooling flow clusters show H$\alpha$ emission \citep{von06,cra99}, an exciting find as it is a signature of current or recent activity in their cores. Diverse and dramatic morphologies of the H$\alpha$ emission, such as long tails of emitting gas \citep{fab01}, as well as highly concentrated emission \citep{don00}, and even more filamentary structures \citep{con02,bla01} have been observed. 

\citet{edw07b} found that it is in the sample of cooling flow BCGs, compared to the sample of BCGs as a whole, where there exists an increased likelihood of H$\alpha$ emission (found in  $\sim$70\% of BCGs in cooling flow clusters, but only $\sim$15\% of BCGs in the sample as a whole). This implies that the cooling flow status of the cluster is an important factor for activity at the center. Furthermore, it suggests that within cooling flow BCGs, cool molecular clouds, warm ionized hydrogen, and the cooling intracluster medium (ICM) are related. This is consistent with the finding of \citet{edg02}, who showed that whenever there are detections of molecular hydrogen in a cooling flow cluster, there are also detections of H$\alpha$ emission. \citet{cra99} obtained optical spectra of 256 X-ray selected BCGs and found that 27\% have emission lines and associate most of these with cooling flow clusters.

In cases where the structure of the ionized gas is well studied, complex and irregular morphologies are often spatially correlated with emission features in other wavelength regimes. As \citet{don00} found for Abell~2597 and PKS-0745-19, the ionized gas is filamentary, and similar in extent and structure to the molecular gas emitting in H$_{2}$ 1-0~S(1) line. Other examples of well known H$\alpha$ structures include Abell~1795, which shows a huge tail \citep{cow83}, as well as NGC~4696 in Centaurus (Crawford et al. 2005b) and NGC~1275 in Perseus \citep{con02,fab08}. Each of the last three show structures spanning tens of kiloparsecs and harboring the same features as seen for the X-ray emitting gas (Fabian et al. 2001; Crawford et al. 2005a,b). Furthermore, the CO(2-1) and H$\alpha$ emissions are found to be associated in NGC~1275 \citep{sal06}. \citet{don00} found the H$_{2}$ 1-0~S(1) line emission peak of NGC~1275 to be co-spatial with the peak of H$\alpha$ emission, although confined to the very center of the BCG. 

\nocite{boh01}
\nocite{pet03}
\nocite{cro06}
Observations from the {\it Chandra} telescope have shown regions of star formation that are associated with bright lumps and filaments of gas whose radiative cooling times are short. In Abell~1795 for example, an excess of blue light is detected from the underlying central dominant galaxy, suggesting a population of hot young stars \citep{mcn96,oeg01,bil08}. Additionally, many rotationally excited transitions of CO been detected, as well as the O~VI line, which indicates that gases at a range of temperatures exist (20$\,$K and 100 000$\,$K, respectively). Nevertheless, the amount of H$_{2}$ emission is inconsistent with gas cooling {\it directly} from X-ray temperatures into cool clouds. Cooling flow models predict more cooled gas than is observed (B\"{o}hringer et al. 2001; Peterson et al. 2003). Possibly, the mass is deposited into molecular clouds which are then reheated by one of several processes - hot stars, shocks, or AGN, for example \citep{wil02}, and only a small fraction of the cooled gas is detected. The ICM and radio sources often appear to be interacting (Croton et al. 2006) and these interactions may form cavities, or bubbles, in the surface brightness of the X-ray gas which move buoyantly through the ICM. In some systems, the bubbles carry enough energy to be able to balance the radiative losses emerging from the center of the clusters in the X-ray band \citep{piz05,bir04,mcn07}. These AGNs are also potential sinks for the cooling gas and can contribute to the ionization of hydrogen gas clouds.  

In an attempt to distinguish between these scenarios, this paper examines detailed maps of the line intensity and morphology of the central regions of 9 BCGs. The line emission properties are obtained using the integral field spectrgraphs. 

The objective is to identify and characterize the current and recent activity in BCGs. In order to develop non-biased conclusions, it is important to study the high X-ray luminosity systems, those of lower luminosity, and those in cooling flows and non-cooling flows. The star formation rate (SFR) is calculated for young stellar populations and compared to the cooling flow mass deposition rate, which is calculated from the X-ray luminosity within the cooling radius \citep{fab94}. Careful attention is paid to signatures associated with processes such as the cooling flow phenomenon, AGN activity and galaxy-galaxy interactions.

The paper is organized as follows. In section \ref{data}, we introduce our galaxy sample selection and outline our observations. In section \ref{results} we report our results:  the emission line morphology and  velocity maps, followed by an investigation into the emission mechanism. In section \ref{conclchap}, we consider the impact of our results on various galaxy and galaxy group formation hypotheses. Throughout our discussion we will compare our results to those of \citet{wil06} and \citet{hat07} who have recently presented integral field spectroscopy of the central regions of cooling flow BCGs. Unless otherwise stated our analysis assumes the values $\Omega_{\mathrm m}$=0.3 for the matter density parameter,  $\Omega_{\Lambda}$=0.7 for the cosmological constant, and $H_0$=70 km$\,$s$^{-1}$$\,$Mpc$^{-1}$ for the Hubble parameter.  L$_{X}$ refers to the bolometric X-ray luminosity
throughout. We will often refer to the thesis of \citet{edwthesis}, within which additional detailed  information can be found\footnote[1]{\it An electronic version of the thesis can found at Scientific Commons http://en.scientificcommons.org/30178658}.

\section{Sample Selection and Data Reduction}\label{data}

\begin{table*}
\begin{center}
\caption{Cluster Properties}
 \footnotesize
 \vspace{0.5cm}
\label{ifuobstab2}
\label{ifuobstab2_oas}
\begin{tabular}{llccccccc}
\hline
\hline
Cluster&BCG& z$_{cl}$ &D&kpc/$^{\prime\prime}$&IFU FOV&CF&MDR&L$_{X}$\\
Name&Name&&Mpc&&kpc $\times$ kpc&Status&M$_{\odot}$$\,$yr$^{-1}$&10$^{37}$W\\
\hline
Abell~1060&NGC~3311&0.0126&53&0.25&1.3 $\times$ 1.8&No\footnotemark[1]&-&0.47\\
Abell~1204&&0.1706&640&2.65&13.5 $\times$ 18.9&Yes\footnotemark[2]&50$^{+40}_{-30}$&6.77$\pm{1.42}$\\
Abell~1668&IC~4130&0.0634&256&1.17&5.9 $\times$ 8.2&No\footnotemark[3]&-&1.59$\pm{0.24}$\\
Ophiuchus&&0.0280&116&0.55&2.7 $\times$ 3.8&No\footnotemark[4],Yes\footnotemark[8]&-&$>$~4\\
MKW3s&NGC~5920&0.0450& 184 & 0.85&4.3 $\times$ 6.0&Yes\footnotemark[5]&45$^{+10}_{-10}$&2.68$\pm{0.29}$\\
Abell~1651&&0.0849&337&1.51& ~7.6 $\times$ 10.6&Yes\footnotemark[4]&231$^{+121}_{-132}$&  8.25\\
Abell~2052&UGC~9799&0.0345& 146 & 0.68&4.4 $\times$ 6.0&Yes\footnotemark[2]&5$^{+1}_{-1}$&2.52$\pm{0.20}$ \\
Abell~2199&NGC~2199&0.0310&125&0.59&4.4 $\times$ 6.0&Yes\footnotemark[6]&12$^{+3}_{-3}$& 3.65$\pm{0.15}$ \\
Cygnus-A&&0.0561&227&1.04& ~7.7 $\times$ 10.8&Yes\footnotemark[7]&$\sim$250& -  \\
\hline
\end{tabular}
\end{center}
\footnotesize
{\it The X-ray luminosity and errors (where available) are from \citet{ebe98} and \citet{ebe96}. References for cooling flow status and MDR calculations:} 
\footnotemark[1]{\it Exosat data from \citet{hay06};}
\footnotemark[2]{\it Chandra and XMM-Newton data from \citet{ode08} and}
\footnotemark[3]{\it {\it ASCA} data from \citet{sal03} and}
\footnotemark[4]{\it \citet{whi00};}
\footnotemark[5]{\it RGS XMM-Newton data from \citet{pet03};}
\footnotemark[6]{\it Chandra data from \citet{joh02}; }
\footnotemark[7]{\it ROSAT data from \citet{rey96} and }
\footnotemark[8]{\it  Suzaku data from \citet{fuj08}.~~~~~~~~}\\

\end{table*}

  \begin{figure*}
  \centering
  \epsfxsize=7in
     \epsfbox{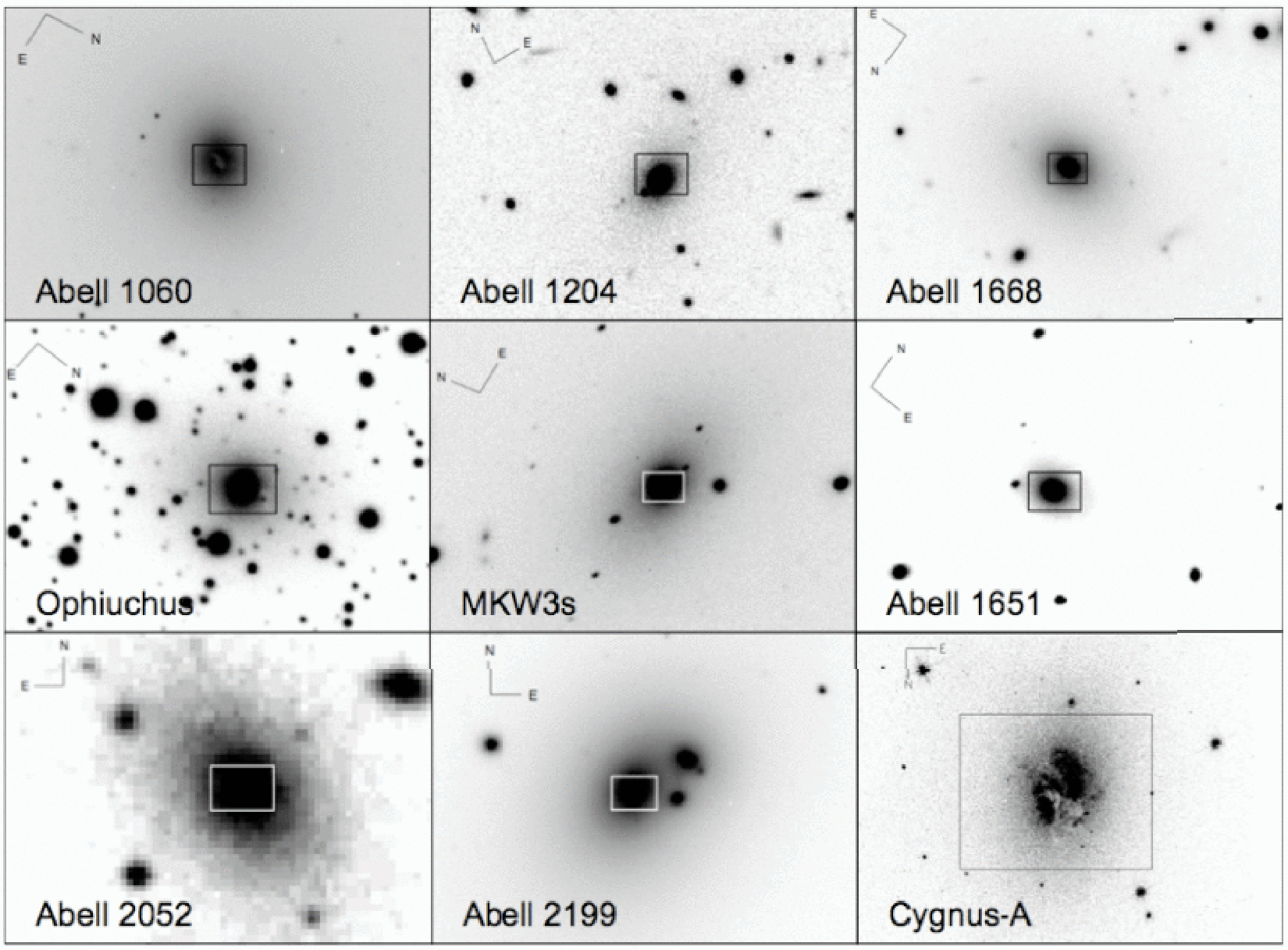}
   \caption[The center of clusters]{\bf Image of the cluster centers and the IFU position. \it The GMOS IFU (5$^{\prime \prime}$$\,$$\times$$\,$7$^{\prime \prime}$) or OASIS (10.3$^{\prime \prime}$$\,$$\times$$\,$7.4$^{\prime \prime}$) field of view is marked as a box centered on the BCG. \underline{Abell~1060}:  Taken in the r-filter around the BCG, NGC~3311. The central dust patch is already noticeable in this image. \underline{Abell~1204}: Taken in the i-filter. A trail of light exists in the direction from the BCG to the small galaxies just East of North. \underline{Abell~1668}:  Taken in the r-filter around IC~4130, the BCG. \underline{Ophiuchus}: Taken in the r-filter. A hint of extinction is seen in the northern corner of the IFU FOV. \underline{MKW3s}: Taken in the r-filter around NGC~5920, the BCG. \underline{Abell~1651}: Taken in the i-filter. \underline{Abell~2052}: Taken in the DSS red filter around UGC~9799, the BCG. \underline{Abell~2199}:  Taken in the Gemini i-filter around NGC~6166, the BCG. \underline{Cygnus-A}: Taken in the HST F622W filter around the BCG. The images are scaled in arbitrary flux units. \label{aquiims}} 
  \end{figure*}

   \nocite{cra05b}
\nocite{con01}
\nocite{wil06}
\nocite{hat07}
Our nine targets are chosen such that line emission in the center of the galaxy is plausible. All BCGs are within 50$\,$kpc of the X-ray center and either previous reports of emission lines exist (most notably from the catalogue of Crawford et al. 1999), or the cluster properties are often associated with emission lines, i.e. a cooling flow (CF) is centered on the BCG and radio emission is detected \citep{edw07b}. The images and spectra are observed using the integral field spectrographs on the Gemini Telescopes (GMOS IFU) and the William Herschel Telescope (OASIS on the WHT). 

The majority of BCGs which have previously been observed with integral field units are in cooling flow clusters (Conselice et al. 2001; Crawford et al. 2005b; Wilman et al. 2006; Hatch et al. 2007). Here, we include non-cooling flow cluster BCGs. The sample is listed in Table \ref{ifuobstab2}. The first column lists the name of the cluster, the second the name of the BCG, and the third column lists the cluster redshift (taken from the NASA Extragalactic Database, hereafter NED). The distance and angular scale, assuming h$_{0}$=0.70, are listed in the fourth and fifth columns, respectively. The dimensions of the integral field unit FOV are given in column~6. The cooling flow status (obtained from the literature) is given in column~7 along with the mass deposition rate (MDR) in  column~8. The values in these last two columns are accompanied by two important caveats. First, the cooling flow status is not uniformly determined throughout the literature. Often a short cooling time will be used to define a cooling flow, however, a high mass deposition rate, or a central temperature drop will also be used to classify the cluster. Overall these methods generally achieve the same results in terms of classifying the clusters, but can lead to discrepancies for the less powerful systems. Second, the mass deposition rates calculated based on the classical cooling flow paradigm are a factor of 10 to 100 times larger than those computed assuming most of the gas does not cool beyond $\sim$1$\,$keV. The values in the table quoted from ROSAT are the erroneous classic values and should be taken as much higher than the actual mass drop out rates. Note the very large errors on all MDR values.  The X-ray luminosity is in column~9 (see the table notes for references). Each BCG and its surrounding local environment is shown in the acquisition images of Figure \ref{aquiims}. The spatial extent covered is drawn on the image and spans 5$^{\prime \prime}$$\,$$\times$$\,$7$^{\prime \prime}$ for the GMOS IFU, and 10.3$^{\prime \prime}$$\,$$\times$$\,$7.4$^{\prime \prime}$ for OASIS.

\begin{table*} 
\begin{center}
\caption{Observational Data}
\vspace{0.5cm}
\footnotesize
\label{ifuobstab}
\label{ifuobstab_oas}
\begin{tabular}{llcccccc}
\hline
\hline
Observatory&Cluster Name&Configuration&$\lambda_{0}$&$\Delta\lambda$&IntTime&E(B$-$V)$_{Gal}$ $^{a}$&E(B$-$V)$_{int}$\\
&&&\AA&\AA&sec&mag&mag\\
\hline
GS&Abell~1060&R400+r&6300&5710-6830&1640&0.079&0.18 \footnotemark[1]$^{,}$\footnotemark[2]\\
GN&Abell~1204&R400+i&7800&5940-6980&3720&0.017& 0 \footnotemark[3]\\
GN&Abell~1668&R400+r&6300&5560-6615&3600&0.032&0.3 \footnotemark[4]\\
GS&Ophiuchus&R400+r&6300/6350&5450-6850&6000&0.591&0.3 \footnotemark[4]\\
GN&MKW3s&R400+r&6300/6350&5450-6700&5400&0.035 &0.3 \footnotemark[4]\\
GS&Abell~1651&R400+i&7800/7850&6460-7700&6000&0.027&0.3 \footnotemark[4]\\
WHT&Abell~2052&MR661&6610&6130-6740&3600&0.037&0.22 \footnotemark[5]\\
WHT&Abell~2052&MR516&5160&4700-5345&1800&0.037&0.22 \footnotemark[5]\\
WHT&Abell~2199&MR661&6610&6430-6770&2400&0.012 &0.10 \footnotemark[5]\\
WHT&Abell~2199&MR516&5160&4700-5345&4800&0.012 &0.10 \footnotemark[5]\\
WHT&Cygnus-A&MR661&6610&5915-6605&3600&0.381&0.6 \footnotemark[6]\\
WHT&Cygnus-A&MR516&5160&4695-5225&3000&0.381&0.6 \footnotemark[6]\\
\hline
\end{tabular}
\end{center}
\footnotesize
$^{a}${\it Taken from NED.~~~~~~~~~~~~~~~~~~~~~~~~~~~~~~~~~~~~~~~~~~~~~~~~~~~~~~~~~~~~~~~~~~~~~~~~~~~~~~~~~~~~~~~~~~~~~~~~~~~~~~~~~~~~~~~~~~~~~~~~~~~~~~~~~~~~~~~~~~~~}\\
{\it References for internal extinction:}
\footnotemark[1]{\it \citet{vas91};}
\footnotemark[2]{\it  \citet{sad85};}
\footnotemark[3]{\it There is no correction for internal extinction in this case since the Balmer decrement could not be calculated in \citet{cra99};}
\footnotemark[4]{\it We use 0.3, average value for the BCGs in the \citet{cra99} sample;}
\footnotemark[5]{\it Taken from \citet{cra99}; and} 
\footnotemark[6]{\it Average value from this data.~~~~~~~~~~~~~~~~~~~~~~~~~~~~~~~~~~~}
\end{table*}
   
   The GMOS IFU observations were completed between February and June, 2006. This was done in queue mode using the two-slit configuration to allow for the largest field of view. The filter and grating pairings were chosen so to observe H$\alpha$ ($\lambda$$_{rest}$6563$\,$\AA) at the redshift of the cluster. Three BCGs, and their standard stars, were observed using OASIS on the nights of June 28-29, and July 2, 2005. The 22mm enlarger was used.  The MR661 configuration was used to obtain H$\alpha$ and the MR516 configuration was used to obtain H$\beta$ at $\lambda$$_{rest}$4861$\,$\AA. Each target was observed close to the zenith in order to reduce the effects of atmospheric absorption. Table~\ref{ifuobstab} gives the instrument configuration, grating central wavelength, rest wavelength coverage and integration time for each galaxy.

 \subsection{GMOS IFU}\label{gmos} 

Several Gemini specific programs from the package {\it gemtools} in the spectral analysis software IRAF were used to reduce and analyze the spectra in the standard fashion (bias subtraction, cosmic ray rejection, flat fielding, wavelength calibration, sky subtraction, and atmospheric extinction-correction). Baseline standard stars were used to perform the absolute flux calibration. The final datacubes of Abell~1651, and most notably Abell~1060, show bright ripples at both sides of the image. These artifacts, were caused by imperfect sky subtraction of the flat field frames during the standard reduction procedures. To make a cleaner continuum image for Abell~1060, the reduction process was slightly modified. Instead of using the pipeline for this case we manually constructed an image of the flat field using only the same wavelength range as the continuum coverage. We then normalized this flat field image and divided it into our continuum image directly. Although this did not result in a perfectly clean frame, it did make for a noticeable improvement: the bright and dark fringes on the right side of the continuum image in Abell~1060 had a flux difference which was 10\% of the average continuum level, which decreased to a 2\% difference in the reworked image. The variation in the high and low continuum level (discounting fringes) stayed constant at $\sim$35\% in each case. These artifacts disappear in the continuum subtracted line images, as both the continuum, and line~+~continuum images contain the fringes. 

Table \ref{ifuobstab}, column~7, includes a value for the reddening expected from the Milky Way \citep{shl98}. This galactic extinction was removed using the IRAF task {\it deredden}. Subsequently, {\it dopcor} and the known cluster redshifts (see Table~\ref{ifuobstab2}) are applied to deredshift the spectra. 

Each hexagonal lenslet is about 0.2$^{\prime \prime}$ and subsampled onto a rectangular grid of 0.1$^{\prime \prime}$ pixels. Each pixel's spectrum is then median averaged with the value of its 8 closest neighbours in order to increase the S/N and better match the seeing (typically, $\sim$0.8$^{\prime \prime}$ to 1$^{\prime \prime}$). The pixels are slightly subsampled with respect to the seeing, however our results rely on the analysis of even larger regions. 

\subsection{OASIS}\label{oasis}

The XOasis reduction package was used for the bias subtraction, spectral extraction, flat fielding, wavelength calibration, as well as for a first cosmic ray subtraction and absolute flux calibration. Further  cosmic rays were removed by median combination of the frames, except when there were less than three exposures taken. In this instance we used IRAF's {\it imreplace} to create a bad pixel mask and the {\it crfix} routine to interpolate across the cosmic ray ruined pixels. To account for accidental pollution created in the housing chamber due to an LED, the sky subtraction itself was preformed using IRAF after the flux calibration was completed in XOasis; the entire sky frame was subtracted from the entire object frame, as the signature of the LED was apparent in both frames (each with the same integration time). Abell~2199 had data taken in the MR516 configuration spread over two nights. After sky subtraction, flux calibration and cosmic ray removal, the datasets were averaged together.

The absolute atmospheric extinction, A(0) for the WHT during the summer months is 0.12, and we used the standard value of E(B-V) of 3.1 for our calculations of the atmospheric extinction. We used the NAOMI AO corrector and the seeing was $\sim$0.7$^{\prime \prime}$, which is greater than the instrumental spatial resolution (0.26$^{\prime \prime}$). Therefore, the surrounding 8 pixels were added to increase the S/N and match the seeing. The corrections of galactic extinction and dereddening were done using the same methods as in the GMOS data.

\subsection{Line Measurements and Internal Extinction}

\begin{figure}
 \subfigure{
\begin{minipage}[c]{0.45\textwidth}
        \centering
        \includegraphics[width=2.5in,angle=0]{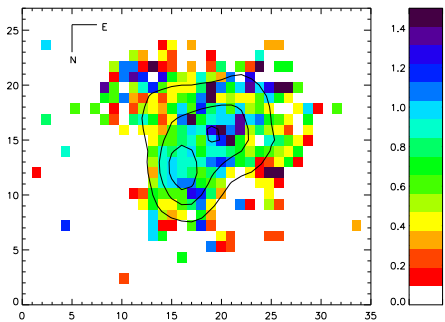}
        \end{minipage}%
    }
      \caption[Extinction map of Cygnus-A]{\bf Extinction map of Cygnus-A. \it The image is a  representation of E(B$-$V) derived from the H$\alpha$ to H$\beta$ line strengths, with the H$\alpha$ emission overlain as contours. One pixel is $\sim$220$\,$pc across. \label{ebvCA}}
\end{figure}

The spectral line characteristics of flux, central position, and line width (FWHM), are mapped in Section~3. We fit multiple Gaussian profiles to the spectrum of each IFU pixel using the IRAF task {\it deblend}. For the nearby H$\alpha$ and [N~{\sc{ii}}]~$\lambda\lambda$~6548,6584 emission lines we find the best fit for all three lines simultaneously. The task works by taking user specified  best guess parameters for the continuum level and location of the line centers. It then varies the parameters of the line model in order to improve the chi square, based on the Levenberg-Marquardt method. Errors quoted on flux levels and central position are also taken from the {\it deblend} fit. They are based on the Poisson statistics of the model of the data and determined by fifty monte carlo simulations which are run to create random Gaussian noise. The difference of the fitted flux to the model flux determines the error.

For most galaxies we are able to measure emission lines in H$\alpha$, [N~{\sc{ii}}]~$\lambda\lambda$~6548,6584 and [O~{\sc{iii}}]~$\lambda$~5007 and the absorption line of NaD at 5890$\,$\AA~with pixel-to-pixel S/N$>$10. In some cases we could also detect [O~{\sc{i}}]~$\lambda\lambda$~6300,6364, and [S~{\sc{ii}}]~$\lambda\lambda$~6716,6734 in emission with S/N$>$5. For the GMOS individual spectra, the minimum detectable level, 1$\sigma$ is $\sim$3$\times$10$^{-19}$$\,$erg$\,$s$^{-1}$cm$^{-2}$ (For the OASIS spectra obtained with the MR661 configuration, 1$\sigma$ is $\sim$1$\times$10$^{-18}$$\,$erg$\,$s$^{-1}$cm$^{-2}$). For the OASIS spectra taken with the MR516 configuration, 1$\sigma$ is $\sim$7$\times$10$^{-18}$$\,$erg$\,$s$^{-1}$cm$^{-2}$ and absorption lines of H$\beta$, Fe~{\sc{i}} and Mg$_{b}$ can only reach a 5$\sigma$ detection by adding the spectra over several individual pixels. Unless otherwise stated, the continuum level in the red has been determined by taking a median of a $\sim$100$\,$\AA-wide region around 6880$\,$\AA~if observed using the R400+i configuration, and around 6450$\,$\AA~if observed in the R400+r configuration (exact continuum regions can be found in Edwards 2007). The rms for each spectrum was calculated in the same spectral windows used to make the continuum image. Average values for the signal-to-noise (S/N) ratio of the continuum near H$\alpha$ are between 10 and 20 (between 8 and 12 for OASIS). For both of the GMOS IFU configurations, R400+r and R400+i, the observed resolution is 3.3$\,$\AA, an average of the measured FWHM of observed sky lines at $\sim$6300$\,$\AA.  The configurations used for OASIS each give an observed FWHM of $\sim$2.5$\,$\AA~for the standard stars. 

In order to calculate qualities such as metallicity and age, and to compare the emission between galaxies, a knowledge of the internal reddening of each galaxy is important. Some elliptical galaxies are known to be dusty \citep{sad85} and this dust can be uniform, filamentary, or patchy \citep{lai03}. Our data allow a map of the internal extinction only for Cygnus-A and we show the work for deriving the pixel to pixel extinction for this case in the following paragraph. For all other galaxies, we only have access to an integrated value yielded from slit measurements obtained from the literature. This is by no means ideal, but as only the central few kpc are covered in the FOV, we resign to using this method. These values are listed in column~8 of Table~\ref{ifuobstab}. In some cases, no known values of the internal extinction for a particular galaxy can be found and so we adopt a value of E(B$-$V)$_{int}$~=~0.3. This value is the average value for X-ray selected BCGs with strong H$\alpha$ emission lines found in \citet{cra99}.

For Cygnus-A, we align the images to the peak in the [O~{\sc{iii}}]~$\lambda$~5007 line image to that in the [N~{\sc{ii}}]~$\lambda$~6584 line image. These two lines are chosen since they are both strong, and both are high ionization lines. The method is validated as the result is a good match to the continuum peak of both images. The extinction map is subsequently created using the following equation on a pixel by pixel basis:

\begin{equation}
\mathrm{E}(\mathrm{B}-\mathrm{V})_{int}=\frac{2.177}{-0.37~\mathrm{R}}\, \times \, \Big( \mathrm{Log}\Big\{\frac{\mathrm{I}_{o\mathrm{H}\alpha}}{\mathrm{I}_{o\mathrm{H}\beta}}\Big\} - \mathrm{Log}\Big\{\frac{\mathrm{I}_{\mathrm{H}\alpha}}{\mathrm{I}_{\mathrm{H}\beta}}\Big\}\Big). \label{eqnExt}
\end{equation}

\nocite{kau03}
\nocite{ost06}

Here, R~=~3.1, I$_{o \mathrm{H}\alpha}$/I$_{o \mathrm{H}\beta}$ is 2.85, the theoretical ratio for Case~B recombination, and I$_{\mathrm{H}\alpha}$ and I$_{\mathrm{H}\beta}$ are the observed values (Osterbrock \& Ferland 2006; Kauffmann et al. 2003). The theoretical I$_{o \mathrm{H}\alpha}$/I$_{o \mathrm{H}\beta}$  ratio of 2.85 may not be the ideal value to use for known Seyfert galaxies, like Cygnus-A, but the actual value is debated. It is often assumed that the H$\alpha$ emission in these systems is enhanced due to collisional processes,  and several authors use a value of 3.1 \citep{gas84,ost06}, although other values have also been determined \citep[calculate a value of 3.4]{bin90}. The extinction map is presented in Figure~\ref{ebvCA} and shows very high values of E(B$-$V)$_{int}$$\simeq$1.0$\,$-$\,$1.2 at the center where the H$\alpha$ emission is maximal. Only pixel values with S/N$>$5 for both H$\alpha$ and H$\beta$ are plotted. Although the values of extinction determined here may be slightly overestimated due to the choice of intrinsic I$_{o \mathrm{H}\alpha}$/I$_{o \mathrm{H}\beta}$ used, the average value in the NW emission peak from our map is 0.75$\pm$0.2. This is not far from the integrated value of 0.69$\pm$0.04 \citep{ost06} for Cygnus-A, also calculated using the Balmer decrement but using an intrinsic I$_{o \mathrm{H}\alpha}$/I$_{o \mathrm{H}\beta}$ ratio of 3.08.

\section{Emission Line Morphology, Kinematics and Diagnostics}\label{results}

\begin{table}
\begin{center}
\caption{Summary of IFU Emission Line Properties}
 \footnotesize
 \vspace{0.5cm}
\label{sumtab}
\begin{tabular}{lcccc}

\hline
\hline
 Cluster &Lines&SFR&Age$_{old}$&Mass $\rho$$_{old}$\\
 Name&&&&$\times$10$^{8}$\\
 &&M$_{\odot}$yr$^{-1}$ & Gyr & M$_{\odot}$$\,$kpc$^{-2}$ \\
\hline
A1060    & SF& 0.02$\pm$0.001&10$\pm$6&2$\pm$1\\
A1204$^{1}$ &SF \& AGN&7.0$\pm$0.4&-&-\\
A1668 &AGN&-&7$\pm$3&3$\pm$2\\
Ophi$_{BCG}$ &No$^{2}$ &-&10$\pm$5&150$\pm$80\\
MKW3s &AGN$^{3}$ &- &$>$12&$>$8460\\
A1651 &No &- &-&-\\
A2052& AGN &-&$>$4 &$>$30\\
A2199& AGN&-&$>$~1&$>$~3\\
Cyg-A&AGN  &-&-&-\\
\hline
\end{tabular}
\end{center}
\footnotesize
{\it Note 1: Possible companions near line emitting gas in A1204.~~}\\
{\it Note 2: There is a clump of gas, Object~B, near to the BCG in Ophiuchus showing LINER or AGN-like emission line ratios.}\\
{\it Note 3: Emission lines in MKW3s are blueshifted with respect to underlying cD population.~~~~~~~~~~~~~~~~~~~~~~~~~~~~~~~~~~~~~~~~~~~~~~~~~}\\
\end{table}

We present the morphology, kinematic structure, and line diagnostics extracted from the continuum emission and line fluxes. Gaussian profiles are fit on a pixel-to-pixel basis, except for the BCGs in Abell~1668 and MKW3s, where single Gaussian fits are impossible and the flux is measured within a window around the position of the H$\alpha$ line. The emission line maps have all been constructed from continuum-subtracted measurements and no absorption correction is made. The region spectra shown are the integrated values of $\sim$ 30 spectra and are labeled on the the H$\alpha$ images. 

In terms of the morphology, we compare the results from the H$\alpha$ and [N~{\sc{ii}}] $\lambda$ 6584 lines to those in the surrounding continuum for each system, and take note of any particular associated galaxy characteristics such as prominent dust features or nearby neighbours. Maps of [S~{\sc{ii}}]~$\lambda$~6716~+~$\lambda$~6734 are available in \citet{edwthesis}, but not  presented here as their morphology mirrors that seen in the [N~{\sc{ii}}] $\lambda$ 6584 maps. H$\beta$, [O~{\sc{iii}}]~$\lambda$~5007, and maps of the continuum near H$\beta$ at $\lambda$$_{rest}$4861\AA~(between 5025 and 5100\AA) are presented where available (Abell~2052 and Cygnus-A). Only pixels for which the flux measurement is $>$5$\sigma$ are used in the maps presented. 

To examine the kinematics, we show the velocity of the emission lines relative to the cluster radial velocity. When available, the NaD absorption line originating from the underlying galaxy is used for comparison. The velocity and FWHM maps also only show the pixels where the line flux measurement is $>$5$\sigma$. 

  \begin{figure*}
  \centering
  \epsfxsize=7in
     \epsfbox{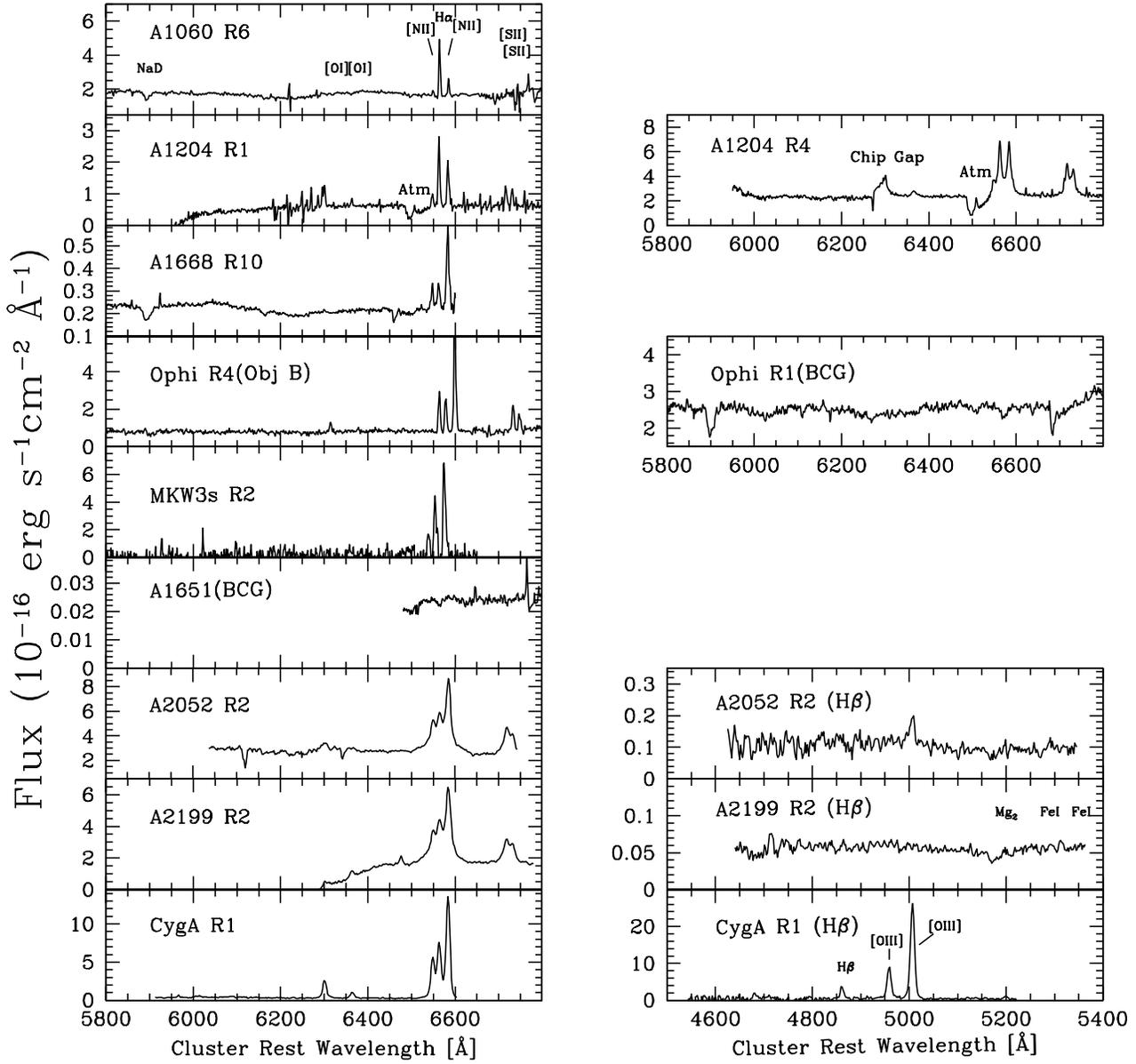}
   \caption[Spectra]{\bf Spectra of representative regions for each BCG. \it The spectra are the integrated values of $\sim$30 individual spectra that make up the labeled regions identified in figures of Section~3. In the case of MKW3s the spectrum has been subtracted by the underlying galaxy spectrum, where absorption lines are shifted by -560$\pm$50$\,$km$\,$s$^{-1}$. Blue spectra for the OASIS clusters are also shown. The size of the region covered for Abell~1060 is 175$\,$pc on either side, 1710$\,$pc for Abell~1204 (both regions), 710$\,$pc for Abell~1668, 330$\,$pc for Ophiuchus (BCG and Object~B), 580$\,$pc for MKW3s, 650$\,$pc for Abell~1651, 710$\,$pc for Abell~2052, 630$\,$pc for Abell~2199, and 890$\,$pc for Cygnus-A. \label{specpap}} 
  \end{figure*}
  
To diagnose the origin of the emission, we present spectra for representative regions of the BCGs (usually at the location of the H$\alpha$ or continuum peak). Generally, we do not have the H$\beta$ and [O~{\sc{iii}}] lines required for constructing a BPT diagram \citep{bpt81} for each galaxy. However, we have plotted  values of  the [N~{\sc{ii}}] $\lambda$ 6584/H$\alpha$ line ratio as a function of H$\alpha$ luminosity for central regions in all the BCGs with emission lines. This figure helps separate the ionization mechanism as a value of ([N~{\sc{ii}}] $\lambda$ 6584/H$\alpha$)~$>$~0.63 is likely to be from an AGN or LINER, rather than star formation \citep{ost06}.

\begin{figure*}
 \subfigure{
\begin{minipage}[c]{0.45\textwidth}
        \centering
        \includegraphics[width=2.1in,angle=0]{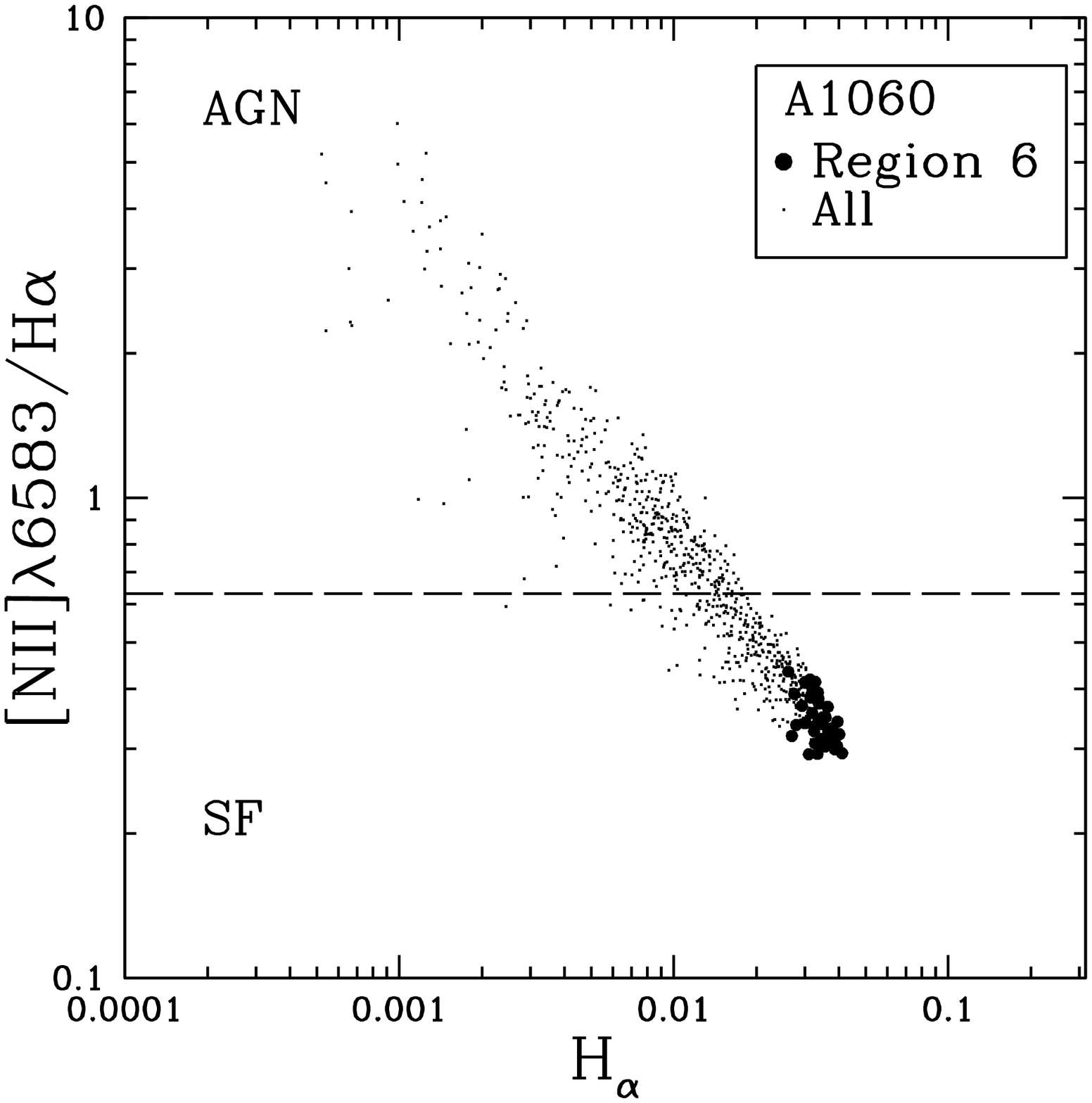}
        \end{minipage}%
    }
    \subfigure{
     \begin{minipage}[c]{0.5\textwidth}
        \centering
        \includegraphics[width=2.1in,angle=0]{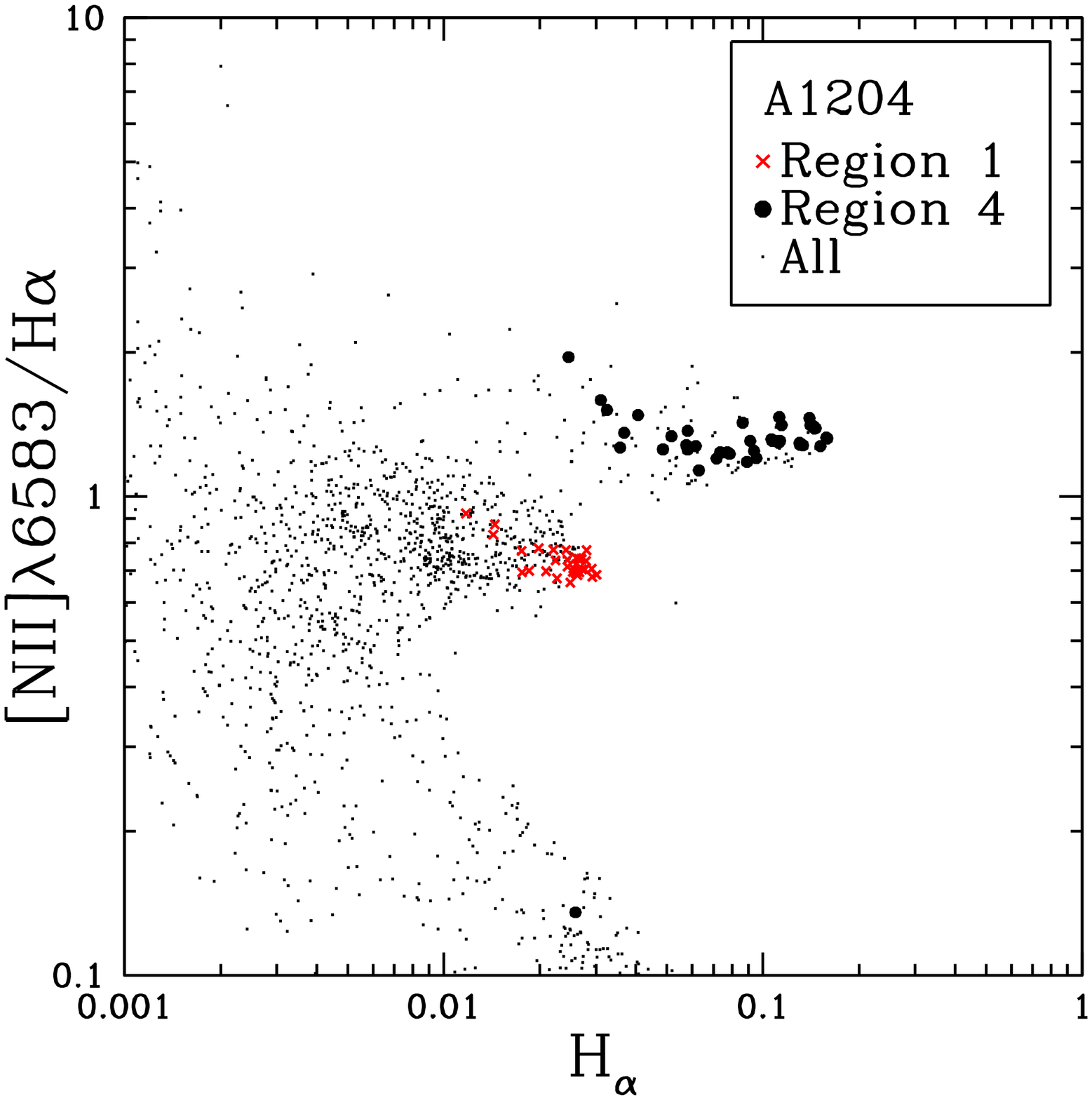}
         \end{minipage}%
    } 
    \subfigure{
\begin{minipage}[c]{0.45\textwidth}
        \centering
        \includegraphics[width=2.1in,angle=0]{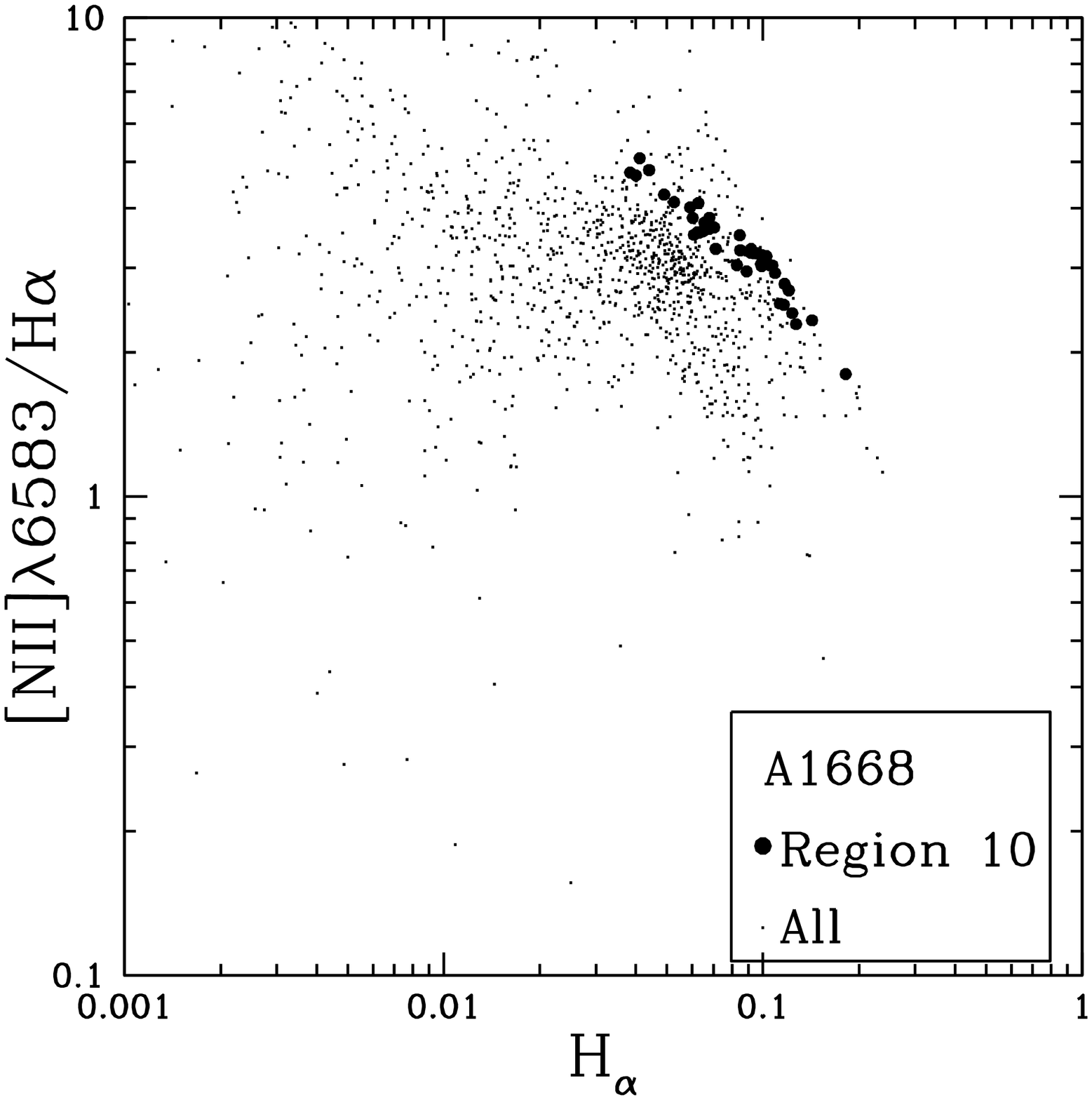}
        \end{minipage}%
    }
    \subfigure{
     \begin{minipage}[c]{0.5\textwidth}
        \centering
        \includegraphics[width=2.1in,angle=0]{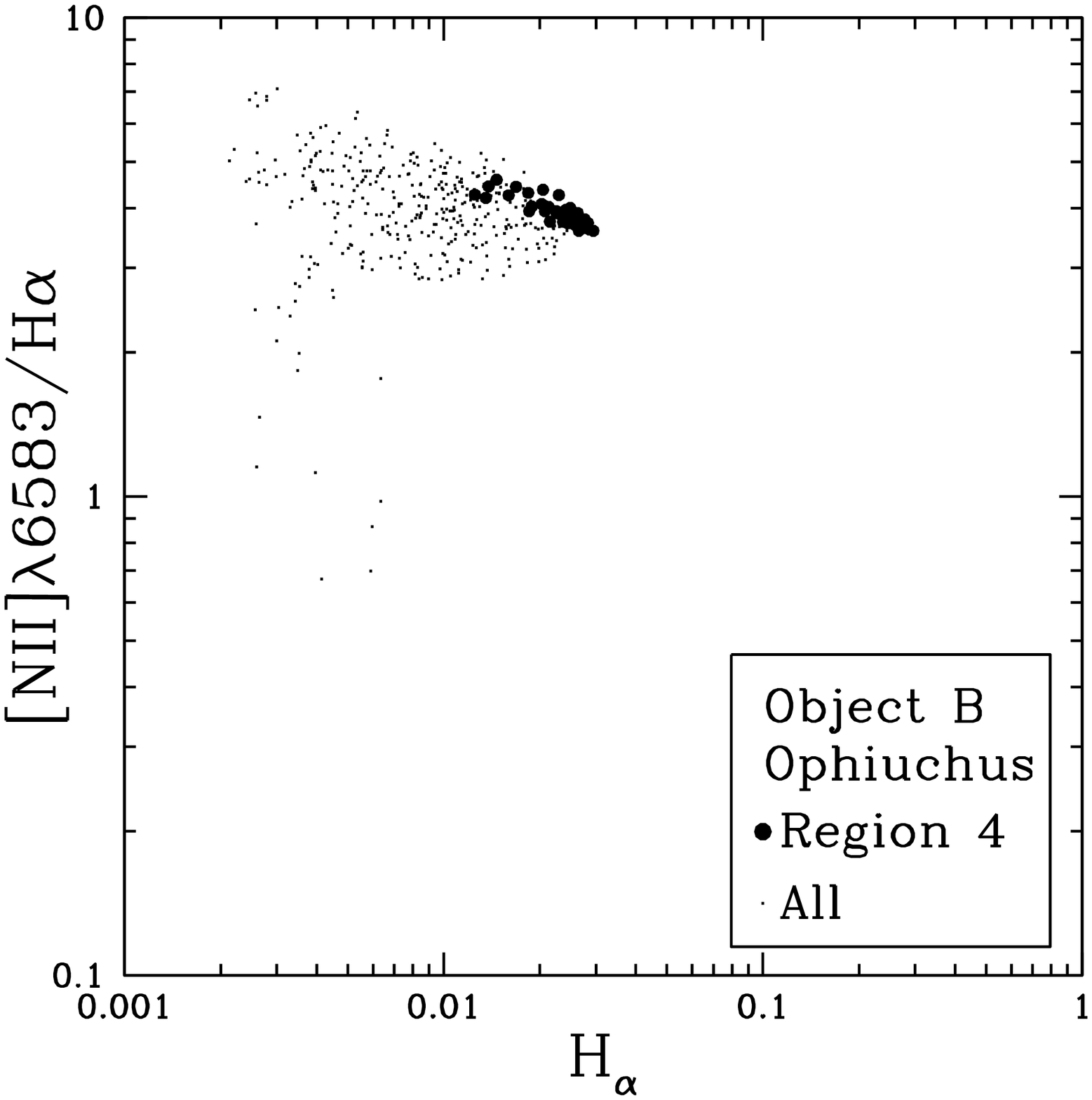}
         \end{minipage}%
    } \subfigure{
\begin{minipage}[c]{0.45\textwidth}
        \centering
        \includegraphics[width=2.1in,angle=0]{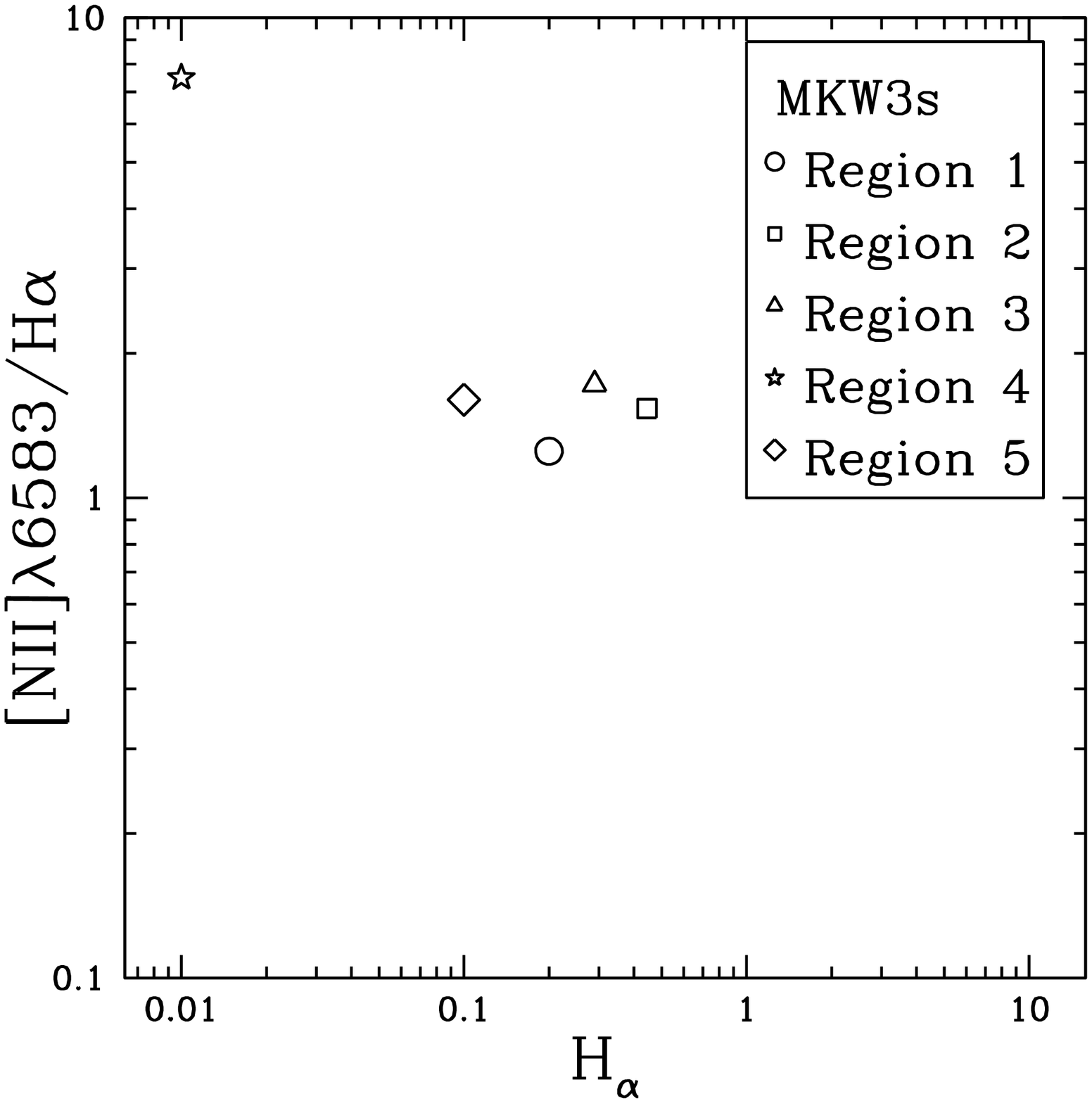}
        \end{minipage}%
    }
    \subfigure{
     \begin{minipage}[c]{0.5\textwidth}
        \centering
        \includegraphics[width=2.1in,angle=0]{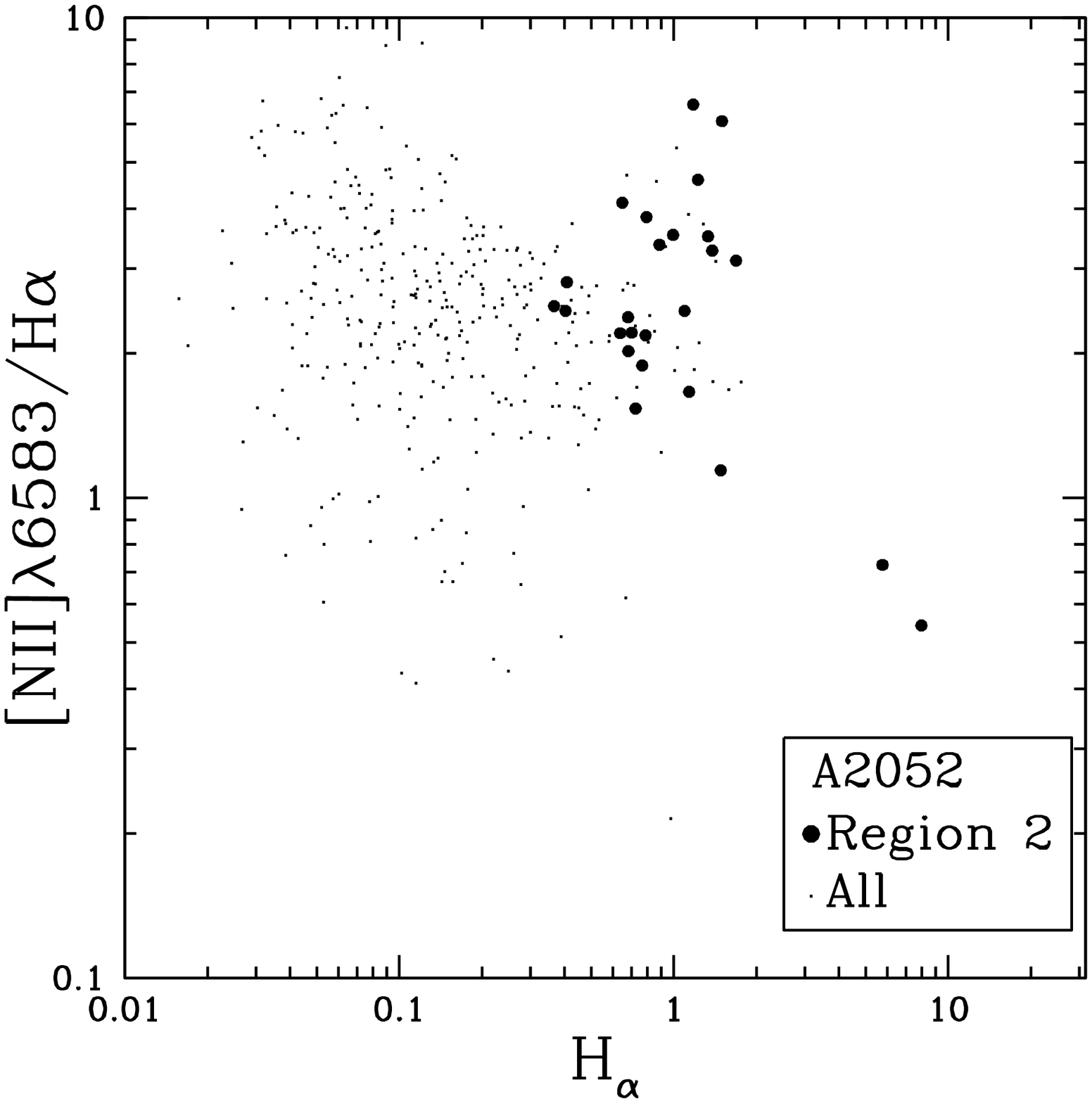}
         \end{minipage}%
    } \subfigure{
\begin{minipage}[c]{0.45\textwidth}
        \centering
        \includegraphics[width=2.1in,angle=0]{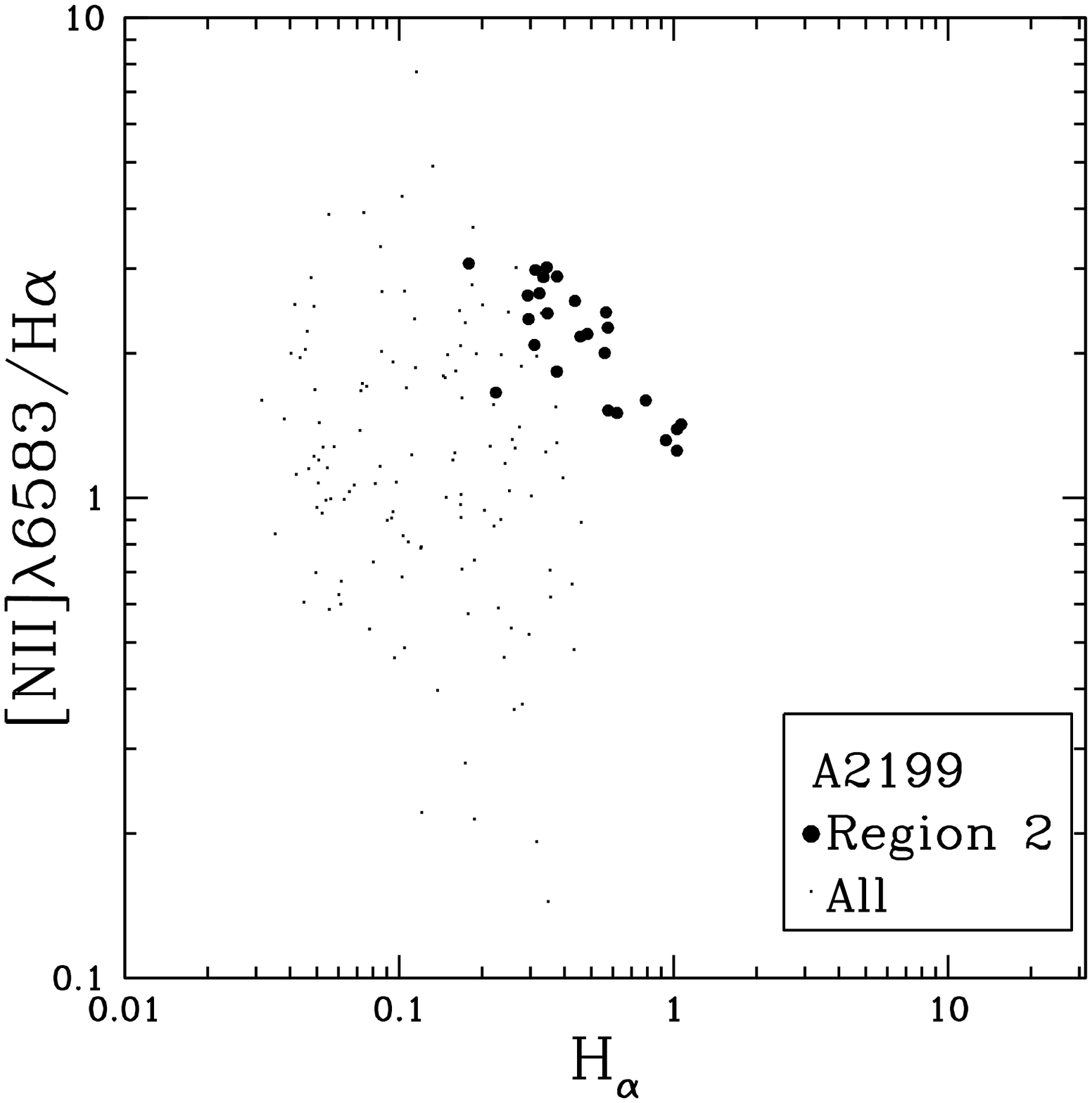}
        \end{minipage}%
    }
    \subfigure{
     \begin{minipage}[c]{0.5\textwidth}
        \centering
        \includegraphics[width=2.1in,angle=0]{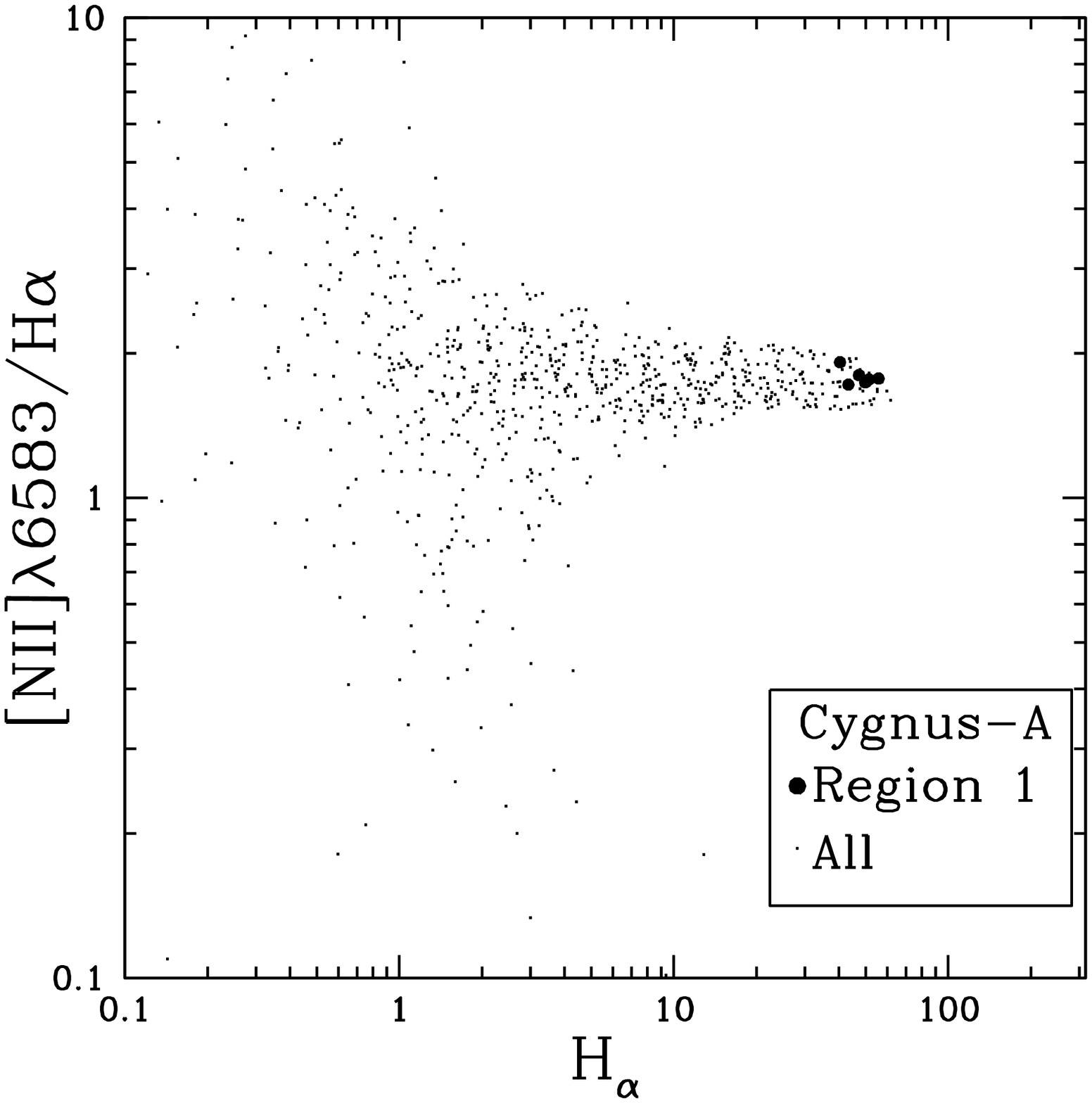}
         \end{minipage}%
    } 
    \caption[Emission mechanism and H$\alpha$ flux]{\bf Emission line mechanism. \it The [N~{\sc{ii}}] ~$\lambda$6584/H$\alpha$ ratio is shown. For NGC~3311 in Abell~1060 one pixel is $\sim$675$\,$pc$^{2}$, for the BCG of Abell~1204 one pixel is $\sim$0.07$\,$kpc$^{2}$\label{a1204n2ha}, for IC~4130 in Abell~1668 one pixel is $\sim$0.014$\,$kpc$^{2}$, for Object~B one pixel is $\sim$3100$\,$pc~$^{2}$\label{n2haophi}, for NGC~5920 in MKW3s individual pixels can not be plotted. The region size is $\sim$7.29$\,$kpc~$^{2}$. For UGC~9799 in Abell~2052 one pixel is $\sim$0.032$\,$kpc$^{2}$, for NGC~6166 in Abell~2199 one pixel is  $\sim$0.023$\,$kpc~$^{2}$, and for Cygnus-A one pixel is $\sim$0.048$\,$kpc$^{2}$. A simplistic cut for AGN vs. SF is at y=0.63 (see also Figure~\ref{bptCA}).\label{pplots} \label{1204lha} \label{o3hbCA}  \label{n2haCA} \label{han22052}\label{n2ha1668}\label{n2ha}\label{han22199}}

\end{figure*}

Before discussing the details of each system individually, we offer the reader a quick summary.
\begin{itemize}
\item All cluster BCGs show smooth continuum emission, except in the case of Abell~1060. Although the continuum emission is usually smooth, the morphology of the line emission is not uniform throughout the sample. Two BCGs show filamentary emission (in Abell~1668 and MKW3s). There are also two clusters for which the BCG harbours extended emission (Abell~1204 and Abell~2199), as well as two in which the BCG line emission is condensed (Abell~2052 and Cygnus-A). In the cases of Abell~1060 (in the BCG) and Ophiuchus (just outside of the BCG), there are large patches of dust and the line emission follows a similar morphology. Two BCGs show no evidence for line emitting gas (that in Abell~1651 and in Ophiuchus).
\item The line kinematics usually vary smoothly in terms of velocity and line width. At times rotation is clearly present (Abell~1060 and Cygnus-A). Also, bulk motions (Abell~1204, Abell~1668, and Ophiuchus) and outflows (MKW3s) are observed. The subtraction of the absorption spectrum for MWK3s allows us to see a clear velocity shift of $\sim$12$\,$\AA~(550$\,$km$\,$s$^{-1}$) for the emission lines.
\item Except for the BCG in Ophiuchus and in Abell~1651, emission lines of H$\alpha$ and [N~{\sc{ii}}]~$\lambda\lambda$~6548,6584,  are prominent (Figure~\ref{specpap}). For the three galaxies observed with OASIS, there also exist spectra around the H$\beta$ line. The 1$\sigma$ noise level usually dominates any emission or absorption in H$\beta$ for Abell~2052 and Abell~2199. However, the [O~{\sc{iii}}]~$\lambda$~5007 emission rises above the 5$\sigma$ level in central region (Region~2). Most of the galaxies show AGN-like spectra as summarized in Table~\ref{sumtab}, with very strong forbidden lines with respect to the Balmer emission. Figure~\ref{pplots} shows this for all galaxies with H$\alpha$ emission. These lines also broaden in some regions of Abell~1204 and Cygnus-A. The BPT diagram of Cygnus-A confirms a Seyfert nucleus in this system. The lines in Object~B of Ophiuchus are well described by ionization from a hard source, such as an AGN, but Object~B is not the BCG. In the case of Abell~1060 and Abell~1204 for which the emission line ratios show very strong H$\alpha$ emission with respect to the [N~{\sc{ii}}]~$\lambda$~6584 lines, we may be detecting a young stellar population. We determine the SFR for these last two.
\end{itemize}

\subsection{NGC~3311 in Abell~1060}

\begin{figure*}
 \subfigure{
\begin{minipage}[c]{0.33\textwidth}
        \centering
        \includegraphics[width=2.5in,angle=0]{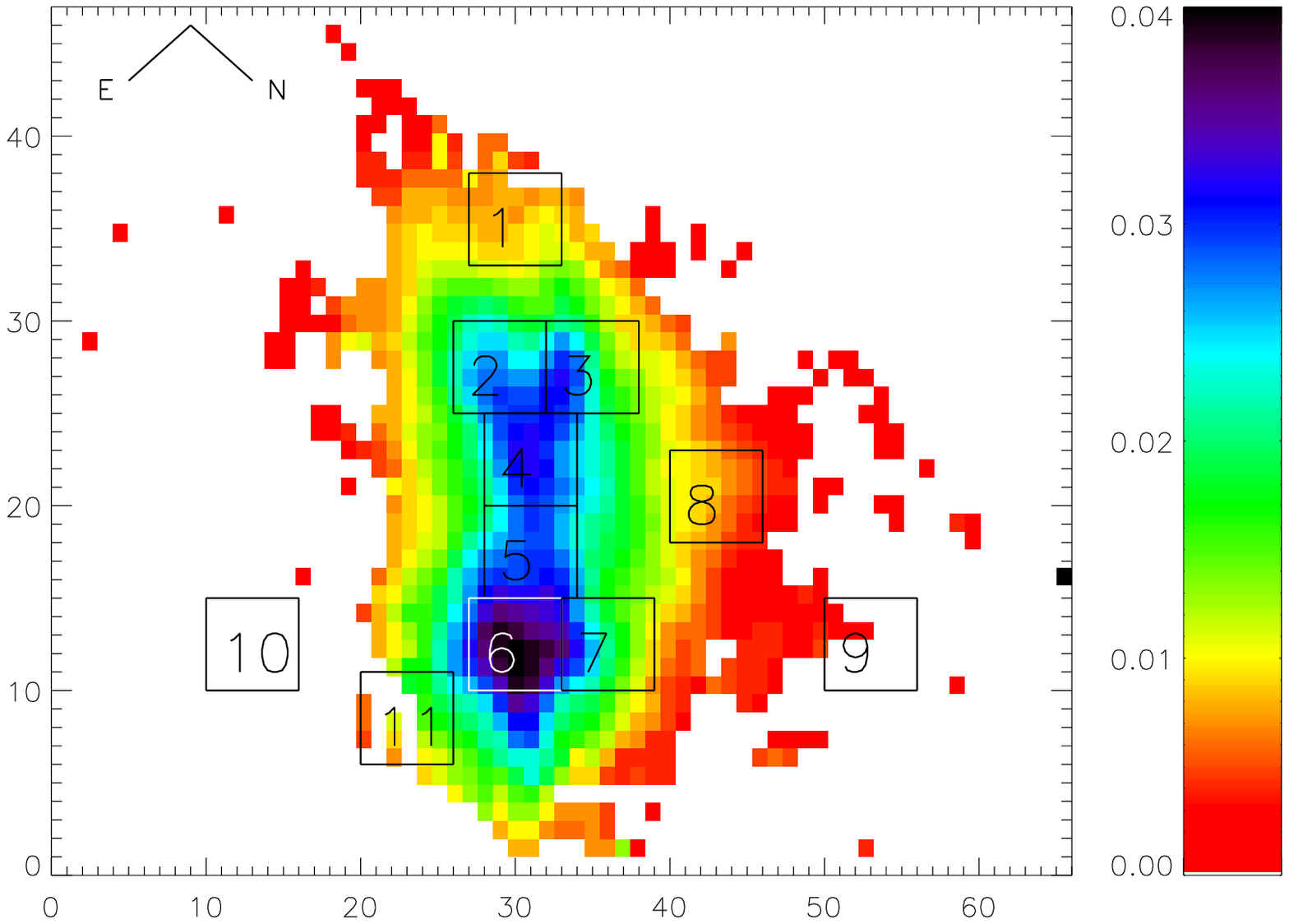}
        \end{minipage}%
    }
    \subfigure{
     \begin{minipage}[c]{0.33\textwidth}
        \centering
        \includegraphics[width=2.5in,angle=0]{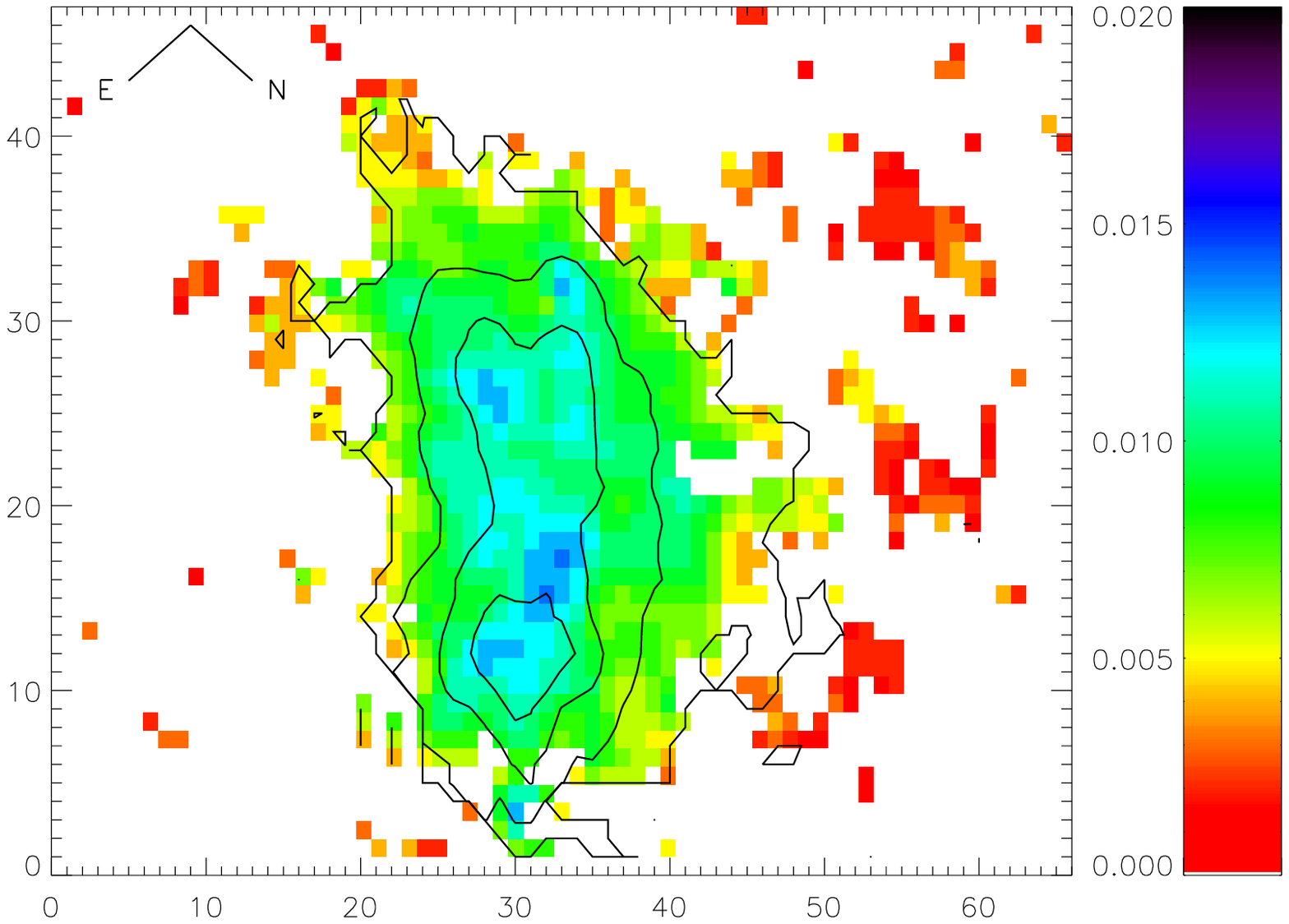}
         \end{minipage}%
    }
        \subfigure{
     \begin{minipage}[c]{0.3\textwidth}
        \centering
        \includegraphics[width=2.5in,angle=0]{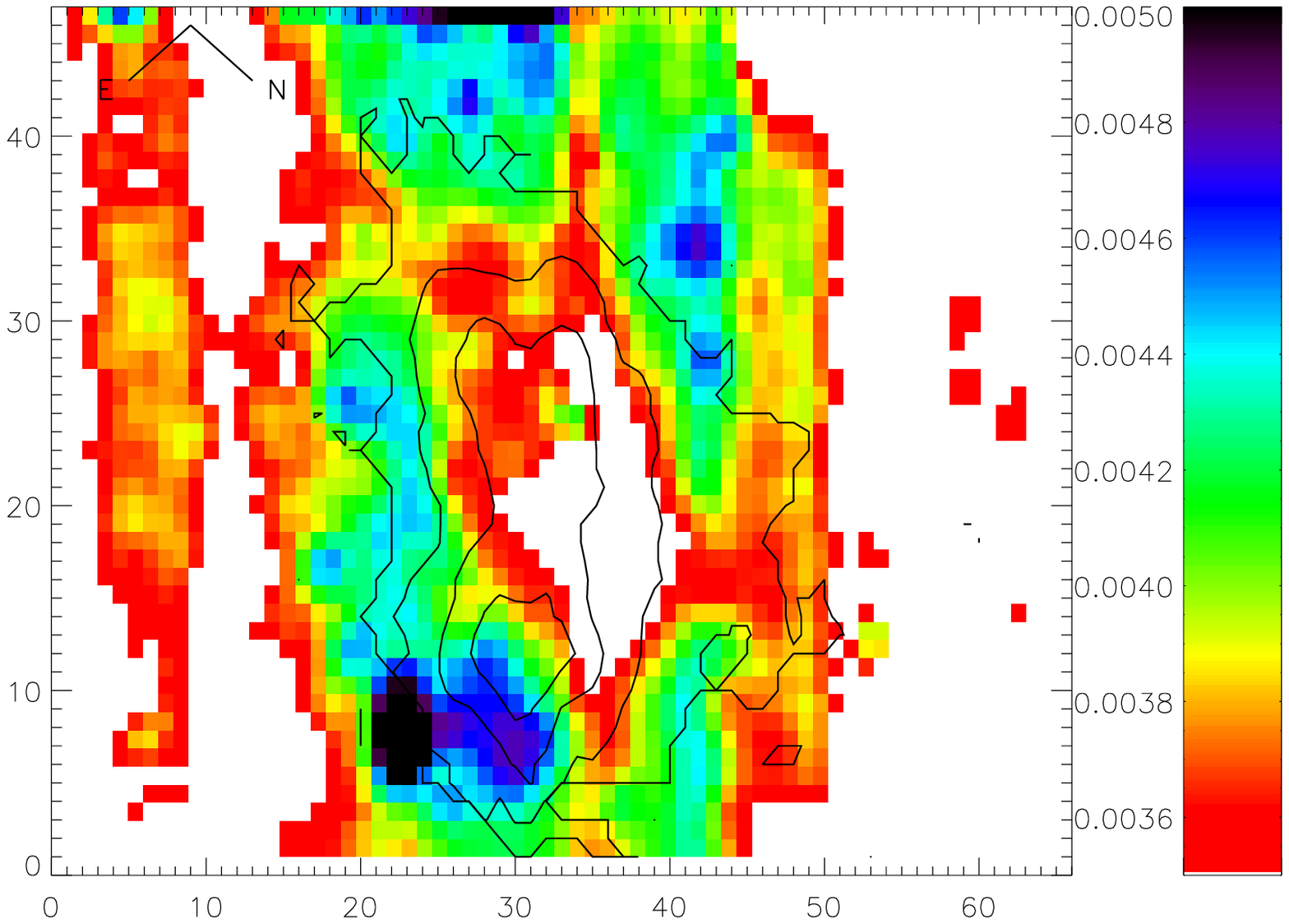}
         \end{minipage}%
    }
\caption{\bf NGC~3311 in Abell~1060. \it From left to right, the continuum subtracted H$\alpha$ emission flux, the continuum subtracted [N~{\sc{ii}}]~$\lambda$~6584 emission flux, and the continuum near the H$\alpha$ emission line. Both emission line maps follow a similar morphology overall and much of the emission is found within regions of strong extinction in the continuum. In the continuum image, a dust patch in the N-S direction is clearly visible. The vertical bands seen on the left and right sides of the figure are associated to fringes. The H$\alpha$ emission is overlain as contours. The regions used in the analysis are represented as boxes. The images are in units of 10$^{-15}$$\,$erg$\,$s$^{-1}$$\,$cm$^{-2}$$\,$\AA$^{-1}$. One pixel is $\sim$26$\,$pc across.  \label{a1060hai}  \label{an2} \label{a1060cnt}}
  \end{figure*}

  {\bf Morphology} - Figure~\ref{a1060hai} shows the images of the H$\alpha$ and [N~{\sc{ii}}] emission lines, as well as the continuum. Note that for this case the [OI] emission lines near 6300$\,$\AA~do not rise above the 1$\sigma$ noise level and therefore we use the median throughout a larger window (6300 and 6500$\,$\AA) to build the continuum map. 
  
 \citet{vas91} and \citet{lai03} both note a large dust patch that corresponds in position to the obscuring feature going from the SW to the NE seen in the continuum image. In comparing the HST I-band image of the center of the BCG from \citet{lai03}, we notice the striking similarity both in extent and morphology between the dust patch they identify, and the regions we identify as strongly emitting in H$\alpha$. The H$\alpha$ contours on the continuum image indicate that much of the H$\alpha$ emission is confined within the dust, and therefore much H$\alpha$ emission could be obscured. Overall, the bright regions in [N~{\sc{ii}}] follow those seen in the H$\alpha$ emission image. The image of [S~{\sc{ii}}] $\lambda$~6716 (not shown) follows the same overall morphology.

\begin{figure*}
 \subfigure{
\begin{minipage}[c]{0.45\textwidth}
        \centering
        \includegraphics[width=2.5in,angle=0]{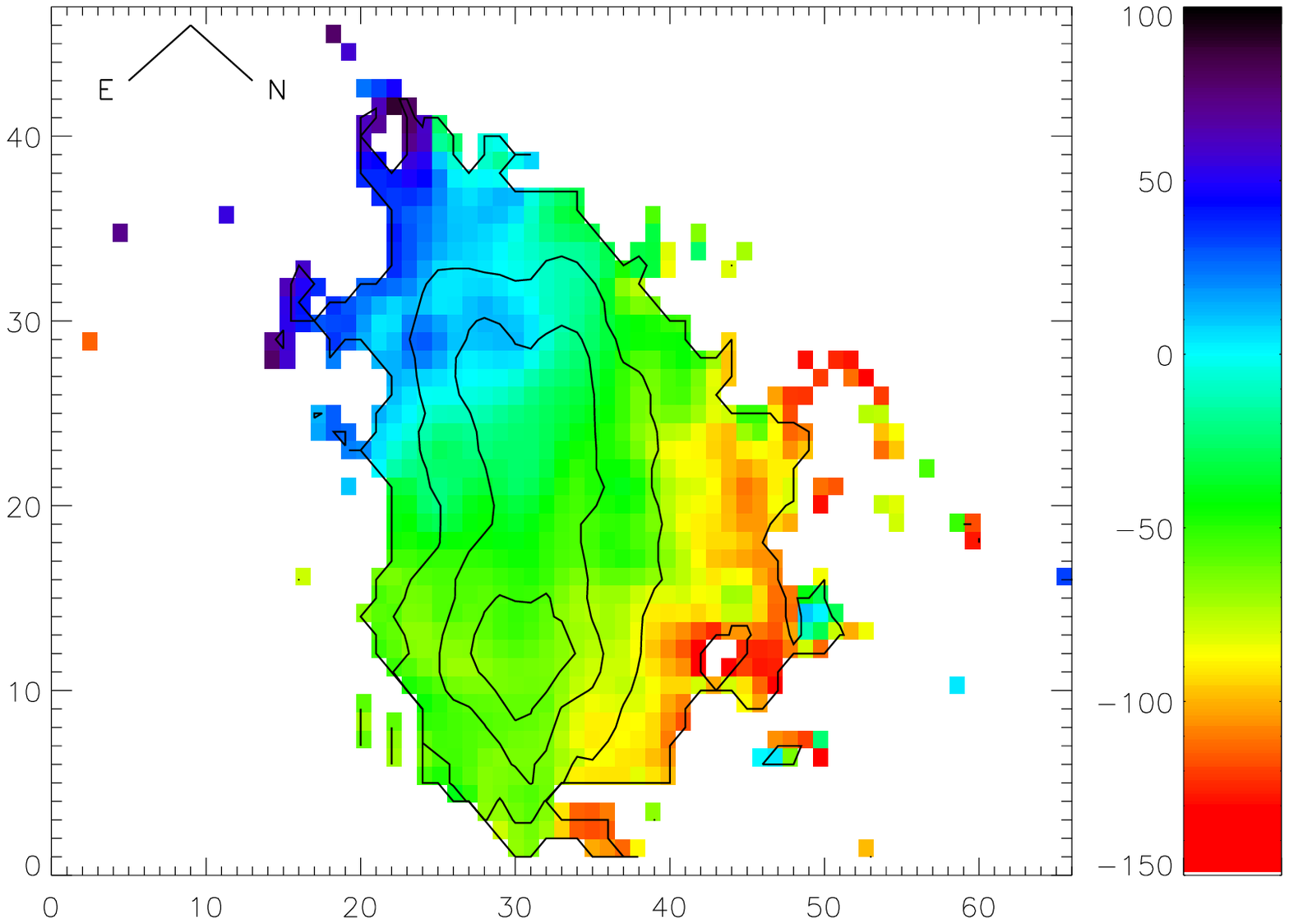}
        \end{minipage}%
    }
    \subfigure{
     \begin{minipage}[c]{0.5\textwidth}
        \centering
        \includegraphics[width=2.5in,angle=0]{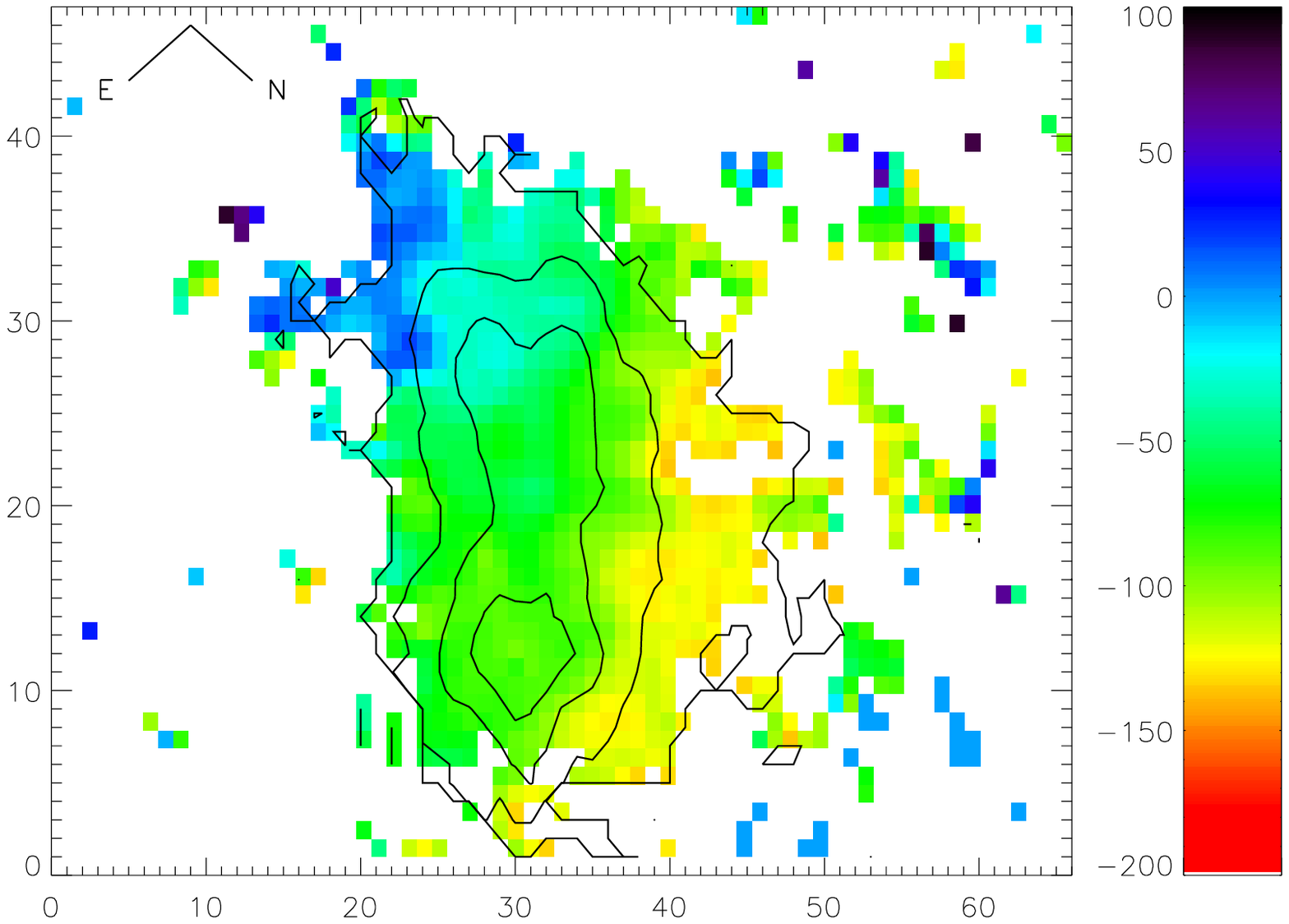}
         \end{minipage}%
    }
\caption[Relative velocities for NGC~3311 in Abell~1060]{\bf Relative velocities for NGC~3311 in Abell~1060. \it \underline {Left Panel:} Map of the H$\alpha$ relative velocity for NGC~3311 in Abell~1060.  \underline {Right Panel:} Map of the [N~{\sc{ii}}] $\lambda$6584 relative velocity for NGC~3311 in Abell~1060. Both maps are consistent with rotating gas.\it The H$\alpha$ emission is overlain as contours and the scale is in units of km$\,$s$^{-1}$. One pixel is $\sim$26$\,$pc across.\label{n2vel}\label{havel}}
\end{figure*}

{\bf Kinematics} - Figure~\ref{havel} shows maps of the relative velocity for the H$\alpha$ and [N~{\sc{ii}}]~$\lambda$~6584 emitting gas. The appearance of both emission lines is smooth, and we measure a velocity shear of 100$\pm$20$\,$km$\,$s$^{-1}$ across $\sim$1.0$\,$kpc. There is a clumpy distribution in the NaD velocity (clumps of -100 to +100$\,$km$\,$s$^{-1}$), originating from the underlying galaxy, but no shear. The emission line widths range from 130$\,$km$\,$s$^{-1}$ to 200$\,$km$\,$s$^{-1}$, close to the resolution, and have a clumpy distribution (available in Edwards 2007). The velocity shear, low velocity values, and lack of structure in line widths are an indication of rotation of the line emitting gas.

{\bf Emission Diagnostics} - It is clear from the spectrum presented in Figure~\ref{specpap} that none of the regions in NGC~3311 show ratios of significant AGN contamination. Figure~\ref{pplots} shows the line ratios of individual pixels and highlights the area of Region~6, where the H$\alpha$ emission is the strongest, and the line ratios are the lowest, falling well below the AGN/SF cut of 0.63.  Based on the line ratios [N {\sc{ii}}] $\lambda$ 6584/H$\alpha$ and also ([S~{\sc{ii}}]~$\lambda$~6716~+~$\lambda$~6731)/H$\alpha$,  most  regions confidently fall into the composite, or star formation section of the BPT diagram.  The ([S~{\sc{ii}}]~$\lambda$~6716~+~$\lambda$~6731)/H$\alpha$ are not shown here, but all regions have a ratio less than 0.25, much below 0.40, which is typical of an AGN or LINER. The image of the [N~{\sc{ii}}]~$\lambda$~6584/H$\alpha$ across the center of the galaxy, see \citet{edwthesis}, shows the central regions to have the lowest [N~{\sc{ii}}]~$\lambda$~6584/H$\alpha$ ratios.  Thus, it is fairly certain that the ionization is not a result of a hard radiation field supplied by an AGN or LINER.

\nocite{fer08}
{\bf Star Formation Rate and Stellar Populations} - Once the possibility of an AGN has been discarded, a common ionization mechanism explored is the effect of a young population of hot stars, which can produce strong Balmer emission \citep{cra99,don00,wil06}.  Another possibility recently calculated by Ferland~et~al.~2008 is non-radiative heating by cosmic rays which can produce the observed molecular hydrogen lines (at least in the thin filaments of Perseus). \citet{fab08} have shown that star formation within the thin filaments of Perseus may be delayed by magnetic fields. However, at the larger distances of our clusters, it is unlikely that the majority of the flux we observe is from such thin surrounding structures, and it is likely dominated by the bright emission at the center of the BCG. Hence, for this paper, we explore the properties of a young population of stars that could excite the observed H$\alpha$ emission.

    \begin{figure}
  \centering
  \epsfxsize=3in
     \epsfbox{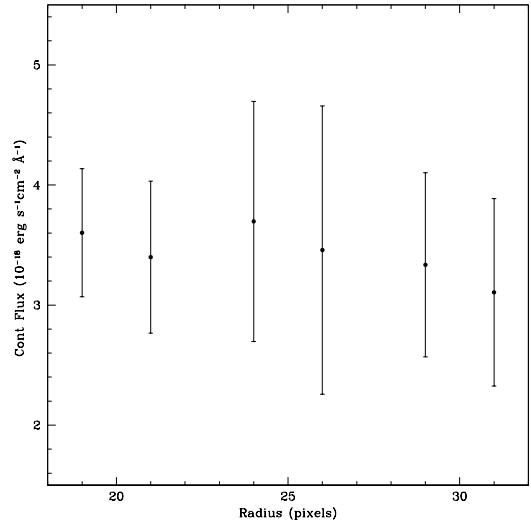}
   \caption[Radial cut of the underlying population for NGC~3311 in Abell~1060]{\bf Radial cut of the underlying population for NGC~3311 in Abell~1060. \it Based on the continuum flux near H$\alpha$. The x-axis shows the distance from the center of the IFU frame. The errorbars show the standard deviation of the continuum values at the distance from the IFU center. There is no evidence for a significant trend with radius and most of the points are equal to each other. \label{1060oldge}}  
\end{figure}

 A lower limit on the total SFR has been calculated using the method of \citet{ken98}. Excluding the dependence on the cosmology chosen, the errors in SFR are most sensitive to the error in H$\alpha$ equivalent width measurement. The intensity of H$\alpha$ emission can be diminished by the presence of H$\alpha$ in absorption. The absorption comes not only from the same ionizing  population of young stars, but also from the underlying and massive older stellar population of the galaxy. Although the absorption is below the 1$\sigma$ level in the individual pixel spectra, it can be removed in the integrated spectrum of the totality of H$\alpha$ emitting pixels. We describe this presently.

\begin{figure*}
 \centering
\epsfxsize=6in
 \epsfbox{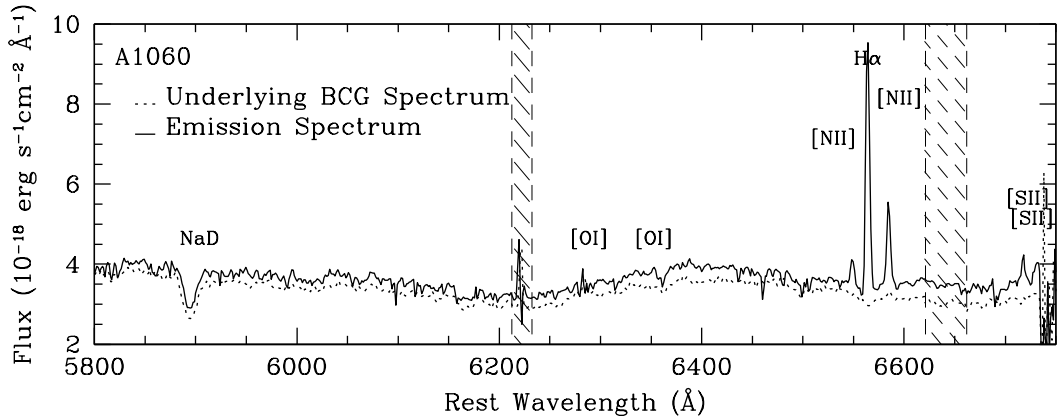}  
   \caption[Average spectra of the emitting and underlying population of NGC~3311 in Abell~1060]{\bf Average spectra of the emitting and underlying population of NGC~3311 in Abell~1060. \it A median average of H$\alpha$ emitting spectra are shown alongside the average spectrum for the underlying population in the regions surrounding the intense emission. The hashed regions are those affected by the chip gap and poorly subtracted sky lines.\label{1060hatots}}
\end{figure*}

{\it Absorption by the Underlying Population} - The emission appears only in the central 20$^{\prime\prime}$$\times$35$^{\prime\prime}$ (500pc$\times$900pc for Abell~1060) of the IFU field of view, thus, a median of the underlying galaxy spectrum can be isolated by ignoring these central pixels with emission. We assume the older population has the same characteristics as (age and metallicity), and is the dominant luminosity source for the overall luminosity profile of the optical image (that is, over the entire r~filter). Therefore, the absorption should scale to this profile. Figure~\ref{1060oldge} shows the average continuum flux in 5 pixel-wide radial bins where the errorbars show the standard deviation of continuum values in the bin. The region 15-35$\,$pixels ($\sim$400 to 900$\,$pc) from the center is plotted. The flux varies less than 10\%, from 3.1 to 3.5$\times$10$^{-18}$$\,$erg$\,$cm$^{-2}$$\,$s$^{-1}$$,$\AA$^{-1}$, showing no (or only a weak) trend with radius.  Therefore, we simply subtract the integrated absorption spectrum form the integrated emission spectrum (scaled to the same projected physical area). The average of both these spectra are shown in Figure~\ref{1060hatots}.

From Figure~\ref{1060hatots} we measure the integrated H$\alpha$ absorption to be 3.2$\pm$0.4~$\times$~10$^{-17}$$\,$erg s$^{-1}$$\,$cm$^{-2}$ per spectrum, on average. With the absorption spectrum in hand, it is straightforward to calculate an age and metallicity of the underlying stellar population by comparing to the results of the population synthesis code of \citet{mol00}.

For Abell~1060 the NaD line is available. From the integrated spectrum of non-H$\alpha$ emitting pixels, an equivalent width of 4.5$\pm$0.4$\,$\AA, for the absorption is measured. Within our model, this strong absorption can only be fit with a super-solar metallicity (2$\,$Z$_{\odot}$). An age of 1.0$\pm$0.6$\,$$\times$$\,$10$^{10}$$\,$yr provides the best match. The error is based on the subset of models which fit the equivalent width within its measurement error. An estimate of the mass is made by scaling the continuum level (per solar mass) at 6400$\,$\AA~of the best model to the observed continuum level of the old population. For Abell~1060, 2.0$\pm$0.8~$\times$~10$^{8}$$\,$M$_{\odot}$$\,$kpc$^{-2}$ are required in order to match the observations. Table~\ref{sumtab} lists the age and mass of the older stellar population for this, and the other clusters where NaD, Fe~I, Mg$_{b}$, or H$\beta$ absorption lines are available (we observe no absorption lines in Abell~1204, Abell~1651, and Cygnus~A; details in the derivation of the old stellar populations for all the BCGs can be found in \citet{edwthesis}). Lower limits for the old population are quoted for Abell~1668 and Abell~2199 as low level H$\alpha$ is observed in every pixel within the field of view.

{\it Star Formation Rate} - We measure a total H$\alpha$ flux of 5.0$\pm$0.3~$\times$~10$^{-15}$$\,$erg~s$^{-1}$$\,$cm$^{-2}$, or an W$_{o}$(H$\alpha$)~=~$-$106$\,$\AA, in the absorption corrected integrated spectrum from all of the H$\alpha$ emitting pixels. This results in a total SFR of 1.4$\pm$0.1~$\times$~10$^{-2}$$\,$M$_{\odot}$$\,$yr$^{-1}$. The SFR density is 1.2$\pm$0.1~$\times$~10$^{-7}$$\,$M$_{\odot}$$\,$yr$^{-1}$$\,$pc$^{-2}$. The SFR derived is surely a lower limit as Figure~\ref{a1060cnt} suggests, and a global measure of the dust absorption in the galaxy has already been accounted for. Assuming that most of the obscured emission is behind this lane, a reasonable estimate on an upper limit to the SFR would be to suppose that all of the obscured emission has the same intensity as in Region~6 which is not behind the dust patch. Therefore, there should be no more than 1.7$\pm$0.1~$\times$~10$^{-2}$$\,$M$_{\odot}$$\,$yr$^{-1}$ in total, or, 1.5$\pm$0.1~$\times$~10$^{-7}$$\,$M$_{\odot}$$\,$yr$^{-1}$$\,$pc$^{-2}$. This small amount of star formation found is less than the one derived for cooling flow cluster BCGs. For example, \citet{hic05} found rates of 0.2-219$\,$M$_{\odot}$$\,$yr$^{-1}$ derived from UV excess for cooling flow BCGs, and \citet{edw07b} found typical SFRs of 0.3~-~1.6$\,$M$_{\odot}$$\,$yr$^{-1}$ for emitting BCGs. The original MDR of 6$\,$M$_{\odot}$$\,$yr$^{-1}$ found for Abell~1060 would be between 0.06 and 0.6$\,$M$_{\odot}$$\,$yr$^{-1}$ (1~-~2 orders of magnitude below the {\it ROSAT} value). However, the {\it Chandra} observations \citep{yam02} show no evidence for any central temperature drop, and so no cooling flow is present in this system. Thus, the fact that we have measured only a very small amount of star formation activity we derive is consistent with this picture.

{\it Properties of the Young Population} - We can further quantify the activity by estimating a metallicity, age and mass for the ionizing population. We start by calculating the metallicity of the gas, using abundance ratios as described in \citet{kew02}, and then we assume the metallicity of the young stellar population is the same as for the gas. Although this may not be exact, it will help us to restrict the solution for the age of the young stellar population using synthesis models.

It is possible to make a rough estimate of the metallicity in the regions by assuming an average ionization parameter, and comparing the ratios of [N~{\sc{ii}}]~$\lambda$~6584/([S~{\sc{ii}}]~$\lambda$~6716~+~$\lambda$~6731) and [N~{\sc{ii}}]~$\lambda$~6584/H$\alpha$. Although, as \citet{kew02} strongly caution, both ratios depend strongly on the ionization parameter and this method is not precise. Assuming the  ionization parameter is between average values of 5$\,$$\times$$\,$10$^{6}$ and 2$\,$$\times$$\,$10$^{7}$ \citep{kew02}, the metallicities from the [N~{\sc{ii}}]~$\lambda$~6584/([S~{\sc{ii}}]~$\lambda$~6716~+~$\lambda$~6731) ratios are compared against those from [N~{\sc{ii}}]~$\lambda$~6584/H$\alpha$. The latter generally has two possible values, therefore constraints from the former enable the correct point to be chosen, although large uncertainties exist for the [S~{\sc{ii}}]~$\lambda$~6731 line measurement. Considering the region within the central 0.60$^{\prime\prime}$~$\times$~0.70$^{\prime\prime}$ ($\sim$175$\,$pc across), we derive a metallicity of 12+log(O/H)~=~9.3$\pm$0.1 (approximately twice solar). This is high, but the mass-metallicity relation of \citet{tre04} shows higher metallicities in more massive galaxies. 

The age of the young stellar population is estimated using the evolutionary synthesis code Starburst99 \citep{lei99}. The H$\alpha$ equivalent width is calculated and matched to results from a run based on an instantaneous burst of star formation using a Salpeter IMF with masses between 1 and 100$\,$M$_{\odot}$. We use a model based on super-solar metallicities (Z=2$\,$Z$_{\odot}$) to best match the metallicity (although a solar metallicity does not significantly affect the age). The best fit model, experimenting with both a mass cut of M$_{up}$~=~100$\,$M$_{\odot}$ and 30$\,$M$_{\odot}$ suggests young ages ($<$10$\,$Myr) as expected for these ongoing bursts in which the youngest stellar populations dominate the spectral energy distribution. 

The Starburst99 code also lists the theoretical spectral energy distribution for a 1$\times$10$^{6}$M$_{\odot}$ burst as a function of wavelength. We convert the model continuum luminosity to a flux and scale to the continuum level of the absorption corrected emission spectrum, 6~$\times$~10$^{-17}$$\,$erg~s$^{-1}$$\,$cm$^{-2}$ (the 3$\sigma$ errorbars are $\pm$2~$\times$~10$^{-17}$$\,$erg~s$^{-1}$$\,$cm$^{-2}$). At 10$\,$Myr, this requires that the young population is 3$\pm$1$\times$10$^{6}$M$_{\odot}$. The masses are lower for younger ages and lower theoretical M$_{up}$, for example, at 5$\,$Myr with M$_{up}$=30$_{\odot}$ the mass is 1.0$\pm$0.3$\times$10$^{6}$M$_{\odot}$.

These results show that if the H$\alpha$ emission does indeed result from a young population of stars, it is consistent with a low level of star formation, which in turn is much less than that predicted by direct cooling from the ICM.

\begin{figure*}
 \subfigure{
\begin{minipage}[c]{0.33\textwidth}
        \centering
        \includegraphics[width=2.5in,angle=0]{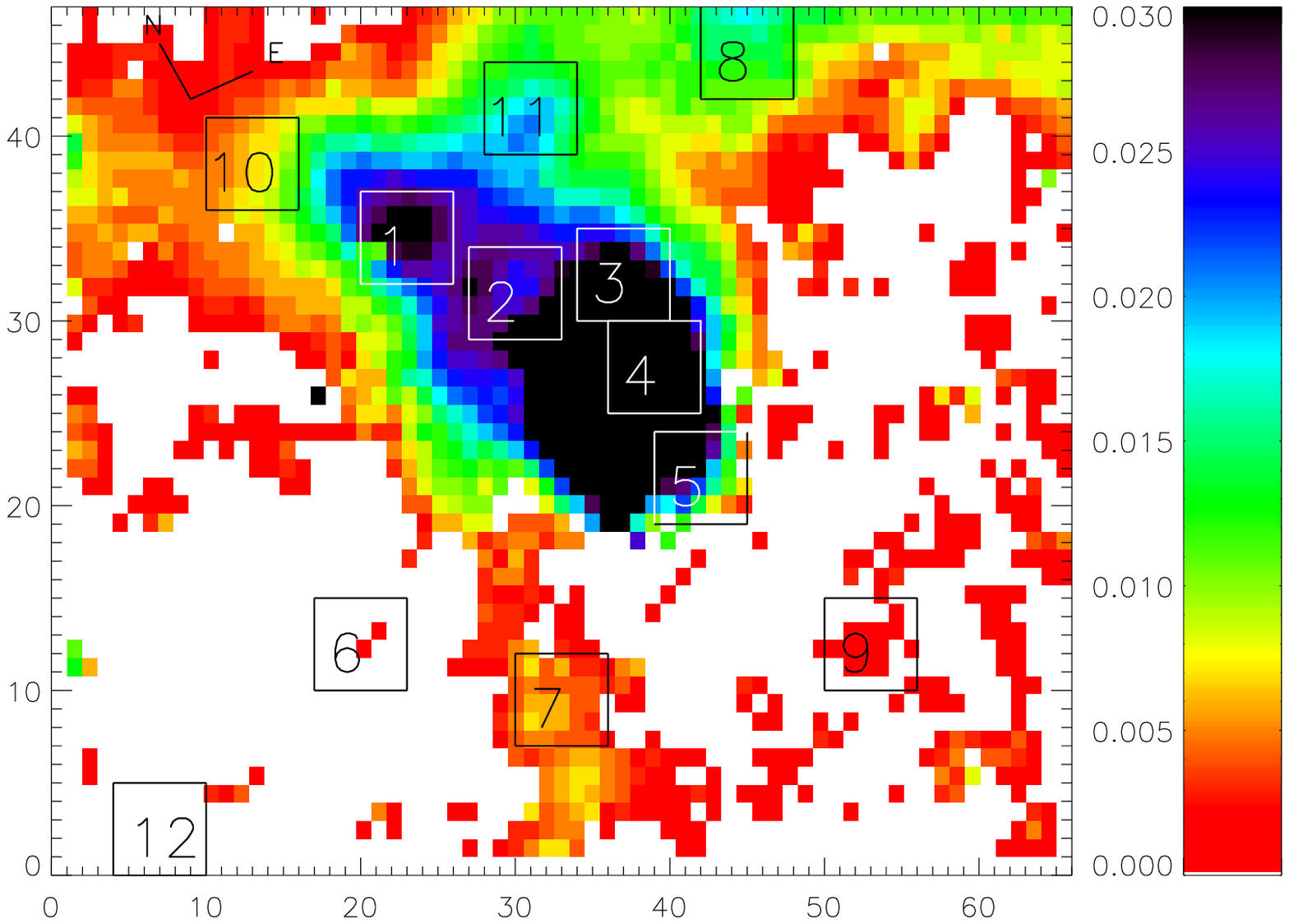}
        \end{minipage}%
    }
    \subfigure{
     \begin{minipage}[c]{0.33\textwidth}
        \centering
        \includegraphics[width=2.5in,angle=0]{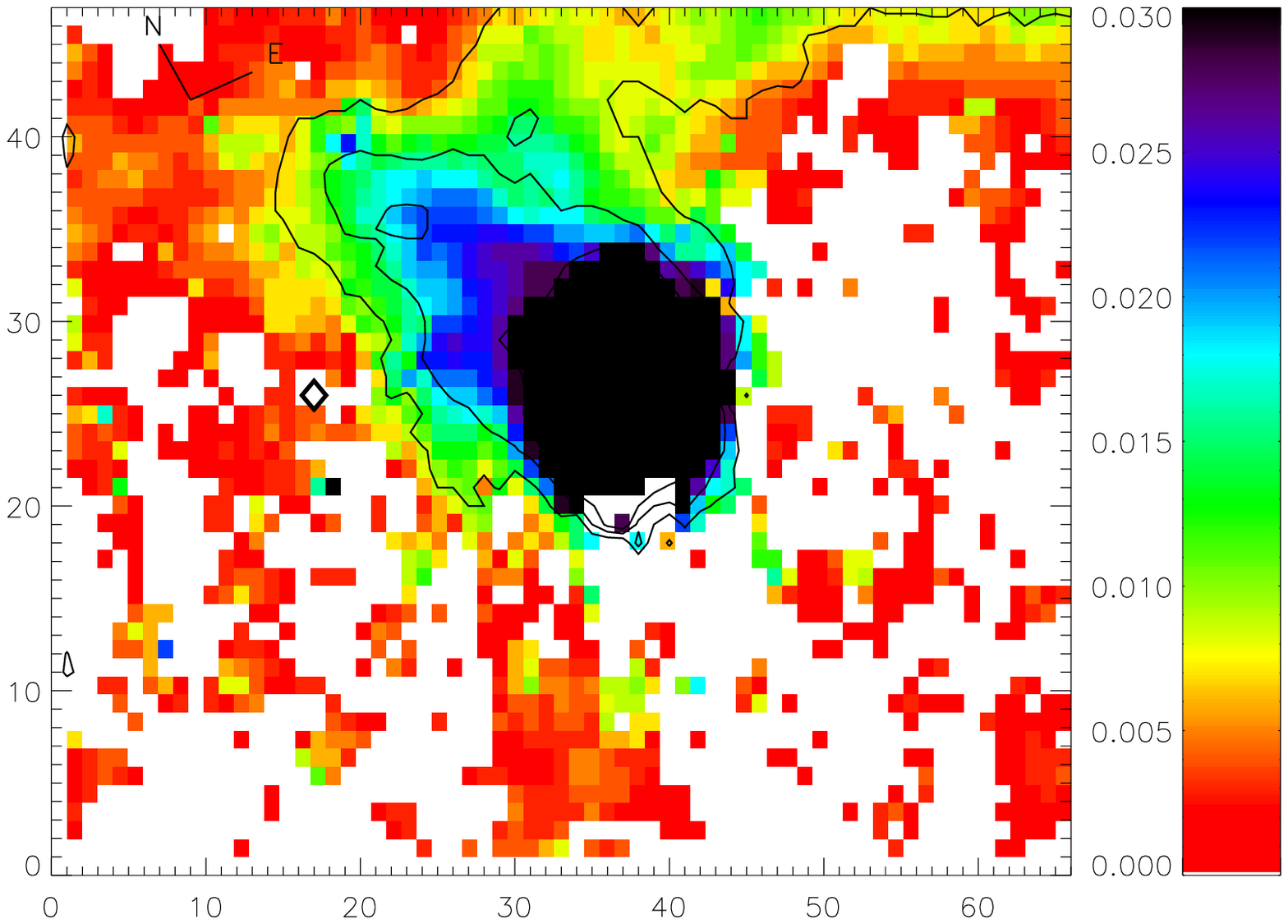}
         \end{minipage}%
    }
        \subfigure{
     \begin{minipage}[c]{0.3\textwidth}
        \centering
        \includegraphics[width=2.5in,angle=0]{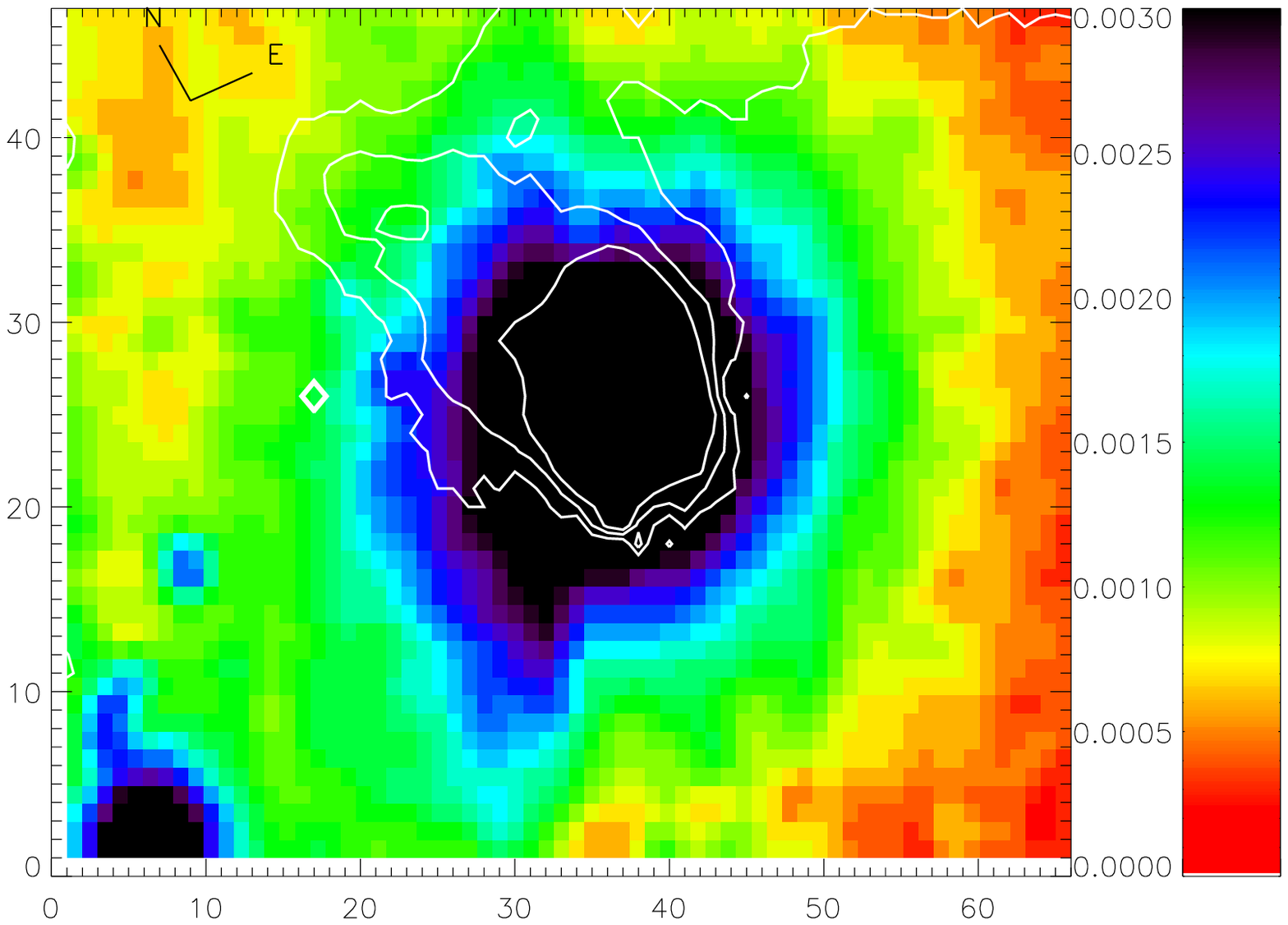}
         \end{minipage}%
    }
\caption{\bf BCG in Abell~1204. \it From left to right, the continuum subtracted H$\alpha$ emission flux, the continuum subtracted [N~{\sc{ii}}]~$\lambda$~6584 emission flux, and the continuum near the H$\alpha$ emission line. The Eastern plume of emission extends towards several possible neighbouring galaxies seen on the acquisition image. The H$\alpha$ emission is overlain as contours. The regions used in the analysis are represented as boxes. The images are in units of 10$^{-15}$$\,$erg$\,$s$^{-1}$$\,$cm$^{-2}$$\,$\AA$^{-1}$. One pixel is $\sim$270$\,$pc across.  \label{a1204reg}\label{a1204n2}\label{a1204s2}}
  \end{figure*}

\begin{figure*}
 \subfigure{
\begin{minipage}[c]{0.45\textwidth}
        \centering
        \includegraphics[width=2.5in,angle=0]{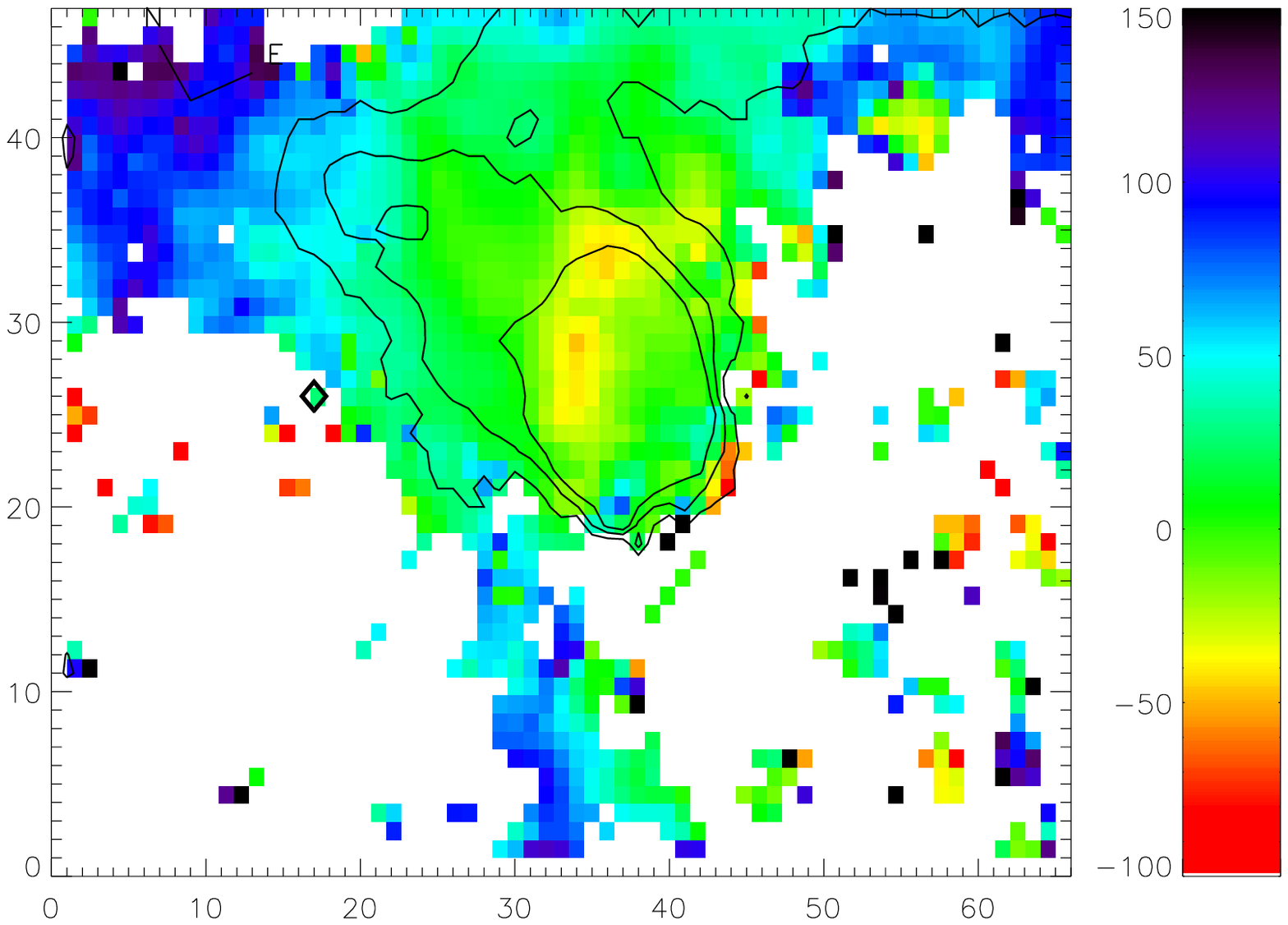}
        \end{minipage}%
    }
    \subfigure{
     \begin{minipage}[c]{0.5\textwidth}
        \centering
        \includegraphics[width=2.5in,angle=0]{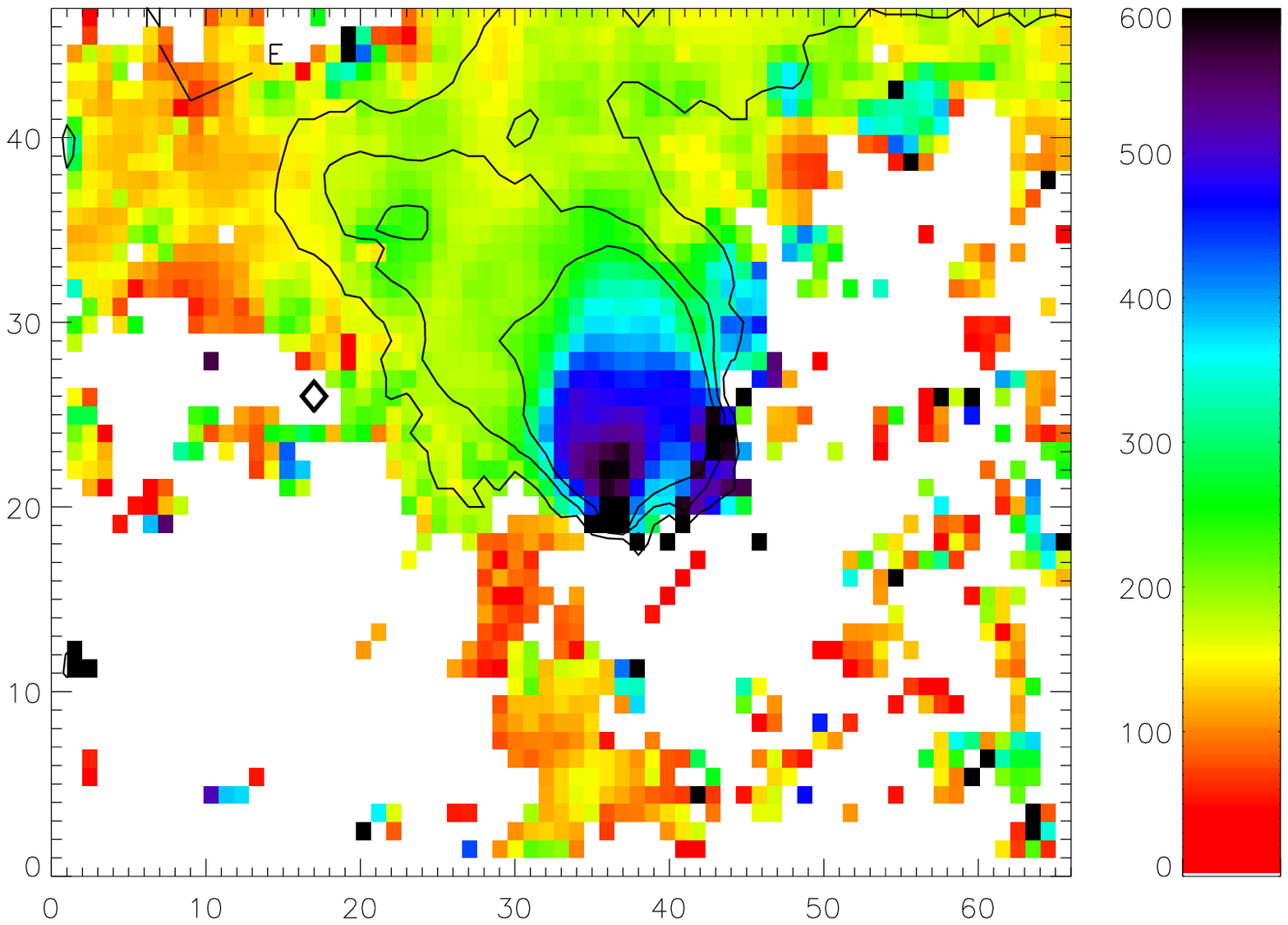}
         \end{minipage}%
    }
\caption[Maps of the kinematics for the BCG in Abell~1204]{\bf Maps of the kinematics for the BCG in Abell~1204. \it \underline {Left Panel:}  Map of the H$\alpha$ relative velocity  \underline {Right Panel:} Map of the H$\alpha$ FWHM. No obvious ordered motion is seen in the velocity map, however emission near the galaxy center is blueshifted with respect to the average velocity. The FWHM increases closer to regions of higher H$\alpha$ to [N~{\sc{ii}}] ratios and closer to the AGN. \it The H$\alpha$ emission is overlain as contours and the scale is in units of km$\,$s$^{-1}$. One pixel is $\sim$270$\,$pc across. \label{havel1204}}
\end{figure*}

\subsection {The BCG in Abell~1204}

{\bf Morphology} - Figure~\ref{a1204reg}, shows the images of the continuum subtracted H$\alpha$, [N~{\sc{ii}}]~$\lambda$~6584, and the continuum near the H$\alpha$ emission line for the BCG. The line images share the position of the peak intensity with that of the continuum emission (including [S~{\sc{ii}}] $\lambda$~6716 which is not shown). They also reveal the existence of a plume of bright H$\alpha$ emission which extends from the central peak to the North, and then towards the East out to the edge of the image. This is in the direction towards several smaller galaxies seen on the acquisition image (Figure~\ref{aquiims}). There is what appears to be a smaller galaxy in the western corner of the continuum image which has no counter part in the emission line images.  

{\bf Kinematics} -  Figure~\ref{havel1204} shows maps of the relative velocity and FWHM for the H$\alpha$ emitting gas (with the H$\alpha$ emission overlain as contours). Regions of slight blueshifting, as well as redshifting are apparent. The physical scale is from $-$100$\,$km$\,$s$^{-1}$ to $+$150$\,$km$\,$s$^{-1}$, with the most negative values echoing the structure of the H$\alpha$ emission. There is no obvious ordered motion such as the rotation seen in NGC~3311 of Abell~1060. There is however some structure in the relative velocity map as the central emission is clearly blueshifted with respect to the rest of the emission (by up to 200$\,$km$\,$s$^{-1}$). The mean FWHM across the field of view is 130$\,$$\pm$50km$\,$s$^{-1}$, but broadens to 570$\,$$\pm$70$\,$km$\,$s$^{-1}$ at the center, in Region~4. The velocity differences are not as high seen in strong outflows, but this region hosts the highest [N~{\sc{ii}}]~$\lambda$~6584/H$\alpha$ ratios, suggesting that the emission lines are ionized by an AGN (to be discussed further below). The line widths would be consistent with the central region being closer to the broad line region of the AGN or LINER. 

\begin{figure*}
 \subfigure{
\begin{minipage}[c]{0.33\textwidth}
        \centering
        \includegraphics[width=2.5in,angle=0]{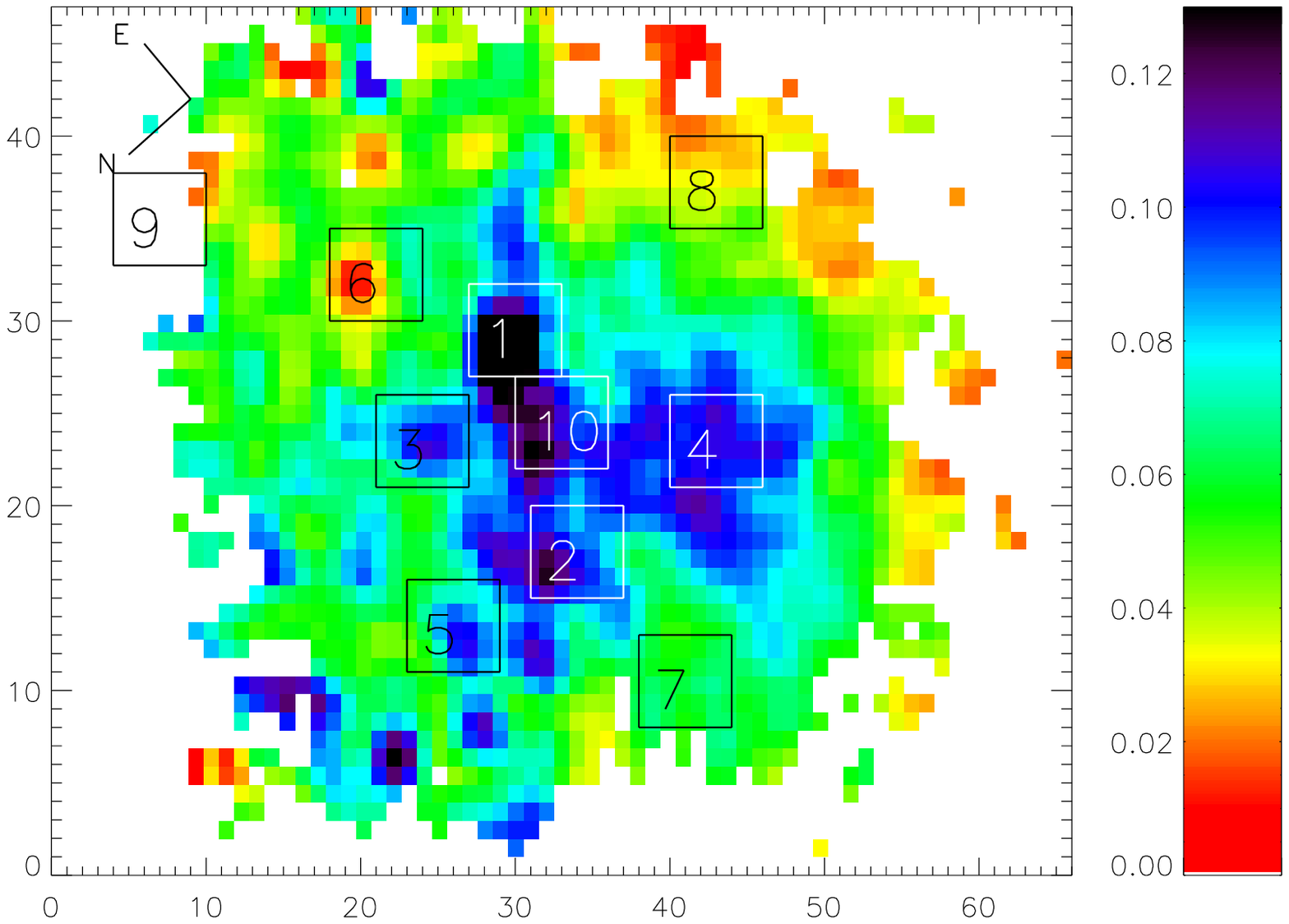}
        \end{minipage}%
    }
    \subfigure{
     \begin{minipage}[c]{0.33\textwidth}
        \centering
        \includegraphics[width=2.5in,angle=0]{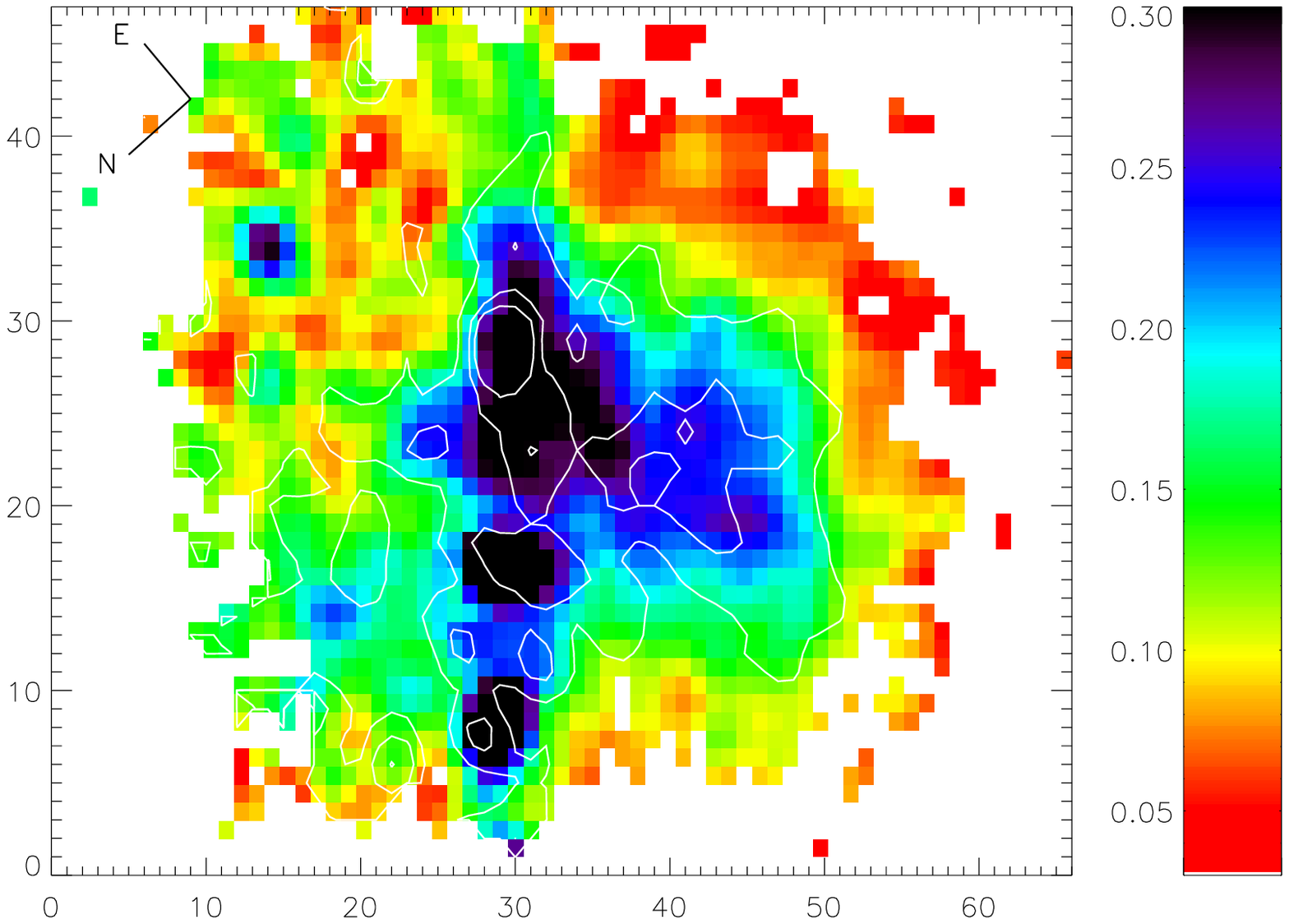}
         \end{minipage}%
    }
        \subfigure{
     \begin{minipage}[c]{0.3\textwidth}
        \centering
        \includegraphics[width=2.5in,angle=0]{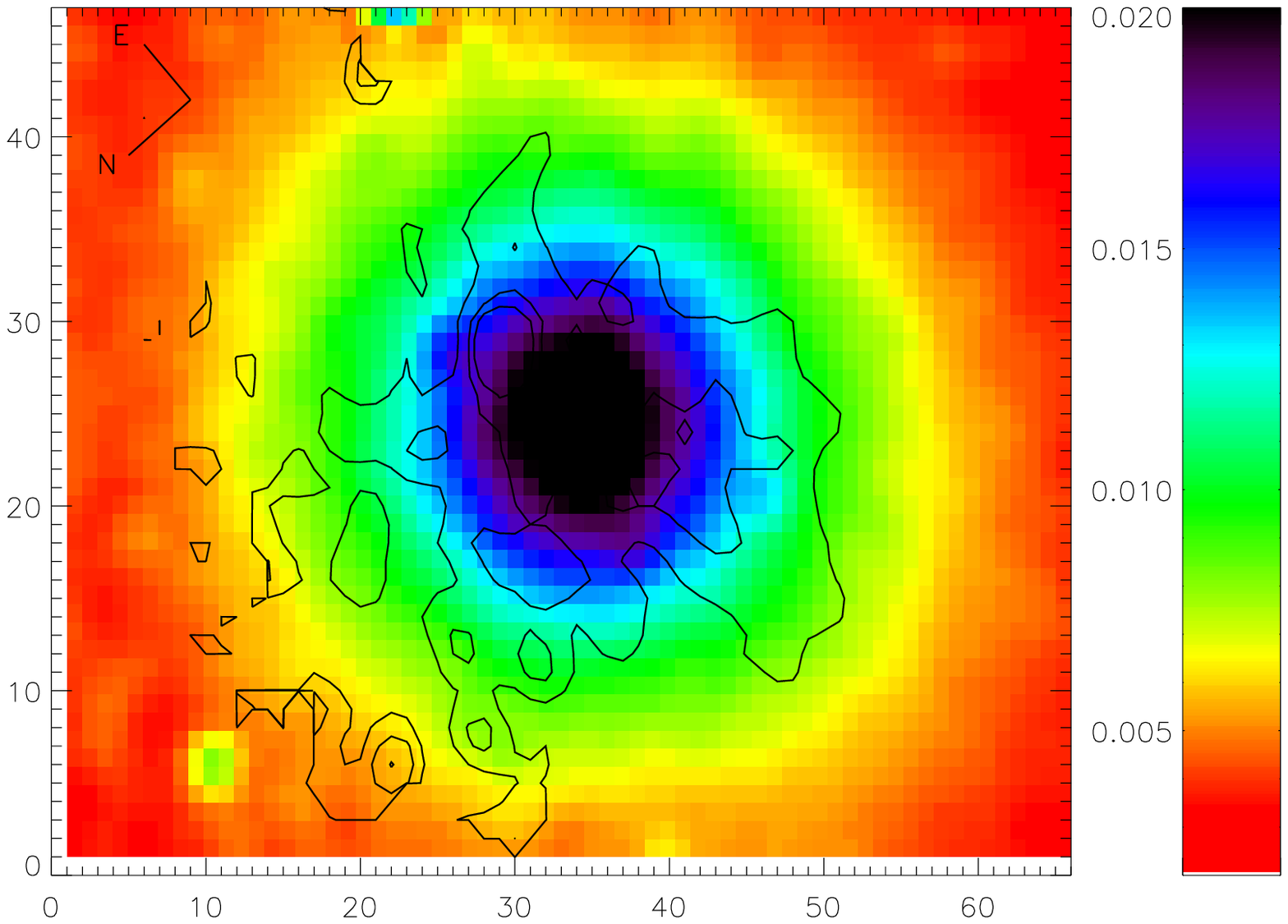}
         \end{minipage}%
    }
\caption{\bf  IC~4130 in Abell~1668.  \it From left to right, the continuum subtracted H$\alpha$ emission flux, the continuum subtracted [N~{\sc{ii}}]~$\lambda$~6584 emission flux, and the continuum near the H$\alpha$ emission line. In this case the emission has not been derived by Gaussian fits but instead includes the integrated flux between 6554 and 6572$\,$\AA~for H$\alpha$ and between 6575 and 6593$\,$\AA~for [N~{\sc{ii}}]. Both lines maps follow the same overall morphology, though the peak is displaced. The H$\alpha$ emission is overlain as contours. The regions used in the analysis are represented as boxes. The images are in units of 10$^{-15}$$\,$erg$\,$s$^{-1}$$\,$cm$^{-2}$$\,$\AA$^{-1}$. One pixel is  $\sim$120$\,$pc across.  \label{aha1668} \label{an21668}\label{acnt1668}}
  \end{figure*}

 \begin{figure*}
 \subfigure{
\begin{minipage}[c]{0.45\textwidth}
        \centering
        \includegraphics[width=2.5in,angle=0]{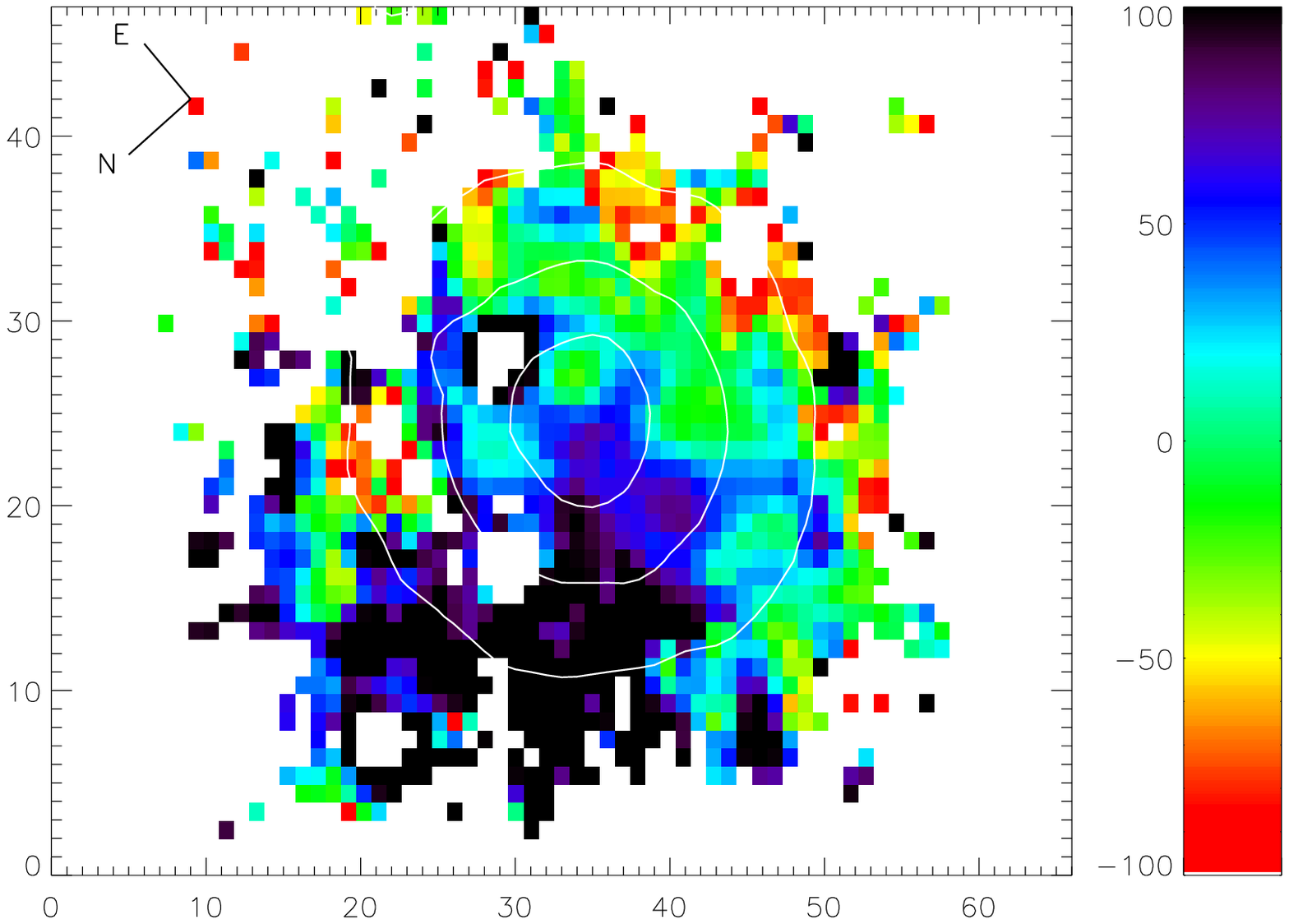}
        \end{minipage}%
    }
    \subfigure{
     \begin{minipage}[c]{0.5\textwidth}
        \centering
        \includegraphics[width=2.5in,angle=0]{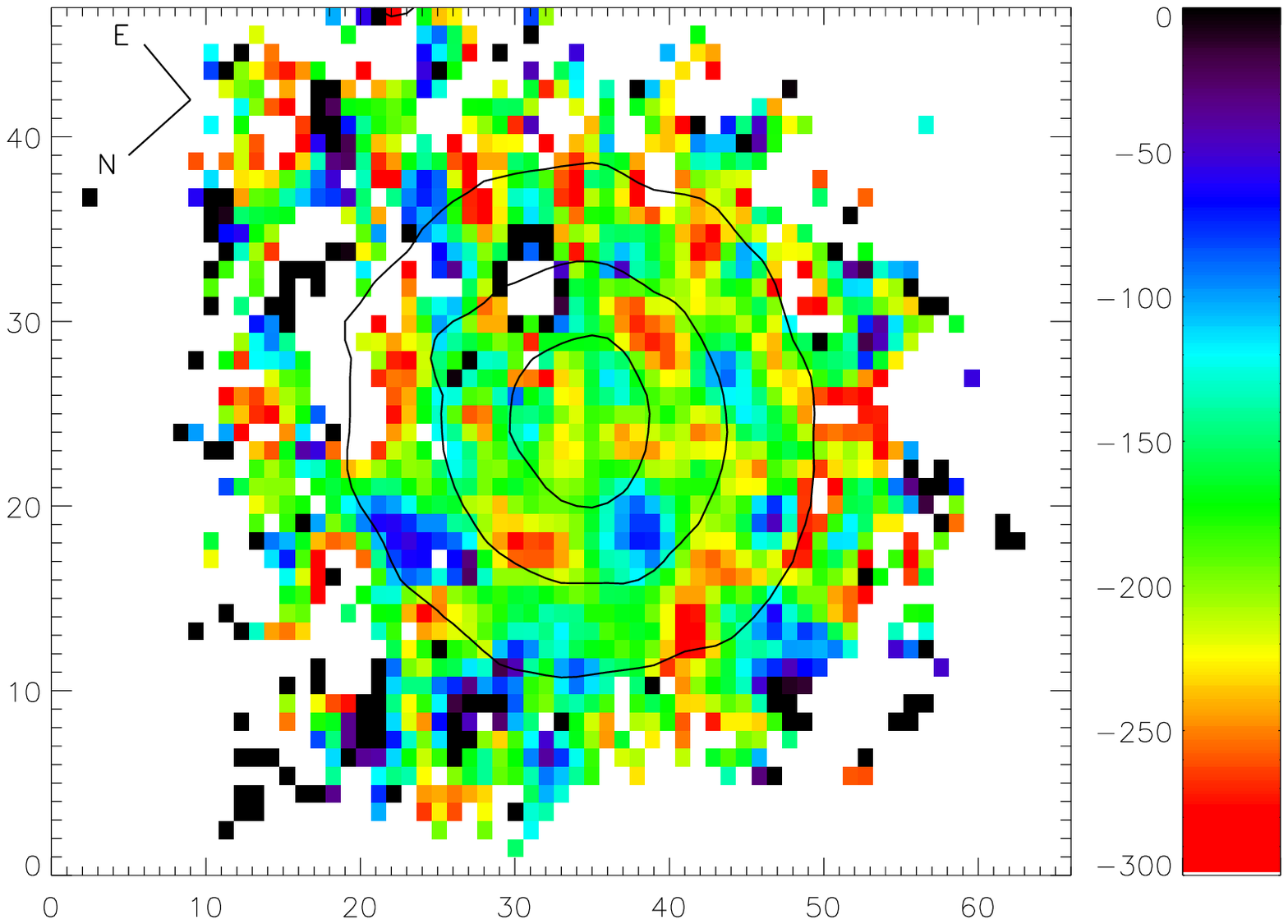}
         \end{minipage}%
    }
\caption[Relative velocities for IC~4130 in Abell~1668]{\bf Relative velocities for IC~4130 in Abell~1668. \it \underline {Left Panel:} Map of the H$\alpha$ relative velocity.  \underline {Right Panel:} Map of the NaD relative velocity. The line emission is not at rest with respect to the underlying galaxy. In addition, the gradient in velocity seen in the line emission is not apparent in the underlying galaxy.\it The scale is in units of km$\,$s$^{-1}$ and is normalized to the redshift of IC~4130. Contours of the continuum emission near H$\alpha$ are overlain and one pixel is $\sim$120$\,$pc across.  \label{havel1668}}

\end{figure*}

{\bf Emission Diagnostics} - Comparing Figures~\ref{specpap} and~\ref{1204lha} we see that [N {\sc{ii}}] $\lambda$ 6584/H$\alpha$ luminosity ratio of star forming Region~1 is lower than line ratio of the AGN affected central Region~4. The line ratios of [N {\sc{ii}}] $\lambda$ 6584/H$\alpha$ and also ([S~{\sc{ii}}]~$\lambda$~6716~+~$\lambda$~6731)/H$\alpha$ suggest that most regions (Regions~1, 2, 7, 8, and 10) are likely a composite of emission from young stellar populations as well as a LINER, while Regions~3, 4, and 5 are from LINER emission only. We cannot distinguish between a Seyfert and a LINER since the ratio of [O~{\sc{iii}}]~$\lambda$~5007/H$\beta$ was not measured.  Integrated along their slit, \citet{cra99} measured a ratio [O~{\sc{iii}}]~$\lambda$~5007/H$\beta$~$>$~1 and [N~{\sc{ii}}]~$\lambda$~6584/H$\alpha$~=~1.4.  These ratios will not be uniform throughout the central regions of the BCG. Nevertheless, their integrated value for [N~{\sc{ii}}]~$\lambda$~6584/H$\alpha$ is between those of our regions. It is therefore useful to keep in mind their result of [O~{\sc{iii}}]~$\lambda$~5007/H$\beta$ as an average value, possibly closer to a LINER. 

{\bf Star Formation Rate} - It is not possible to calculate a SFR for Regions 3, 4, and 5 for example, as they are clearly dominated by AGN signatures. It is also possible that the SFR calculated in the other regions are affected by the AGN, so in this way the rates presented should be interpreted as upper limits, contaminated by AGN ionization. No strong absorption lines are present, so there is no absorption correction to the integrated H$\alpha$ equivalent width. A total SFR is derived from the combined spectrum of the brightest 255  H$\alpha$ emitting pixels and yields a flux value of 14$\pm$1$\,$$\times$$\,$10$^{-15}$$\,$erg~s$^{-1}$$\,$cm$^{-2}$. This yields a SFR of 7.0$\pm$ 0.4$\,$M$_{\odot}$$\,$yr$^{-1}$, or a SFR density of 2.3$\pm$0.1$\,$$\times$$\,$10$^{-8}$$\,$M$_{\odot}$$\,$yr$^{-1}$$\,$pc$^{-2}$.

This upper limit is not inconsistent with the revised value of 50$^{+40}_{-30}$$\,$M$_{\odot}$$\,$yr$^{-1}$ for the MDR assuming the gas does not cool below 0.1keV found by \citet{ode08}, as not all the molecular gas will convert to stars, and as the current estimates of MDRs are an order of magnitude {\it below} the previously derived rates. Our total SFR is very close to that derived by \citet{ode08} from infrared luminosities inside of a 12.2$^{\prime\prime}$ aperture \citep{qui08}, especially considering our effective aperture is slightly smaller. Our results further suggest that although not the dominant source, part of the luminosity in the O'Dea measures may arise from the central AGN.

We do not attempt to predict a valid age and mass estimate for this source, as contamination from the AGN is important, and the SFR is only an upper limit. The interested reader may refer to Edwards 2007 for an examination into the ages in select less contaminated regions.

\subsection{IC~4130 in Abell~1668}

{\bf Morphology} -  There are several pixels which are not well fit by single Gaussian profile fits for this target because of the lower signal, and here, the flux is simply added within a specific window. In this case, Figure~\ref{aha1668} shows the flux added between 6554 and 6572$\,$\AA~for H$\alpha$, and between 6575 and 6593$\,$\AA~for [N~{\sc{ii}}] ~$\lambda$~6584. The figure also shows the continuum near H$\alpha$. Unlike the smooth elliptical distribution of the continuum image, both the H$\alpha$ and [N~{\sc{ii}}] ~$\lambda$~6584 flux images show much more patchy and filamentary structure throughout. The bright regions on the [N~{\sc{ii}}] ~$\lambda$~6584 image correspond to bright regions on the H$\alpha$ image, and the two emission lines share the same overall structure. However, the peak emission is displaced in the images.
\nocite{kew01a}  

 \begin{figure*}
 \subfigure{
\begin{minipage}[c]{0.33\textwidth}
        \centering
        \includegraphics[width=2.5in,angle=0]{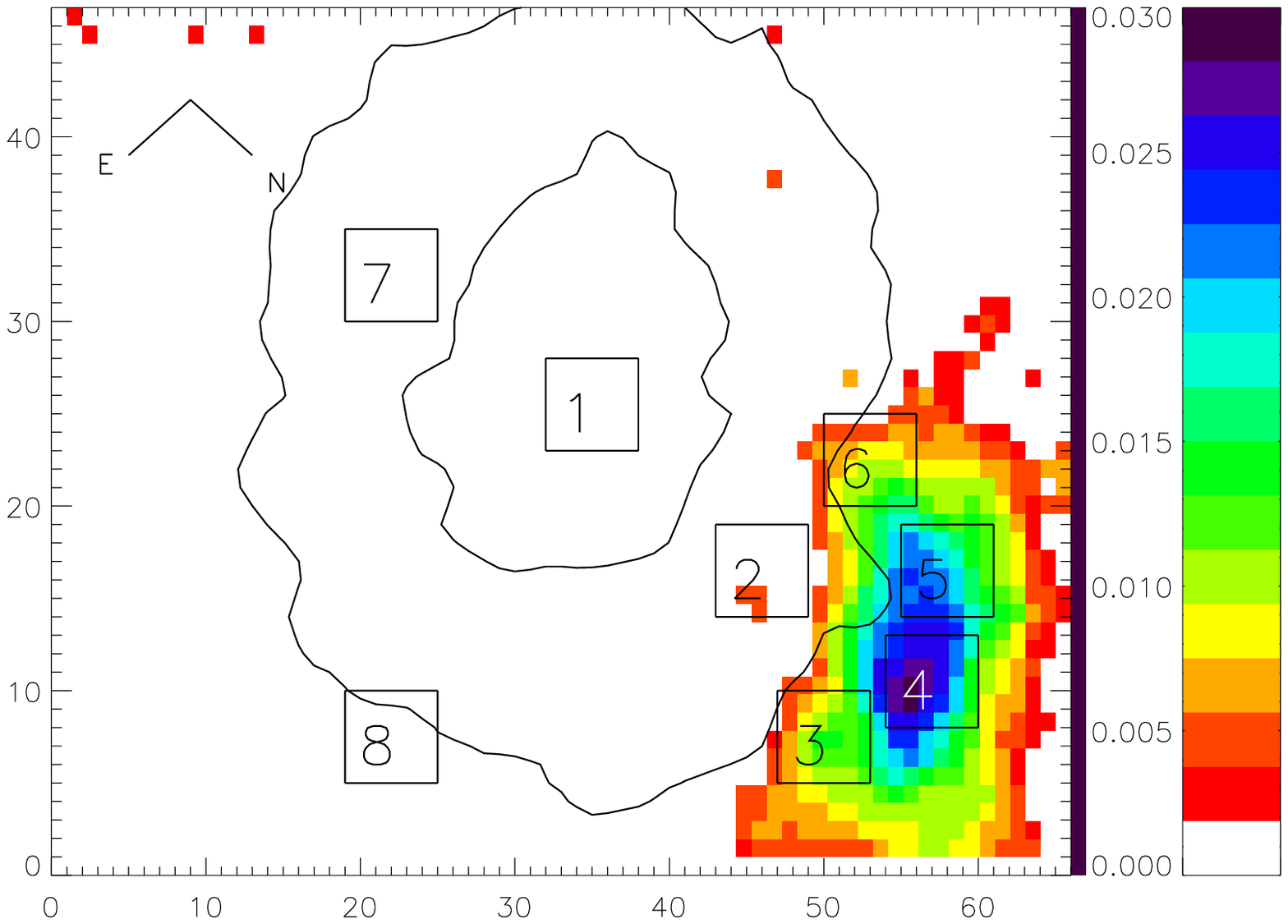}
        \end{minipage}%
    }
    \subfigure{
     \begin{minipage}[c]{0.33\textwidth}
        \centering
        \includegraphics[width=2.5in,angle=0]{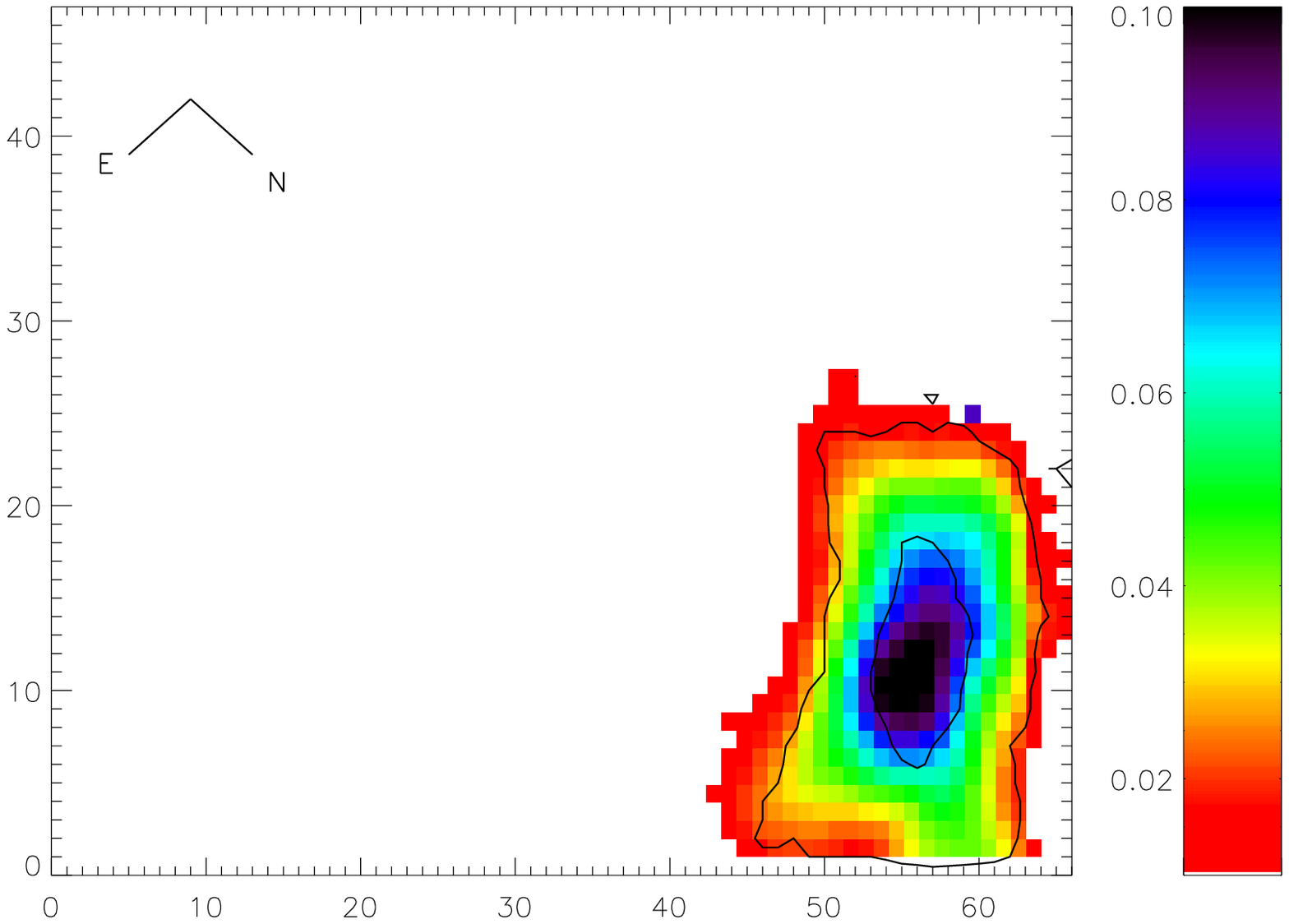}
         \end{minipage}%
    }
        \subfigure{
     \begin{minipage}[c]{0.3\textwidth}
        \centering
        \includegraphics[width=2.5in,angle=0]{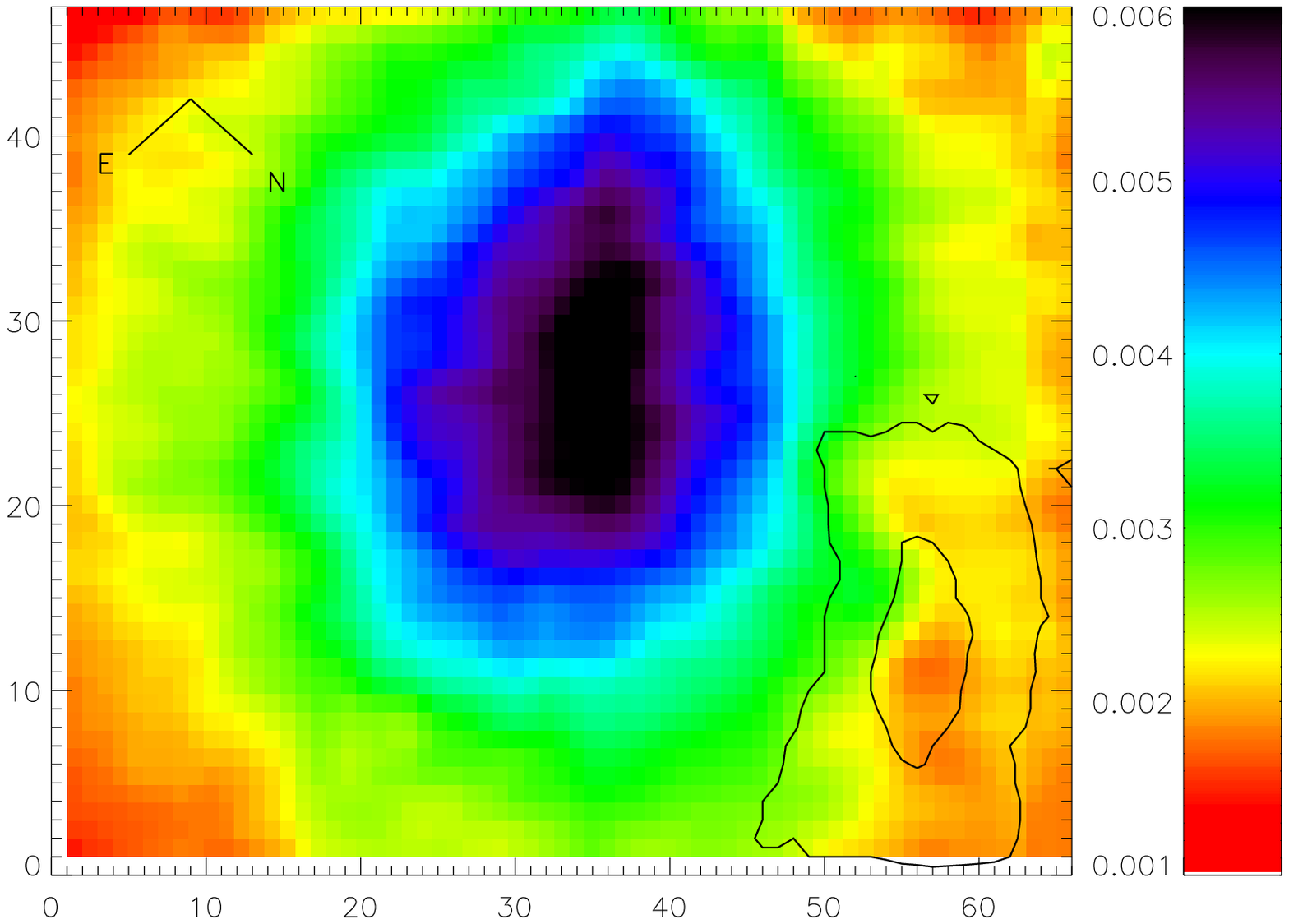}
         \end{minipage}%
    }
\caption{\bf BCG in Ophiuchus. \it From left to right, the continuum subtracted H$\alpha$ emission flux with the continuum emission overplotted as contours, the continuum subtracted [N~{\sc{ii}}]~$\lambda$~6584 emission flux with the H$\alpha$ emission overlain as contours, and the continuum near the H$\alpha$ emission line also with the H$\alpha$ emission overlain as contours. Both emission lines show similar morphology and both are displaced from the center of the BCG, at the location of an extinction patch. The regions used in the analysis are represented as boxes. Extinction is clearly seen in the Northern corner of the continuum image. The images are in units of 10$^{-15}$$\,$erg$\,$s$^{-1}$$\,$cm$^{-2}$$\,$\AA$^{-1}$. One pixel is $\sim$60$\,$pc across.  \label{ahaophi}\label{an2ophi} \label{aophiconti}}
  \end{figure*}

{\bf Kinematics} -  A map of the relative H$\alpha$ line velocities is shown in Figure~\ref{havel1668}. It shows a clear gradient from positive values North of the center ($\gtrsim$~100$\,$km$\,$s$^{-1}$), to negative values ($\lesssim$~$-$100$\,$km$\,$s$^{-1}$) at the South with the zeropoint near the center of the continuum.  The low S/N values at the edge of the figure, as well as a few central patches where emission lines could not be fit as single Gaussians are plotted as having zero relative velocity.  Also shown in the figure is the velocity of the NaD lines relative to the average velocity of IC~4130. This map traces the motion of the underlying galaxy, as opposed to the line emitting gas. It shows a smaller variation in magnitude ($\sim\pm$70$\,$km$\,$s$^{-1}$), no smooth gradient, and all of the velocities are negative. This implies that the line emitting gas is not at rest with respect to the underlying galaxy.

{\bf Emission Diagnostics} - All regions show stronger [N~{\sc{ii}}]~$\lambda$6584 line fluxes than those for H$\alpha$ as can be seen in Figures~\ref{specpap} and~\ref{pplots}. Region~8 has the smallest ratio of all the regions, but the value is 1.7$\pm0.2$, including the measurement error from both of the lines. This is still well described by AGN or LINER emission. The facts that the [N~{\sc{ii}}]~$\lambda$6584/H$\alpha$ ratio does vary across the image and with H$\alpha$ luminosity, and that the scaled underlying spectrum has a lower continuum than the emitting spectrum (although some amount of continuum is surely from the nuclear emission) suggest a young population may still exist. However, its lines are completely masked by the presence of the AGN or LINER, and therefore, it is not possible to use the H$\alpha$ emission line to characterize a SFR or an age for the younger stellar populations.

\subsection{Ophiuchus}

{\bf Morphology} -  The continuum subtracted H$\alpha$ flux image for the BCG is shown in Figure~\ref{ahaophi}. The continuum subtracted [N~{\sc{ii}}]~$\lambda$~6584 flux image is also shown, but the contours represent the H$\alpha$ flux. There is no line emission at the center of the BCG, rather it is concentrated North of the continuum emission. This concentration of line emission is hereafter referred to as Object~B.

\begin{figure}
 \subfigure{
\begin{minipage}[c]{0.45\textwidth}
        \centering
        \includegraphics[width=2.5in,angle=0]{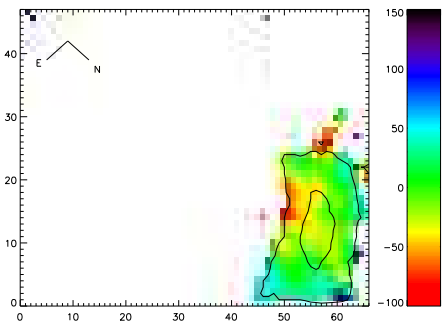}
        \end{minipage}%
    }
\caption[Map of the H$\alpha$ relative velocity for Object~B in Ophiuchus]{\bf Map of the H$\alpha$ relative velocity for Object~B in Ophiuchus. \it The velocity scale here is with respect to the average velocity in Object~B but this is offset from the BCG velocity by 600$\,$km$\,$s$^{-1}$ suggesting Object~B is in front of the BCG and falling onto it. The H$\alpha$ flux is overlain as contours. The scale is in units of$\,$km$\,$s$^{-1}$ relative to the average Object~B velocity. One pixel is $\sim$60$\,$pc across. \label{havelophi}}
\end{figure}

    \begin{figure*}
 \subfigure{
\begin{minipage}[c]{0.45\textwidth}
        \centering
        \includegraphics[width=2.5in,angle=0]{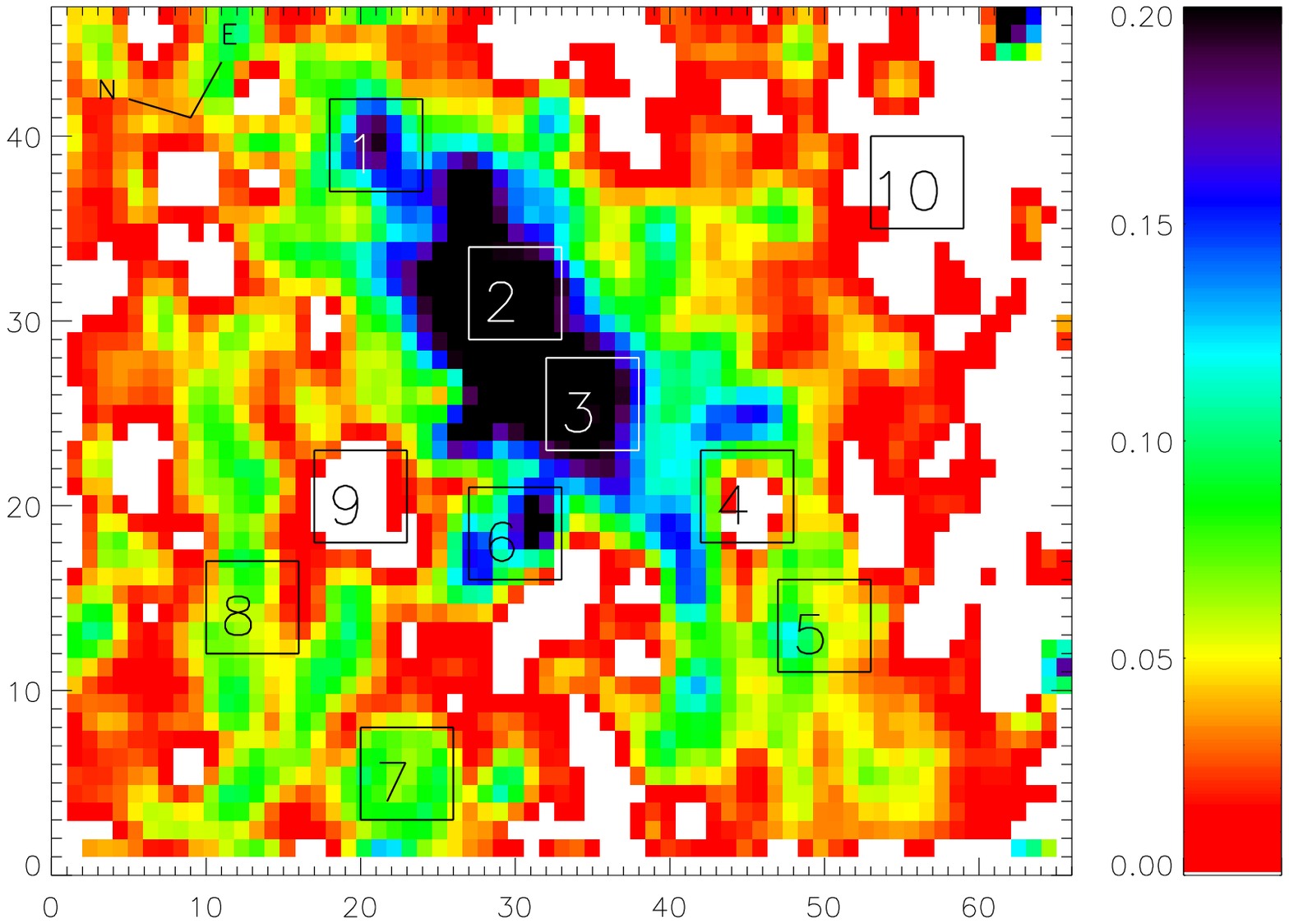}
        \end{minipage}%
    }
    \subfigure{
     \begin{minipage}[c]{0.5\textwidth}
        \centering
        \includegraphics[width=2.5in,angle=0]{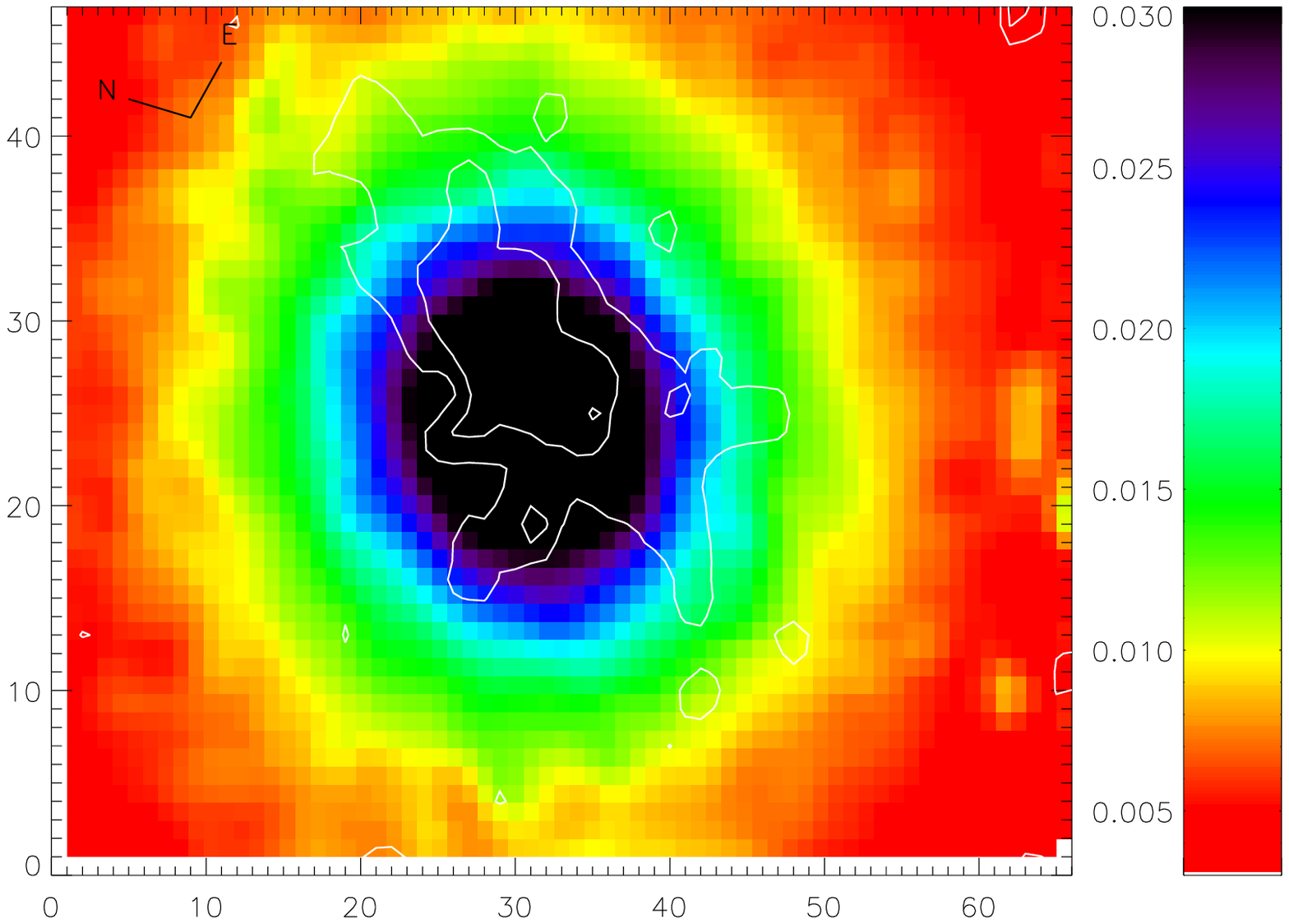}
         \end{minipage}%
    } 
\caption{\bf  NGC~5920 in MKW3s. \it From left to right, H$\alpha$~+~[N~{\sc{ii}}]~$\lambda\lambda$~6548,6584 and the continuum near H$\alpha$. In this case the line
strength is measured by adding all the flux within 6530 and 6590$\,$\AA. An elongated emission feature that runs across the center of the galaxy can be observed. The
H$\alpha$ emission is overlain as contours. The regions used in the analysis are represented as boxes. The images are in units of 10$^{-15}$$\,$erg$\,$s$^{-1}$$\,$cm$^{-2}$$\,$\AA$^{-1}$. One pixel is $\sim$90$\,$pc across.  \label{ahamkw} \label{acntmkw}}
  \end{figure*}

 \begin{figure*}
 \subfigure{
\begin{minipage}[c]{0.45\textwidth}
        \centering
        \includegraphics[width=2.5in,angle=0]{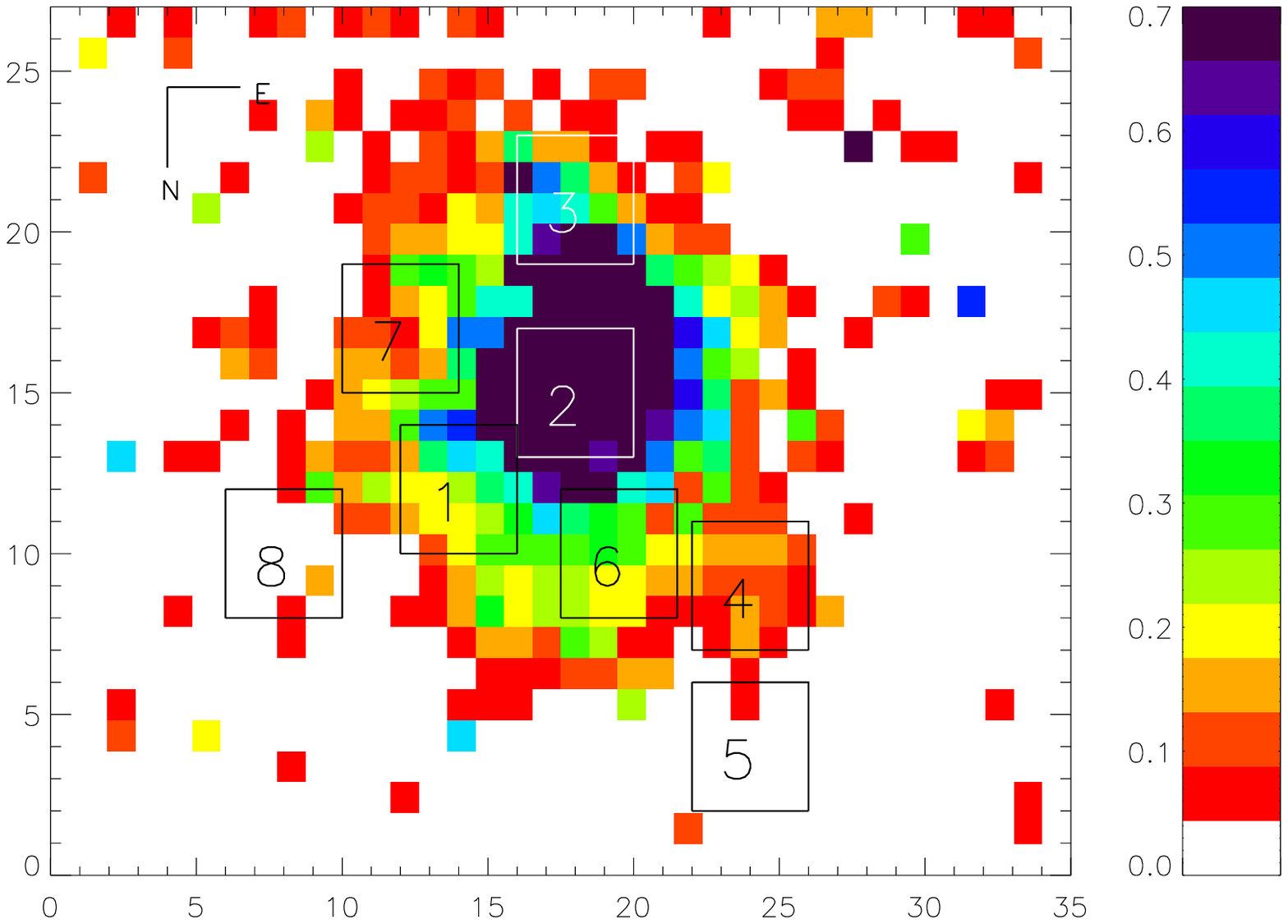}
        \end{minipage}%
    }
    \subfigure{
     \begin{minipage}[c]{0.5\textwidth}
        \centering
        \includegraphics[width=2.5in,angle=0]{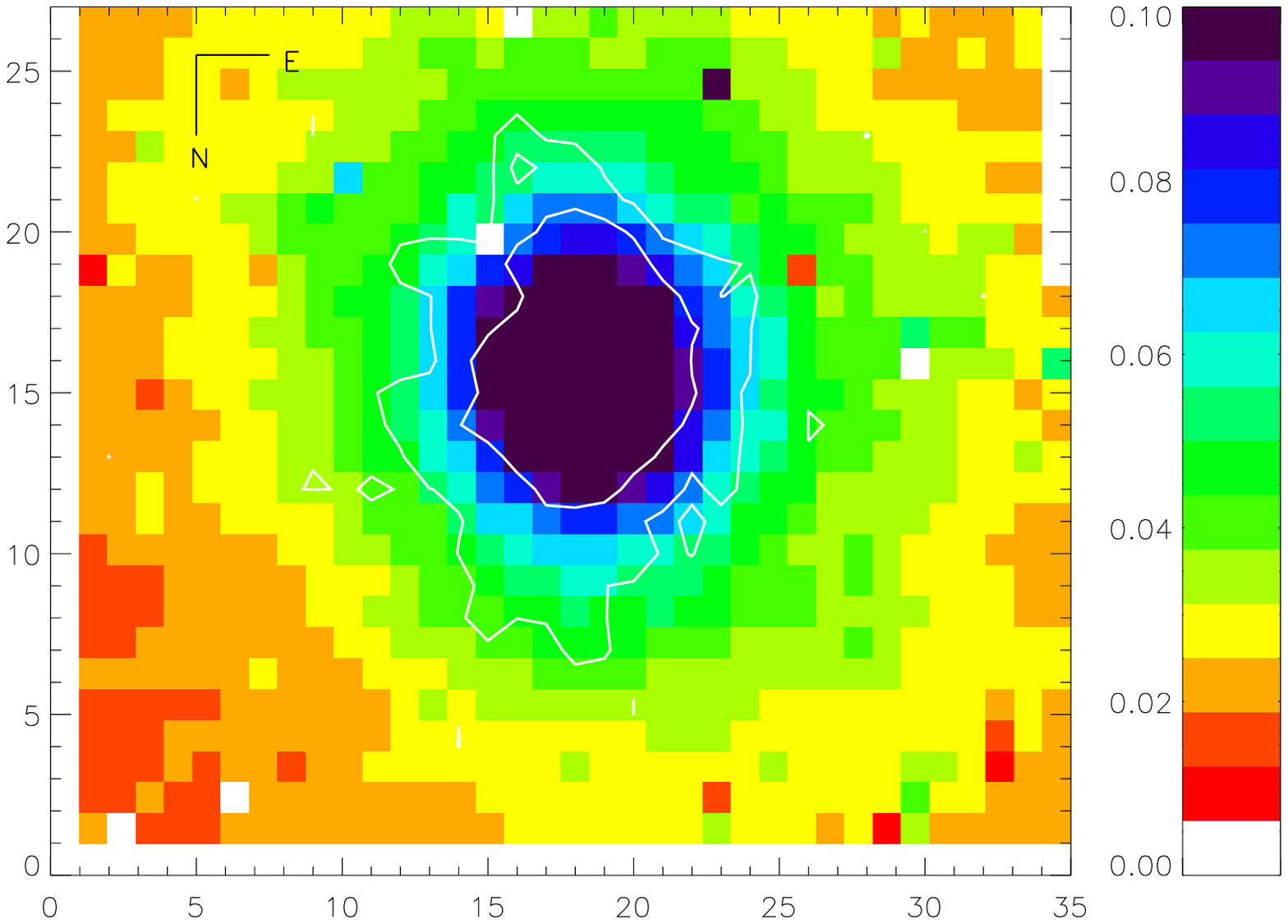}
         \end{minipage}%
    } 
\caption{\bf UGC~9799 in Abell~2052.\it From left to right, the continuum subtracted H$\alpha$ emission flux, the continuum near the H$\alpha$ emission line. Both maps show smooth, centrally condensed emission. The H$\alpha$ emission is overlain as
contours. The regions used in the analysis are represented as boxes. The images are in units of 10$^{-16}$$\,$erg$\,$s$^{-1}$$\,$cm$^{-2}$$\,$\AA$^{-1}$. One pixel is $\sim$150$\,$pc across.  \label{as2052} \label{aha2052} \label{an22052} \label{acnt2052}}
  \end{figure*}

There is a drop in the continuum levels of the BCG to the North side, seen in Figure~\ref{aophiconti}. This is also apparent in Figure~\ref{aquiims} where the dust is seen as a decrease in flux just to the side of the bright line-emitting Object~B. Both these figures show a clear lack of emission at the same projected position as Object~B, probably caused by dust.  Since the dust feature appears to be very localized relative to the BCG and it is absorbing the galaxy continuum we conclude that the dust patch is in front of the BCG.

{\bf The Nature of Object~B} - By comparing the average redshifted positions of the emission lines in Object~B to the average velocity of the BCG NaD absorption line we measure a velocity difference of $+$600$\pm50$$\,$km$\,$s$^{-1}$ between Object~B and the BCG, see Figure~\ref{specpap}. This is well within the typical velocity dispersion of a massive galaxy cluster. There is also a difference of $+$150$\,$km$\,$s$^{-1}$ between the BCG and the cluster velocity (derived from the cluster redshift measurement of z=0.028 from \citet{lah89}), though this difference is close to the error of the cluster radial velocity measurement. 
  
Three scenarios can explain the nature of Object~B: 1) that it is infalling onto the BCG, 2) that it is a background galaxy, and 3) that it is being ejected from the BCG. The first scenario is supported by the absorption seen in the continuum and acquisition images as well as the relative velocity difference between Object~B and the BCG. Scenario 2 is supported by the velocity difference but the dust absorption of the BCG in the region of Object~B would have to be a coincidence. If Object~B were being ejected away from us, the third scenario could produce the observed velocity difference, but would not explain the dust absorption. Finally, if Object~B were being ejected towards us, the dust feature would be explained, but the velocity difference would not.

The hypothesis for which Object~B is in front of the BCG and falling onto the BCG or spiraling around it most completely explains our observations.  Figure~\ref{havelophi} shows the H$\alpha$ velocity map measured with respect to the average Object~B velocity. This map could indicate some rotation of Object~B (where the side near the BCG is rotating in our direction) perhaps perturbed due to some tidal effect with the BCG. In this scenario, Object~B could be a small galaxy (or a large molecular cloud) where the interaction with the BCG is compressing the gas resulting in part of the optical and X-ray emission observed. 

The integrated spectrum of Object~B has [N~{\sc{ii}}]~$\lambda$6584/H$\alpha$~=~3.2$\pm0.3$, a value which is similar to those in its individual regions of (Figure~\ref{1204lha}), and which indicates an AGN or LINER ionizing source.
 
Furthermore, [O~{\sc{i}}]~$\lambda$6300/H$\alpha$~=~0.3$\pm0.05$, and [S~{\sc{ii}}]~$\lambda$6716~$+$~$\lambda$6731/H$\alpha$~=~0.9$\pm0.3$ are consistent with the AGN region of the BPT diagrams, though less clear. The ([S~{\sc{ii}}]~$\lambda$6716~$+$~$\lambda$6731)/H$\alpha$ for instance, varies in the different regions, from 0.72 to  1.26, which is nearer to but on the border of the AGN or LINER side of the BPT diagram. 

\begin{figure}
  \centering
  \epsfxsize=2.5in
        \epsfbox{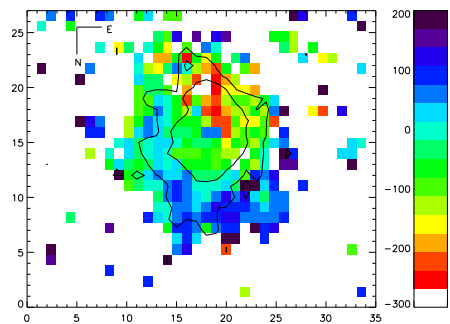} 
\caption[Map of the H$\alpha$ relative velocity for UGC~9799 in Abell~2052]{\bf Map of the H$\alpha$ relative velocity for UGC~9799 in Abell~2052. \it There exists a
gradient in the velocity, however it is not smooth enough to be able to differentiate between rotation or an outflow. The H$\alpha$ flux is overlain as contours. The
scale is in units of km$\,$s$^{-1}$, and values for pixels with bad H$\alpha$ line
fits are plotted as having zero relative velocity. One pixel is $\sim$150$\,$pc across. \label{havel2052}}
\end{figure}
  
\subsection{NGC~5920 in MKW3s}

{\bf Morphology} - Figure~\ref{acntmkw} shows the image of the continuum subtracted H$\alpha$~+~[N~{\sc{ii}}]~$\lambda\lambda$~6548,6584. Deep absorption lines from the underlying galaxy are shifted in velocity with respect to the emission lines. This superposition renders single Gaussian fits to the emission lines impossible. Therefore, the image shown is the addition of all flux between 6530 and 6590$\,$\AA. In contrast with smooth spherical continuum emission, the line emission shows a strong elongated  feature which crosses the center of the cD galaxy from the NE to the SW.

{\bf Kinematics} - In Figure~\ref{specpap}, we present the spectrum of the central region (Region 2) for which, in this case, the absorption spectrum is subtracted. Only here, do we subtract the absorption, as in this case as it is deep and shifted in velocity from the emission lines. We measure the absorption by building a spectrum of the underlying galaxy, as in Section~3.1. After the subtraction, the emission lines become easier to distinguish in the integrated spectrum of Region~2. They are clearly blueshifted from the underlying galaxy by $-$560$\pm50$$\,$km\,s$^{-1}$. This is a known radio galaxy and the velocities are probably indicative of an outflow in our direction.

{\bf Emission Diagnostics} - Individual pixels are of much too low S/N to plot on Figure~\ref{1204lha};  we therefore plot the values of the (absorption corrected) integrated regions.  The average ratios are also calculated for Regions 1, 3, 4, and 5, there are close to or larger than for Region~2. The values are 1.3, 1.3, 2.6, and 1.7. All emitting regions are thus likely ionized by a hard source, AGN or LINER-like emission.

\subsection{The BCG in Abell 1651}

\begin{figure*}
 \subfigure{
\begin{minipage}[c]{0.33\textwidth}
        \centering
        \includegraphics[width=2.5in,angle=0]{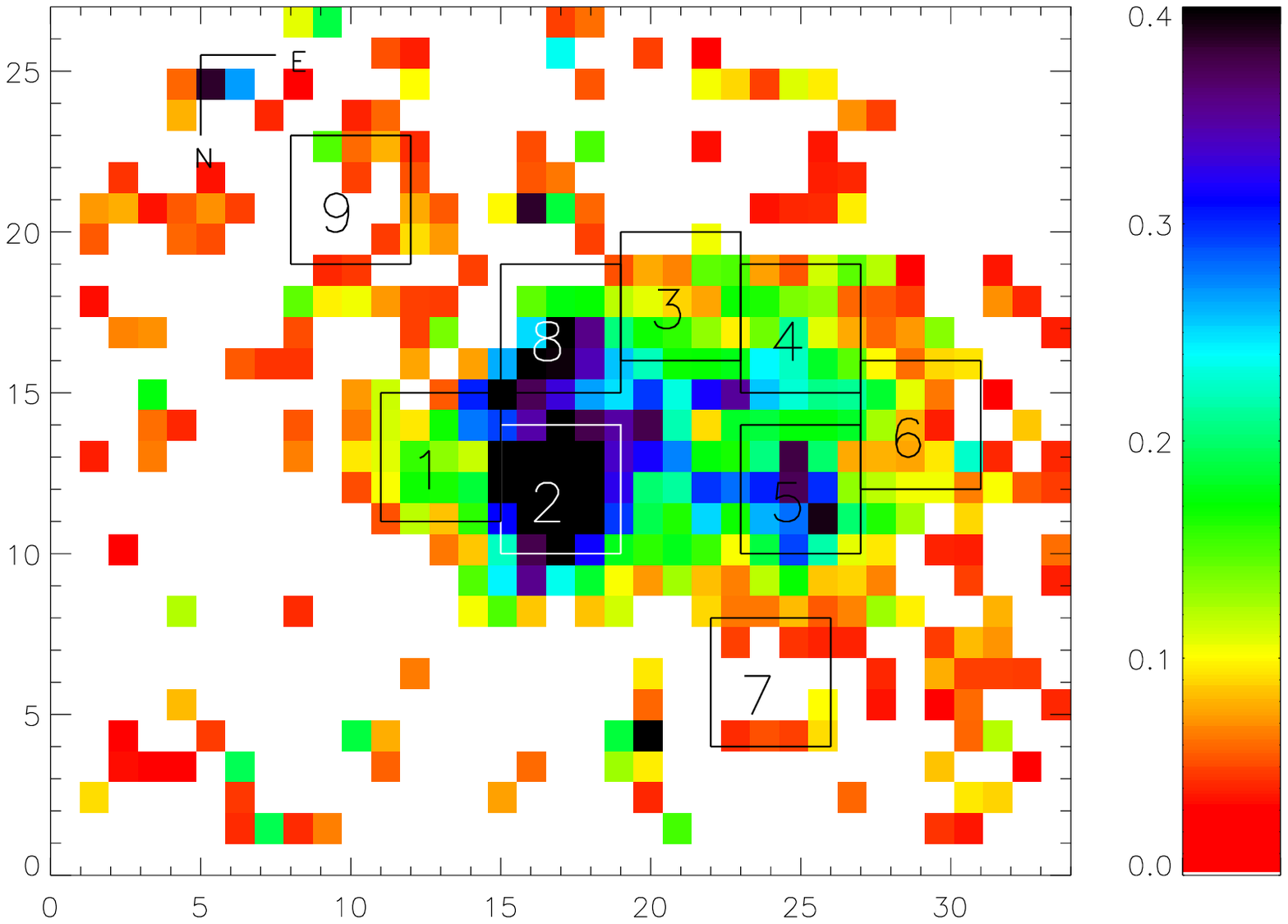}
        \end{minipage}%
    }
    \subfigure{
     \begin{minipage}[c]{0.33\textwidth}
        \centering
        \includegraphics[width=2.5in,angle=0]{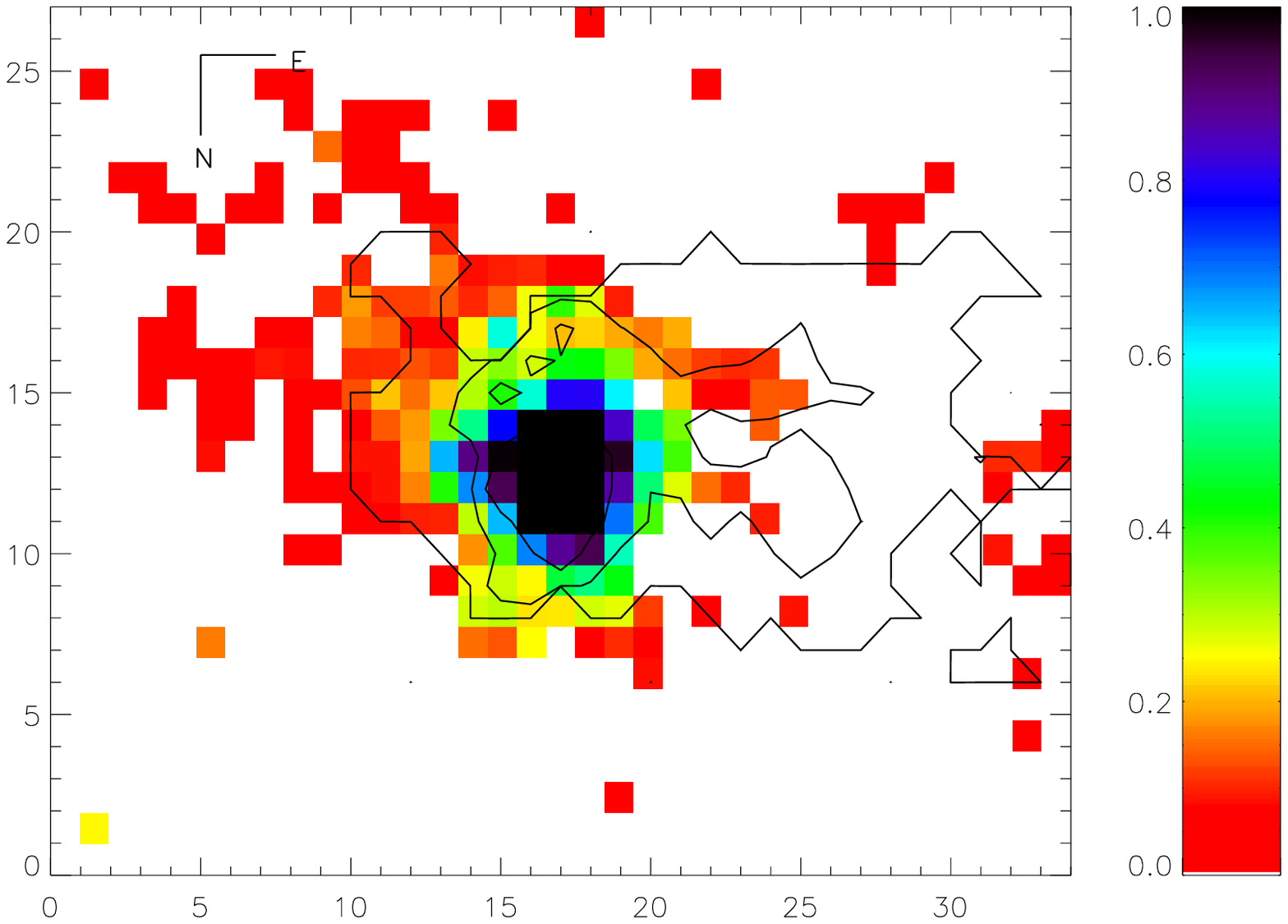}
         \end{minipage}%
    }
        \subfigure{
     \begin{minipage}[c]{0.3\textwidth}
        \centering
        \includegraphics[width=2.5in,angle=0]{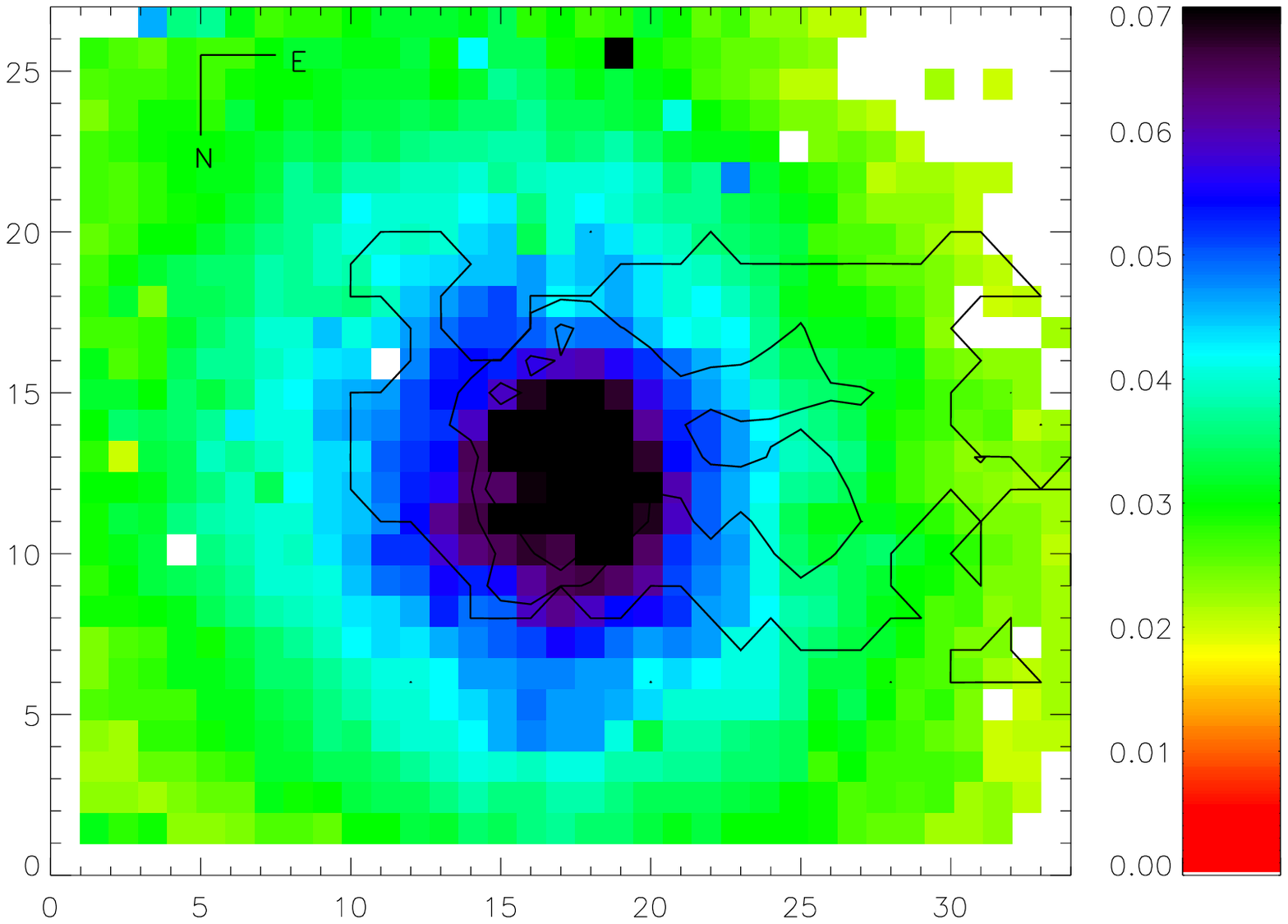}
         \end{minipage}%
    }
\caption{\bf NGC~6166 in Abell~2199. \it From left to right, the continuum subtracted H$\alpha$ emission flux, the  continuum subtracted [N~{\sc{ii}}]~$\lambda$~6584 emission flux, and the continuum near the H$\alpha$ emission line on the right. Neither emission line shows emission as condensed as seen in the continuum. The H$\alpha$ emission is overlain as contours. The regions used in the analysis are represented as boxes. The images are in units of 10$^{-16}$$\,$erg$\,$s$^{-1}$$\,$cm$^{-2}$$\,$\AA$^{-1}$. One pixel is  $\sim$150$\,$pc across.  \label{aha2199} \label{an22199} \label{as2199} \label{acnt2199}}
  \end{figure*}

 The integrated spectrum shown in Figure~\ref{specpap} has no emission lines but only H$\alpha$ in absorption. Thus, although this cluster has a cooling core, there are neither signs of gas ionized by an AGN, nor by a population of hot young stars at the center of the BCG.

\subsection{UGC~9799 in Abell~2052} 

{\bf Morphology} -  Figure~\ref{aha2052} shows the images of the continuum subtracted H$\alpha$ flux, and the continuum near H$\alpha$ (between 6380 and 6430$\,$\AA). These images  show smoothly varying emission and share the same peak location. We do not detect any H$\beta$ emission above the 1$\sigma$ level of the noise, the upper limit in the central Region~2 being 4$\times$10$^{-17}$$\,$erg$\,$s$^{-1}$cm$^{-2}$. Lines of  [N~{\sc{ii}}]~$\lambda$~6584, [S~{\sc{ii}}]~$\lambda$~6716~+~$\lambda$~6734 and [O~{\sc{iii}}]~$\lambda$~5007, and the continuum around H$\beta$ (which are not shown) display a similar  morphology as H$\alpha$.  

{\bf Kinematics} - The relative velocity map of H$\alpha$, Figure~\ref{havel2052}, has a range of $-$250$\,$km$\,$s$^{-1}$ to $+$150$\,$km$\,$s$^{-1}$. The figure shows negative velocities to the South of the center of the emission, and more positive velocities North of the center of the line emitting region.  The the line widths (map not shown) vary by 450-700$\,$km$\,$s$^{-1}$ across the galaxy, but show no clear structure. However, compared to cases with clear signs of rotation where the velocity gradient develops smoothly from positive to negative along an axis, in this case, it is less clear as to whether there is rotation or an outflow.

{\bf Emission Diagnostics} - Figures~\ref{specpap} and \ref{1204lha} show ratios characteristic of ionization due to AGN or LINER activity in most of the pixels. Regions~1, 2, and 6 have [O~{\sc{iii}}]~$\lambda$~5007 line fluxes of 8.5, 8.2, and 2.4$\times$10$^{-16}$$\,$erg$\,$s$^{-1}$cm$^{-2}$, respectively. There are the only three regions where the line flux is greater than 5$\sigma$ (2.0$\times$10$^{-16}$$\,$erg$\,$s$^{-1}$cm$^{-2}$). Therefore, in these regions, the [O~{\sc{iii}}]~$\lambda$~5007 emission is necessarily stronger than the H$\beta$ emission and the [O~{\sc{iii}}]~$\lambda$~5007/H$\beta$ ratio is $\gtrsim$~6. This helps to constrain the harder emission source as a Seyfert nucleus, over a LINER ([O~{\sc{iii}}]~$\lambda$~5007/H$\beta$~$>$3.2; also, see Crawford et al. 1999 who obtain an O~{\sc{iii}}]~$\lambda$~5007/H$\beta$ ratio of 9.5). The ([S~{\sc{ii}}]~$\lambda$~6716~+~ [S~{\sc{ii}}]~$\lambda$~6731)/H$\alpha$ line ratio is above 1.10 for all regions and well within the AGN side of the diagram.  Therefore, it is not possible to use the H$\alpha$ emission line to search for any young stellar population. However, observing the scatter in the individual pixels of the central  Region~2 on Figure~\ref{1204lha} does suggest that there may be 2 cases: 1) for H$\alpha$ luminosities that increase as the ratio decreases, and hence are affected by star-forming regions, and 2) where the ratios are higher and the H$\alpha$ luminosities are $\sim$1. 

\subsection{NGC~6166 in Abell~2199}

{\bf Morphology} - Images of the continuum subtracted H$\alpha$ flux along with the  continuum subtracted [N~{\sc{ii}}] ~$\lambda$~6584, and the continuum near H$\alpha$ are shown in Figure~\ref{aha2199}. The continuum image which includes flux between 6650 and 6700$\,$\AA, increases smoothly in brightness towards the center, without any prominent dust features. The peak of the H$\alpha$ emission does coincide with the peak of the continuum flux, however, a second bright peak, not noticeable in any of the other line images is seen to the South. The H$\alpha$ line flux continues to extend towards the East, a morphology that is mirrored in the [S~{\sc{ii}}]~$\lambda$~6716~+~$\lambda$~6734 emission line image (not shown), although in this case the [N~{\sc{ii}}] ~$\lambda$~6584 line image differs in that it is much more condensed. The elliptical morphology of the continuum in the blue, between 5025 and 5100$\,$\AA~(not shown) is similar to that seen for the red continuum, though it is less condensed. The individual pixels of the blue configuration show no strong H$\beta$ or [O~{\sc{iii}}] ~$\lambda$~5007 line emission.

\begin{figure}
  \centering
  \epsfxsize=2.5in
        \epsfbox{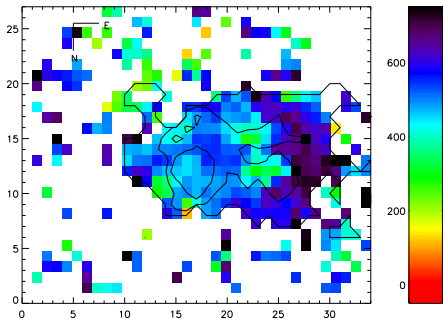}
 \caption[Map of the H$\alpha$ relative velocity for NGC~6166 in Abell~2199]{\bf Map of the H$\alpha$ relative velocity for NGC~6166 in Abell~2199. \it The increasing and decreasing velocities in the Eastern part of the emission may be explained if the low velocity blobs are closer and therefore subject to less extinction. The H$\alpha$ flux is overlain as contours. The scale is in units of km$\,$s$^{-1}$. One pixel is $\sim$150$\,$pc across. \label{havel2199}}
\end{figure}

\nocite{dev91}  
{\bf Kinematics} -  The relative velocity map of H$\alpha$ emission is shown in Figure~\ref{havel2199}. The velocities are calculated with respect to the rest frame of the central galaxy and plotted as such in the figure. However, one can see that clearly the emission lines are not at rest with respect to the central galaxy (z$_{BCG}$=0.03035$\pm$0.00003; de~Vaucouleurs et al. 1991), and that the velocity difference within the line emitting gas itself goes from  $-$200$\,$km$\,$s$^{-1}$ to $+$200$\,$km$\,$s$^{-1}$. The low level flux of the Eastern side of the image has a complex structure. It is the location of both the Region~6 with the largest velocity, 700$\pm$50$\,$km$\,$s$^{-1}$, and Region~5 with the lowest velocity, 580$\pm$50$\,$km$\,$s$^{-1}$. The pixels which have the most intense line emission are those with the lowest velocities, it could be that they are closer to us and hence subject to less extinction. If this were the case, one would then expect the H$\beta$ line emission to be more smooth, however hardly any emission lines are seen in the image (only Region~8 has detectable H$\beta$ emission). However, the H$\beta$ emission lines are quite possibly effected by absorption which cannot be seen at the S/N level of this data. In this case we would be seeing a curved jet with relative velocity $\sim$500$\,$km$\,$s$^{-1}$. A higher S/N H$\beta$ image would help verify this scenario.

{\bf Emission Diagnostics} - The spectra in the red and blue are shown in Figure~\ref{specpap}. The [N~{\sc{ii}}]~$\lambda$~6584/H$\alpha$ ratios (see Figure~\ref{1204lha}) for the different regions in this galaxy vary from 1.7 to 10. The [O~{\sc{iii}}]~$\lambda$~5007/H$\beta$ ratio, measurable only for Region~8, has a value of 1.8$\pm0.5$. This is consistent with the ratios from \citet{cra99} derived from a 6$\,$$^{\prime\prime}$ slit.  It is possible that stellar absorption affecting H$\alpha$ and H$\beta$ could be artificially raising the ratios, and hiding the effects of a young population. However, this would not be a significant effect since no absorption is seen in the integrated spectra from the outskirts \citep{edwthesis}.
\nocite{kew01b}

The only region for which an H$\beta$ emission strength is measurable is Region~8, which shows an H$\alpha$/H$\beta$ ratio of 2.4$\pm0.4$. This is within the theoretical lower limit of 2.8 to within our measurement  errors.  The ratio is much lower than the integrated value of \citet{cra99}, H$\alpha$/H$\beta$~=~9.3. However, as H$\alpha$ is strong in several regions and H$\beta$ is only present in Region~8, the integrated ratio is naturally much larger. Nevertheless H$\beta$ is clearly detected in Region~8 which suggests that the extinction level is lower in this region. It is therefore reasonable to suggest that Region~2 may be closer to the dusty torus of an AGN. It may also be that Region~8 has a population of young stars and that H$\beta$ absorption could be lowering this ratio.

  \begin{figure*}
 \subfigure{
\begin{minipage}[c]{0.33\textwidth}
        \centering
        \includegraphics[width=2.5in,angle=0]{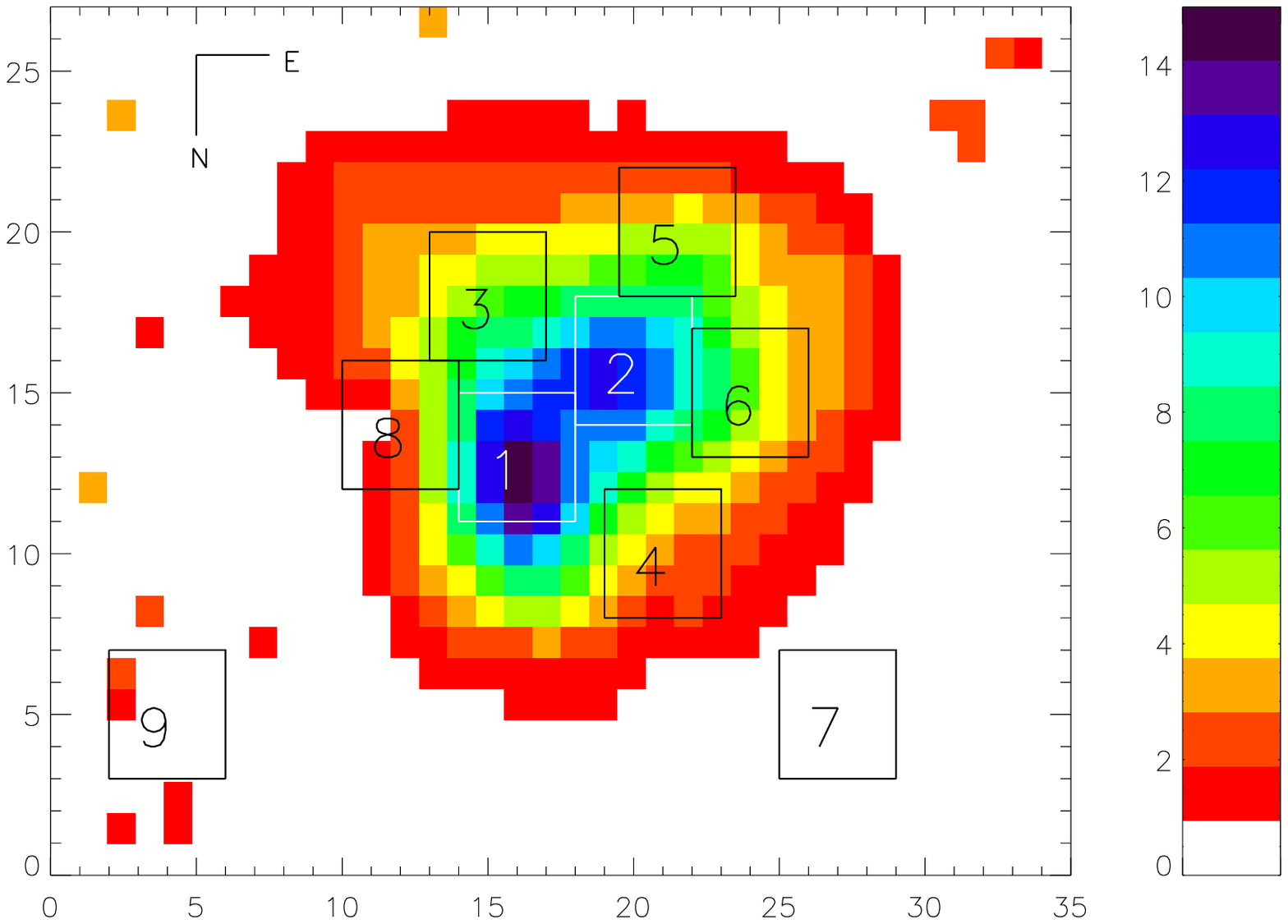}
        \end{minipage}%
    }
    \subfigure{
     \begin{minipage}[c]{0.33\textwidth}
        \centering
        \includegraphics[width=2.5in,angle=0]{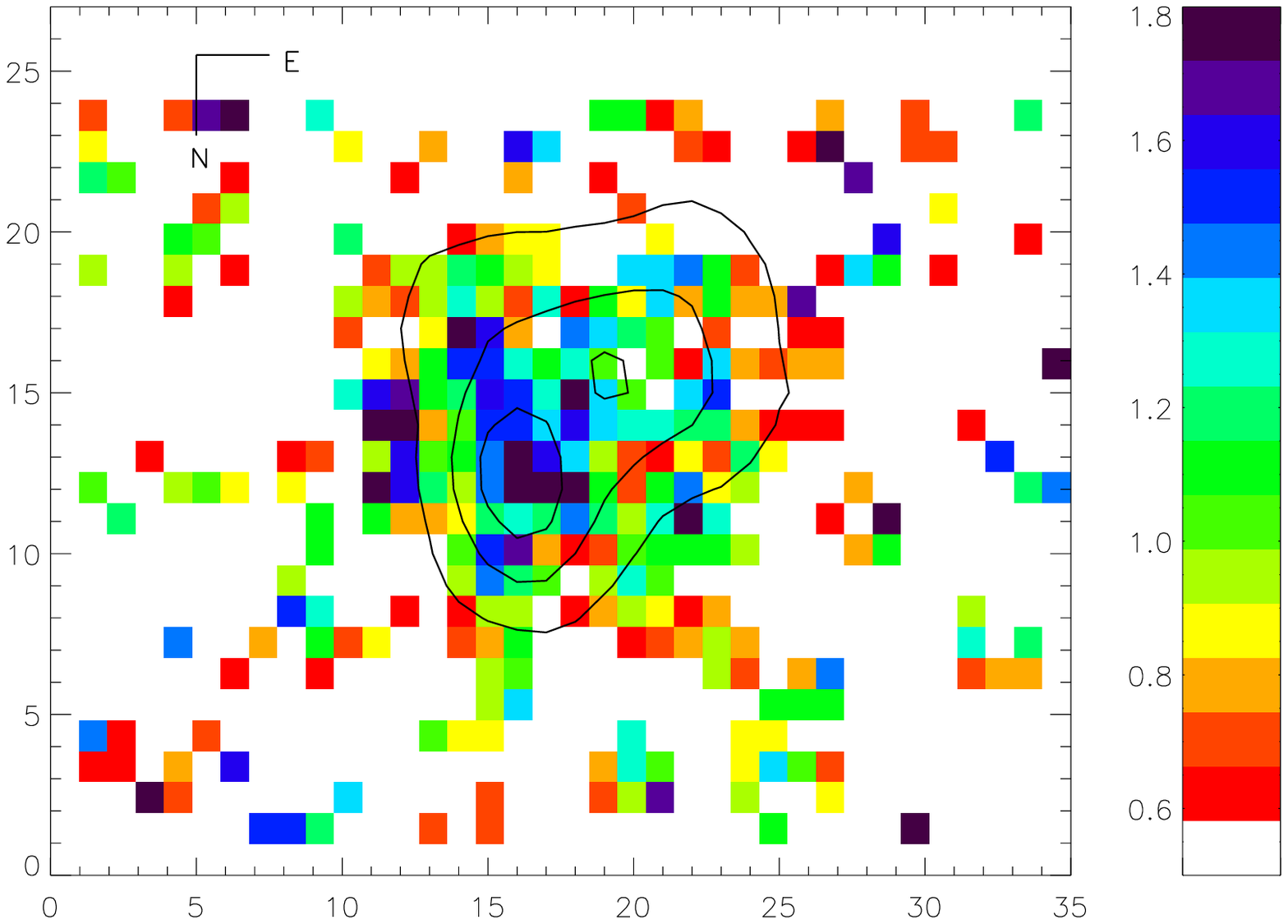}
         \end{minipage}%
    } 
        \subfigure{
     \begin{minipage}[c]{0.3\textwidth}
        \centering
        \includegraphics[width=2.5in,angle=0]{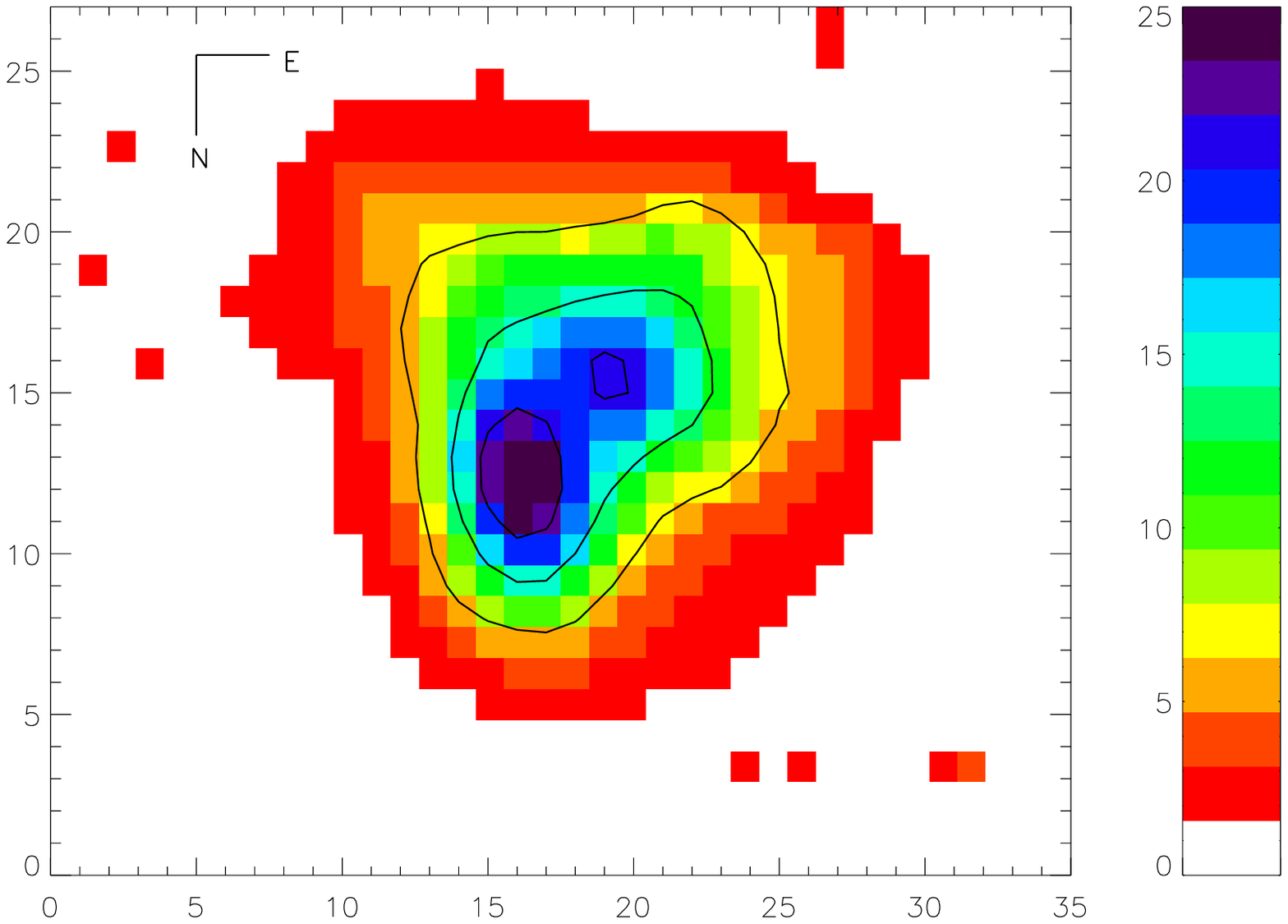}
         \end{minipage}%
    } 
    \subfigure{
     \begin{minipage}[c]{0.33\textwidth}
        \centering
        \includegraphics[width=2.5in,angle=0]{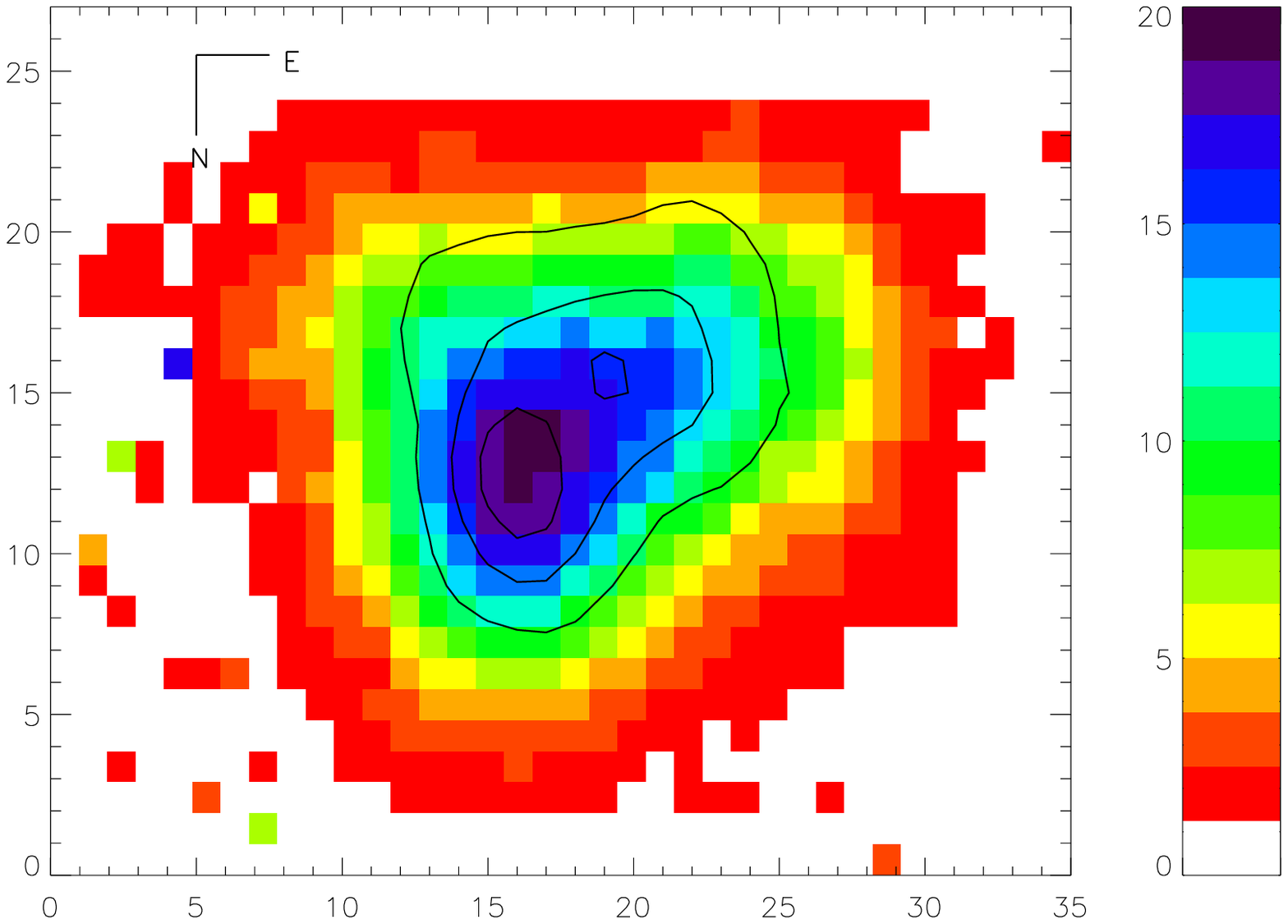}
         \end{minipage}%
    } 
        \subfigure{
     \begin{minipage}[c]{0.33\textwidth}
        \centering
        \includegraphics[width=2.5in,angle=0]{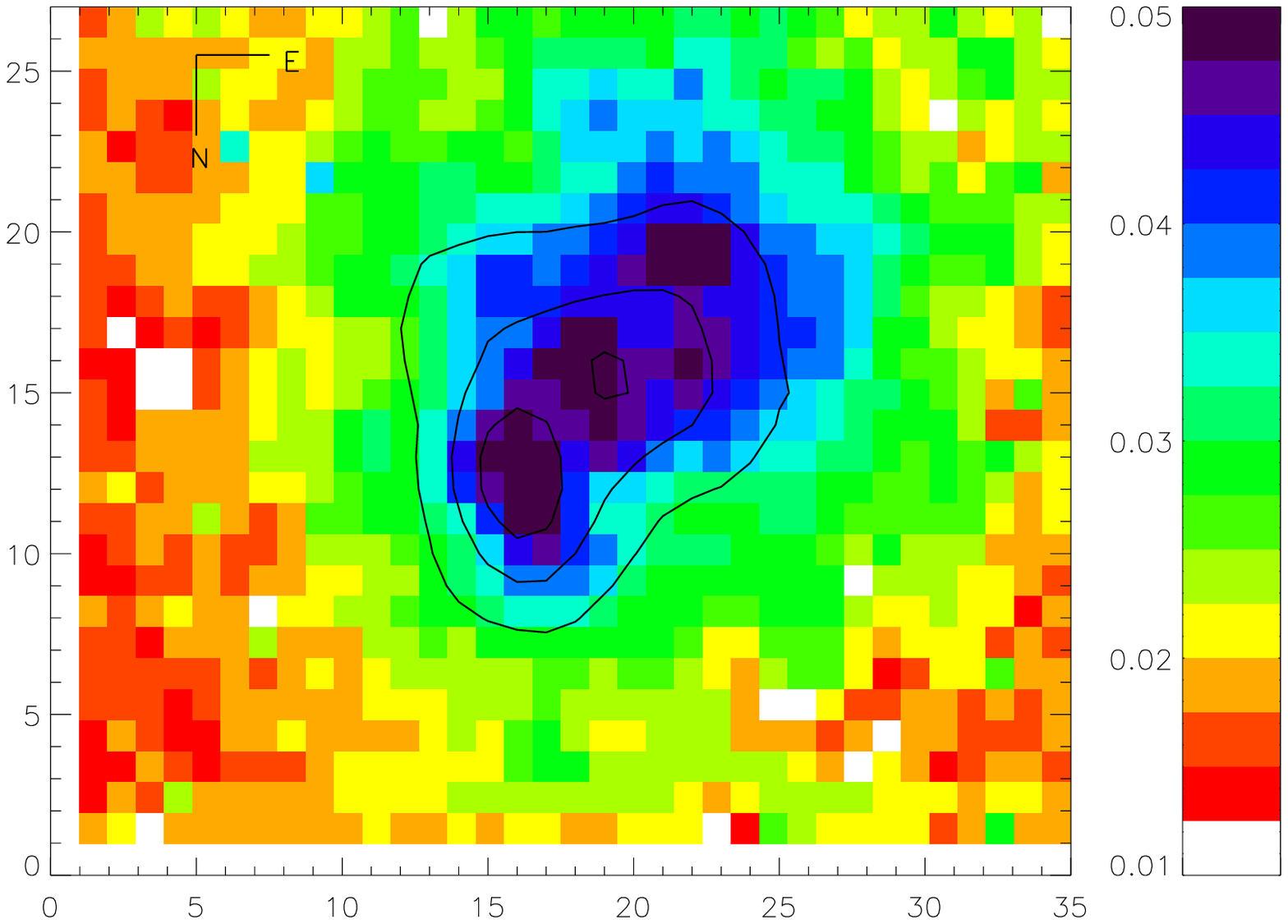}
         \end{minipage}%
    } 
       \subfigure{
     \begin{minipage}[c]{0.3\textwidth}
        \centering
        \includegraphics[width=2.5in,angle=0]{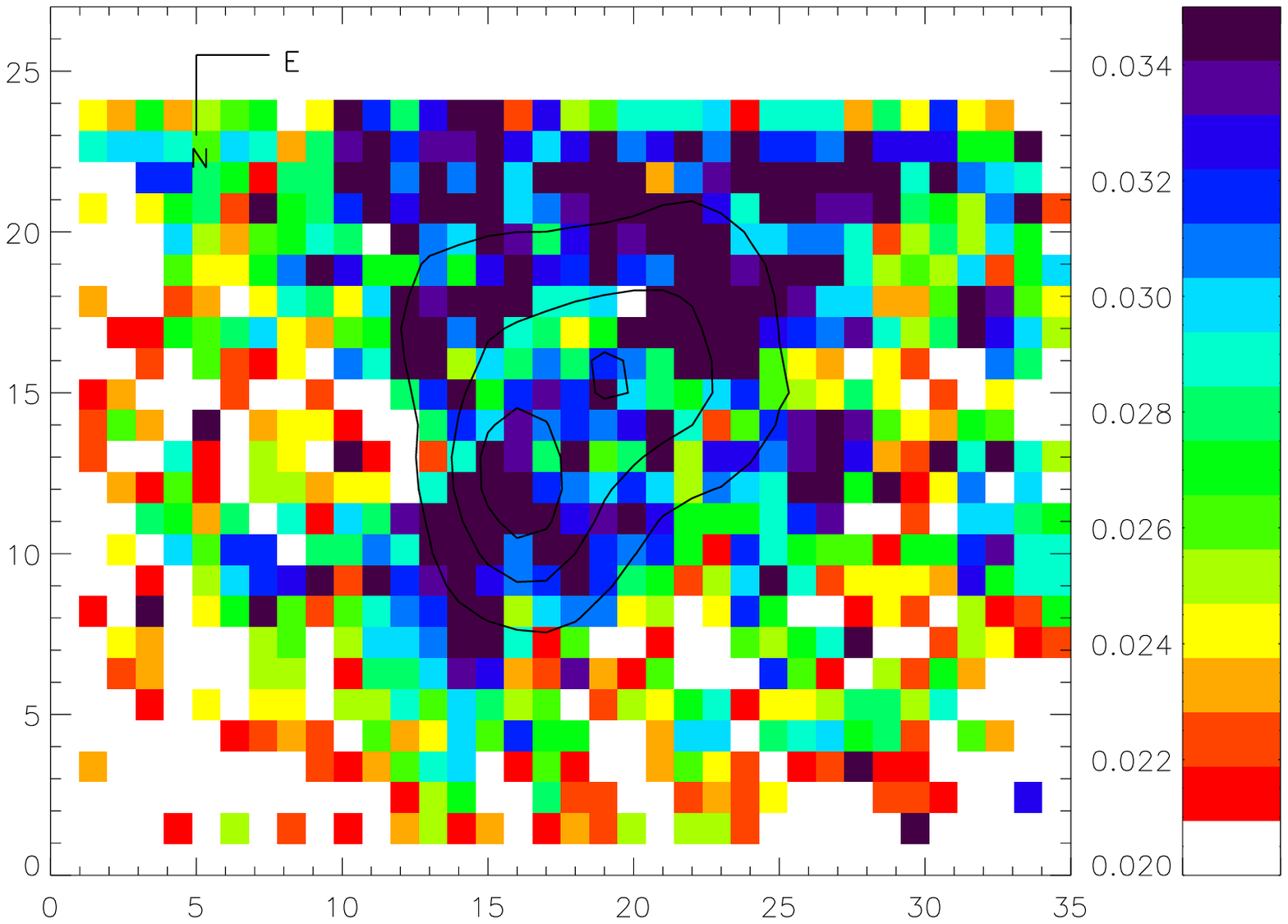}
         \end{minipage}%
    } 
\caption{\bf Cygnus-A. \it Top Row: From left to right, the continuum subtracted H$\alpha$ emission flux, the continuum subtracted H$\beta$ emission flux, and the continuum subtracted [N~{\sc{ii}}]~$\lambda$~6584 emission flux.  The red emission lines show very similar morphology, however, the [N~{\sc{ii}}] emission is much stronger, we are looking at gas clearly ionized by an AGN. Bottom Row: From left to right, the continuum subtracted [O~{\sc{iii}}]~$\lambda$~5007 emission flux, the continuum near the H$\alpha$ emission line and the continuum near the H$\beta$ emission line. The central peak seen in the red continuum is not apparent in the blue continuum, as it is probably hidden by dust extinction. The H$\alpha$ emission is overlain
as contours. The regions used in the analysis are represented as boxes. The image is in units of 10$^{-16}$$\,$erg$\,$s$^{-1}$$\,$cm$^{-2}$$\,$\AA$^{-1}$. One pixel is $\sim$220$\,$pc across.  \label{aoIIIbCA}  \label{acontCA}  \label{acontbCA} }
  \end{figure*}

\subsection{Cygnus-A}

{\bf Morphology} - Figure~\ref{aoIIIbCA} presents the continuum subtracted H$\alpha$, H$\beta$, [N~{\sc{ii}}] ~$\lambda$~6584, [O~{\sc{iii}}] ~$\lambda$~5007 emission, and the continuum near H$\alpha$ and near H$\beta$.  In the line emission images, it is the NW and the central components (Regions~1 and 2) that have the highest flux. In the image of the continuum near H$\alpha$ (between 6380 and 6430$\,$\AA), the brightest emission is seen in four blobs: one central, one East of center, one to the SE and another to the NW. The image of the continuum near H$\beta$ (between 4720 and 4830$\,$\AA) shows bright emission in the SE and NW blobs, but the central peak is not seen, likely due to intense dust extinction \citep{jac98}. 

\begin{figure*}
 \subfigure{
\begin{minipage}[c]{0.33\textwidth}
        \centering
        \includegraphics[width=2.5in,angle=0]{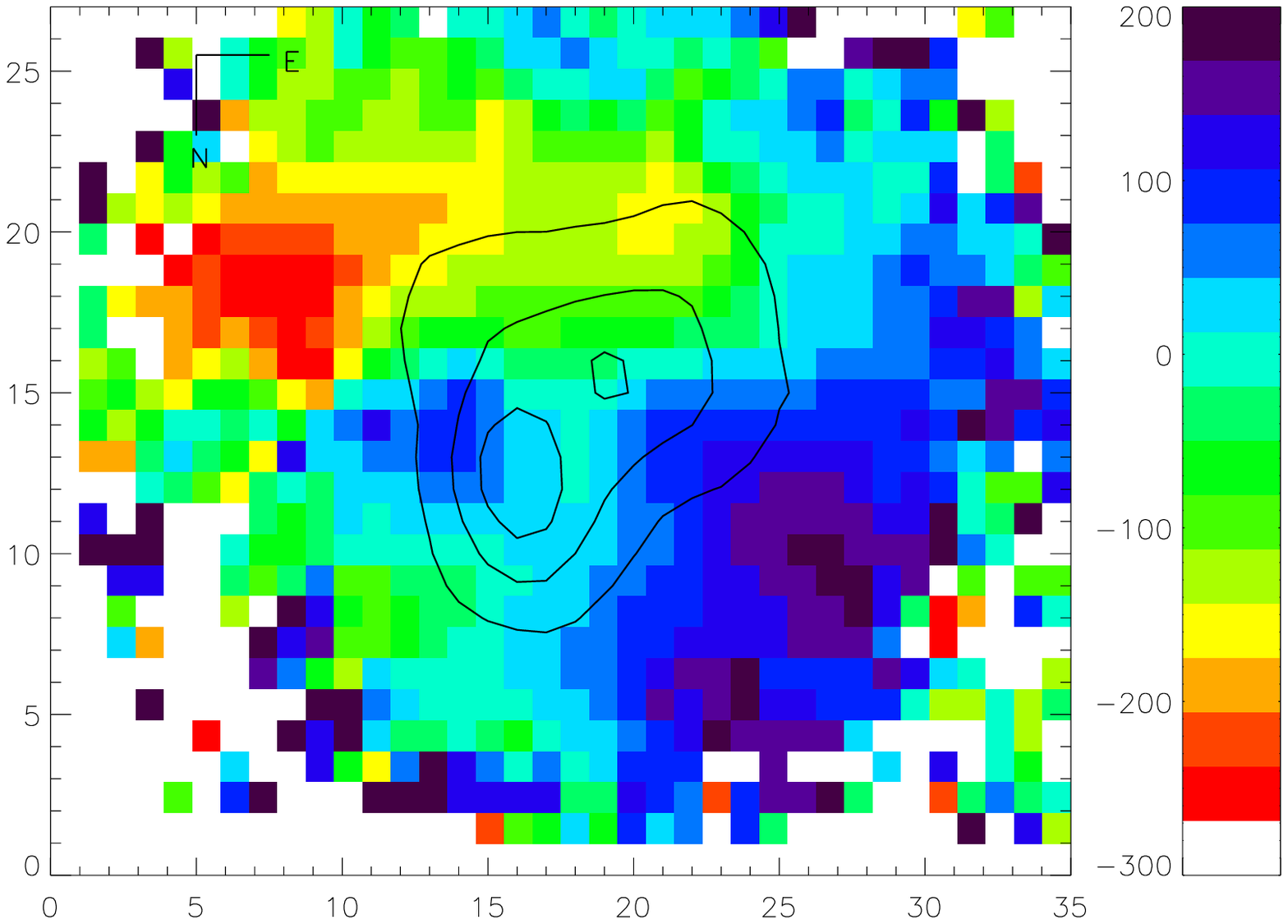}
        \end{minipage}%
    }
    \subfigure{
     \begin{minipage}[c]{0.33\textwidth}
        \centering
        \includegraphics[width=2.5in,angle=0]{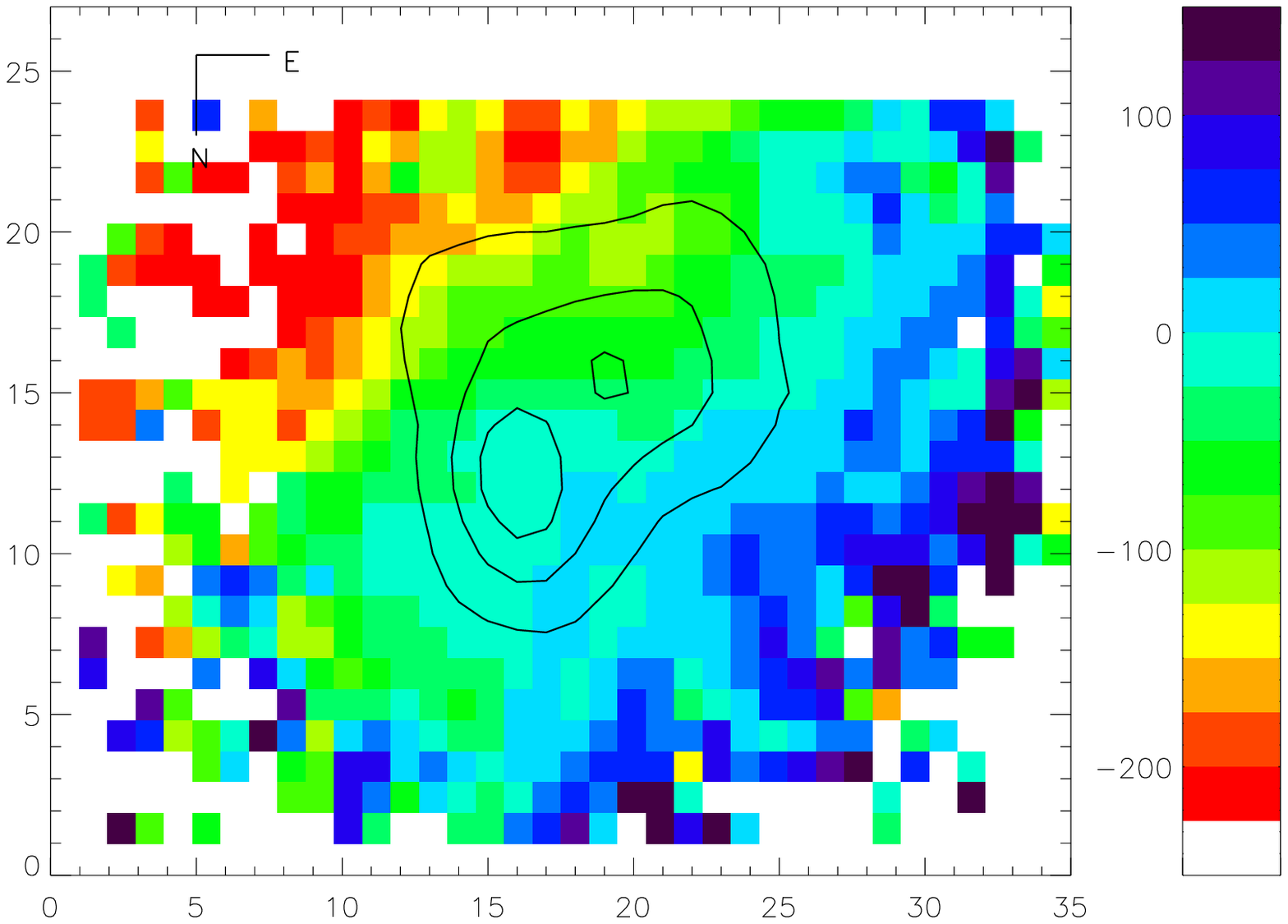}
         \end{minipage}%
    } 
        \subfigure{
     \begin{minipage}[c]{0.3\textwidth}
        \centering
        \includegraphics[width=2.5in,angle=0]{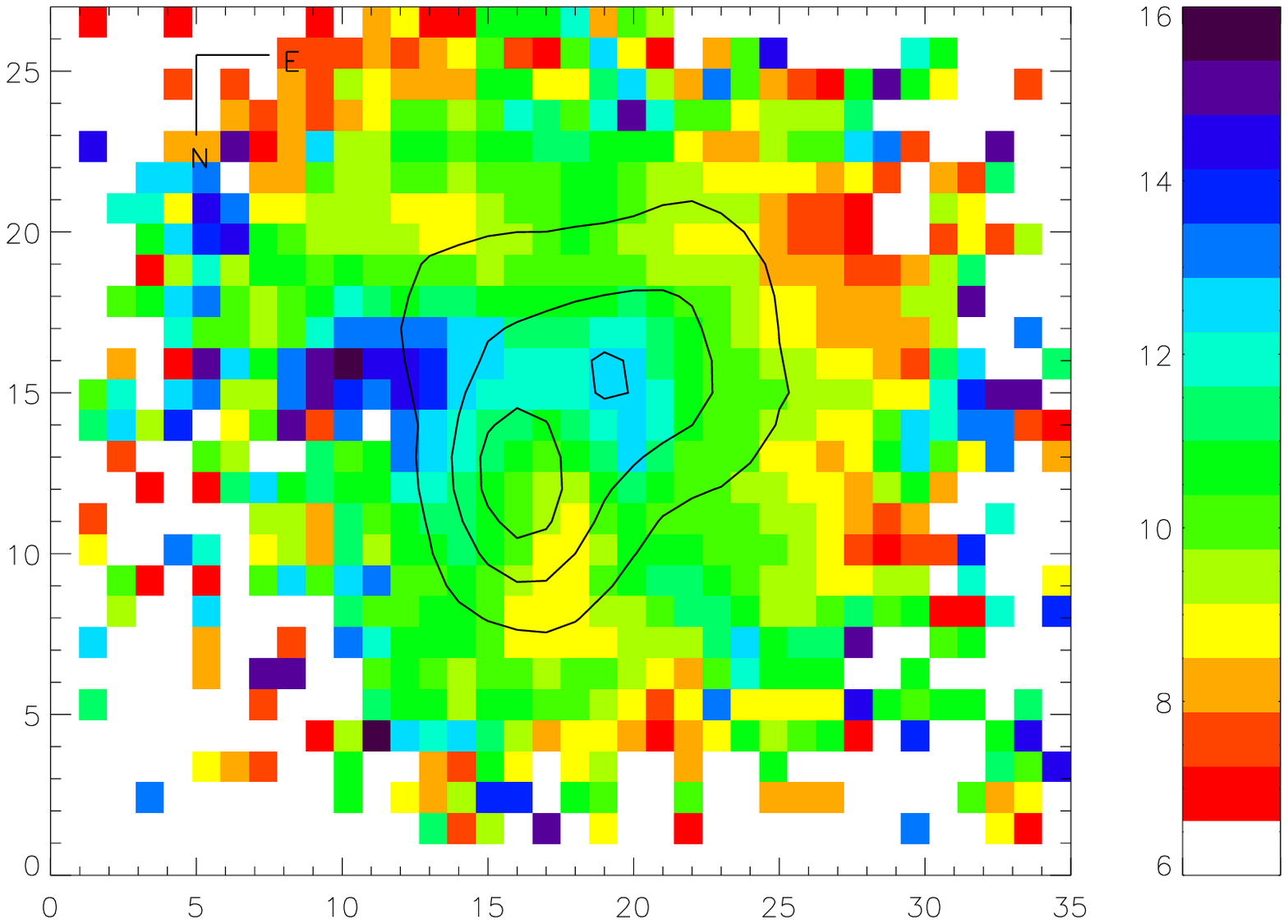}
         \end{minipage}%
    } 
\caption[Map of the H$\alpha$ relative velocity in Cygnus-A]{\bf Map of the line kinematics in Cygnus-A. \it From left to right - the H$\alpha$ relative velocity, the  [O~{\sc{iii}}]~$\lambda$~5007 relative velocity, and the H$\alpha$ FWHM. Large scale velocity gradients are due to rotation. The largest values of FWHM are found at
the center (Region~2) near the central black hole and also between Regions~3 and 8, near the location of the jet. The H$\alpha$ emission overlain as contours. Notice
that the [O~{\sc{iii}}] image needs to be shifted North a few pixels to align properly to the H$\alpha$ image, and in the central panel the white pixels at the top represent regions of no data. The scale is in units of km$\,$s$^{-1}$. One pixel is $\sim$220$\,$pc across.  \label{havelCA}\label{oIIIbvCA}\label{hafwCA}}
\end{figure*}

{\bf Kinematics} -  The relative velocity maps of the H$\alpha$ and [O~{\sc{iii}}]~$\lambda$~5007 emission lines are shown in Figure~\ref{havelCA}. The relative velocity map of the [N~{\sc{ii}}]~$\lambda$~6584 emission is not shown as it harbours very similar  morphology and magnitude as to that for the H$\alpha$ relative velocity map. Large scale velocity gradients of $\pm$200$\,$km$\,$s$^{-1}$ are seen for H$\alpha$, and of $\pm$100$\,$km$\,$s$^{-1}$ for [O~{\sc{iii}}]~$\lambda$~5007, both in the SW and NE direction. On this scale, the velocity gradients in both lines are reminiscent of rotation.

The FWHM of the H$\alpha$ emission line is also shown in Figure~\ref{hafwCA}. It reveals large widths corresponding to the central emission peak, the location of the central radio point source \citep{tad03}. There is also an area of large width that extends west of this peak (that is, west of Region~2). This second area extends to Region~8, the direction of the previously detected jet. The large width could be caused by disturbance from the jet. 

\begin{figure}
  \centering
  \epsfxsize=3in
    \epsfbox{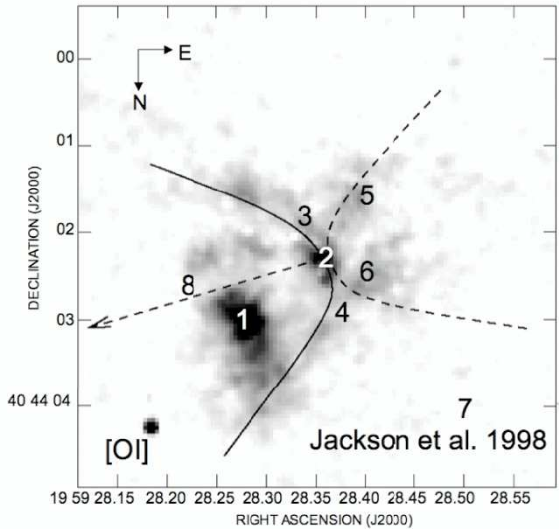}
  \caption[Cygnus-A Geometry]{\bf Cygnus-A Geometry. \it This image is adapted from Figure~4 in \citet{jac98} and shows the geometry of Cygnus-A. The position and direction of the opening cone and jet are shown on the a figure of the [OI] line emission. The numbers refer to the regions studied in this paper. The scale is in arbitrary flux units. \label{cygAgeom}}
  \end{figure}

The morphology and velocity seen in these maps are consistent with the picture given by \citet{jac98}. Figure~\ref{cygAgeom} summarizes their geometry and highlights an opening cone that passes from the NW side (Regions 1 and 2) of the image through the central blob to the SW side of the image.  These authors also point out blue condensations (Region~6) and the jet that emanates from the center westwards, passing along south side of the NW blob. Our largest velocity gradients are located perpendicular to the opening cone as well as to the direction of the jet identified by \citet{jac98}.

{\bf Emission Diagnostics} - Figure~\ref{specpap} shows the red and blue spectra for the central Region~1. As suspected, strong [O~{\sc{iii}}]~$\lambda$~5007 and [N~{\sc{ii}}]~$\lambda$~6584 relative to H$\beta$ and H$\alpha$. Figure~\ref{1204lha} shows that most pixels have [N~{\sc{ii}}]~$\lambda$~6584/H$\alpha$ ratios higher than $\sim$ 1.4, placing them on the AGN side of the BPT diagram. The highest ratios are in two clumps almost directly South and North of the center of the image, however there appears to be no relationship between the H$\alpha$ luminosity and the value of the line ratio. The lowest line ratios are East of the center, at Region~6. This is the location of bright, blue, condensed clumps of continuum emission (observed using the F622W HST filter) previously discussed in \citet{jac98}. The spatial resolution here is not as high as for HST, and neighbouring pixels may be washing out signs of young stars. Indeed, \citet{jac98}  are unable to identify any emission of [O~{\sc{iii}}]~$\lambda$~5007 or H$\beta$ within these blue condensations.  

\begin{figure}
  \centering
  \epsfxsize=3in 
   \epsfbox{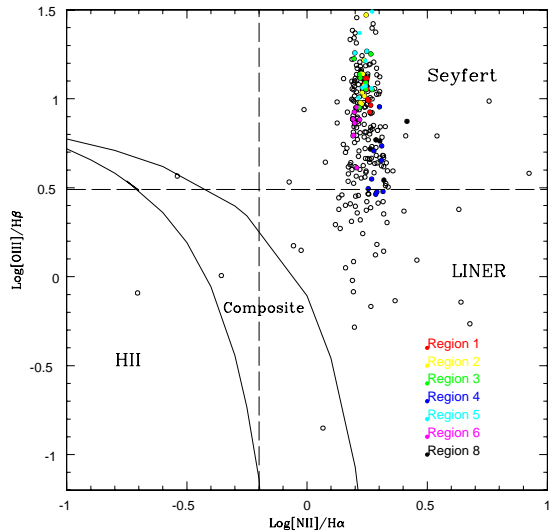} 
      \caption[BPT diagnostic diagram for Cygnus-A]{\bf BPT diagnostic diagram for Cygnus-A. \it Individual pixels are plotted as open circles and pixels within the analyzed regions are plotted as filled circles.  Almost every pixel shows exhibits the well known Seyfert status of Cygnus-A. This figure uses the definitions of H~{\sc{ii}}~region, Composite and AGN ionizing sources from Kauffmann et al. (2003). The designations between LINER and Seyfert are from \citet{ost06}. \label{bptCA}}
\end{figure}

This is the only BCG for which we can constrain the emission mechanism using all four diagnostic lines. The [O~{\sc{iii}}]~$\lambda$~5007/H$\beta$ map (not shown) is affected by the poorer quality of the H$\beta$ image, but shows no clear structure immediately apparent in the map. However, all of the central pixels show ratios of [O~{\sc{iii}}]~$\lambda$~5007/H$\beta$ $>$8, disclosing the nature of this AGN to be a Seyfert.  The BPT diagram shown in Figure~\ref{bptCA} classifies virtually every pixel as ionized  by AGN emission, and there is a smooth transition from pixels at the Seyfert corner down to the LINER side. This agrees with the well established classification of Cygnus-A being a narrow line radio galaxy. 
\nocite{kew02}
Two additional diagnostics are available for Cygnus-A, the ratio of  [O~{\sc{i}}]~$\lambda$~6300/[O~{\sc{iii}}]~$\lambda$~5007 and [O~{\sc{iii}}]~$\lambda$~5007/H$\alpha$. The former is a good tracer of the ionization parameter (q, as defined in Kewley \& Dopita 2002; q=u$\times$c, a measure of the number of ionizing photons per atom at the boundary layer). Both lower and higher ionization states are present. The [O~{\sc{iii}}]~$\lambda$~5007/H$\alpha$ ratio also shows a region of higher ionization along the edge of emission, North of the NW blob (Region~1), which was noted in the observations of \citet{jac98}.

\section{Discussion} \label{conclchap}

\subsection{Overview of Results}

We have investigated the emission line morphologies, dynamics and ionization state in a sample of nine BCGs. Condensed (Abell~2052, Cygnus-A), filamentary and patchy (Abell~1060, Abell~1668, and  MKW3s), and extended (Abell~1204, Abell~2199) morphologies are observed and the H$\alpha$ and [N~{\sc{ii}}] emitting gas usually follows the same morphology. Also of note are two cases, Abell~1060 (a non-cooling flow) and Ophiuchus (at best a low level cooling flow), where the line emitting gas morphology echos that of prominent dust features seen in the optical images. By using BPT diagrams to diagnose the ionization mechanism, we found that in two out of seven emitting BCGs, hot stars are likely perpetrators - NGC~3311 in the non-cooling flow cluster Abell~1060 and the BCG in cooling flow cluster Abell~1204. In these cases, SFRs and ages for the young populations are derived. However, the presence of an optical AGN, in both cooling flow as well as non-cooling flow systems, was strong. Two of the nine BCGs observed had no emission lines. This includes the BCG of Abell~1651 and the BCG Ophiuchus. Unexpectedly in Ophiuchus, it is another source within the cluster (Object~B), rather than the BCG that shows line emission. In most cases, the relative velocities are in the neighborhood of $\pm$100~-~200$\,$km$\,$s$^{-1}$, though some higher relative velocities indicative of large inflows (Object~B in Ophiuchus) or outflows (MKW3s) are observed. 

The properties and main results for each cluster BCG are highlighted in Table~\ref{sumtab} and discussed briefly below.

\begin{itemize}
\item Abell~1060: Previous detections of star formation in this dusty \citep{lai03} non-cooling flow \citep{hay06} BCG have been described by \citet{vas91}. We characterize several regions of star formation, note that the morphology of the star forming region follows that of the dust, and present a smooth velocity gradient reminiscent of rotation.
\item Abell~1204: This cooling flow cluster \citep{bau05} has the highest redshift of our sample (z~=~0.1706) and thus a larger extent of the cD galaxy is viewed. Regions of ionization due to AGN are separated from those indicative of a young stellar population. The latter are found further from the center in a plume that extends towards a chain of smaller cluster galaxies. The relative velocities are modest ($\pm$200$\,$km$\,$s$^{-1}$) and show no signs of rotation. An interaction with the nearby galaxies may be important.
\item Abell~1668: This is a non-cooling flow cluster \citep{sal03}, yet strong lines are present. These appear in a filamentary distribution with line ratios signifying ionization from an AGN. The relative velocity of the line emitting gas shows that it is not at rest with the underlying cD.
\item Ophiuchus: This cluster was recently observed to have a cool core, even though the X-ray temperature of the cluster itself is very hot at 9-10$\,$keV \citep{fuj08}. The BCG shows no emission lines. However, emission lines are seen in Object~B, at a projected distance of $\sim$2$\,$kpc away from the center of the BCG, which is also at the position of the X-ray centroid. Object~B is also cospatial with a dust feature seen on the acquisition image, and the emission lines have a relative velocity of $+$750$\,$km$\,$s$^{-1}$ with respect to the BCG. We attribute this to infall onto the BCG. 
\nocite{kaa04}
\item MKW3s: This cooling flow cluster (Kaastra et al. 2004) has previously been observed to have a UV excess and attributed star formation by \citet{mcn89} and by \citet{hic05}. We find the emission lines in the GMOS IFU image to be filamentary in morphology, yet well described by ionization from an AGN. The lines are blueshifted by $+$560$\,$km$\,$s$^{-1}$ with respect to the BCG, suggesting an outflow.
\item Abell~1651: This cooling flow cluster \citep{whi00} BCG shows no emission lines.
\item Abell~2052: This cooling flow cluster \citep{bla03,kaa04} has patchy dust in the center of the BCG \citep{lai03} and \citet{hic05} and \citet{bla03} have deduced star formation in the BCG from excess UV-IR and U-band continuum emission. We find that the Seyfert signature overwhelms any star formation in the central few arcseconds of the BCG. The morphology of the emission lines here is point-like, in that it shows no asymmetric or extended features. The relative velocities vary on the scale of $\pm$250$\,$km$\,$s$^{-1}$, but are not smooth enough to warrant a classification of rotation.
\item Abell~2199: Another cooling flow \citep{joh02} BCG with previous calculations of star formation based on UV excess \citep{mcn89}. We find the AGN signature in the emission lines is dominant throughout the extent of the OASIS image. The lines are brightest in the center, but exhibit an extended morphology towards the East. The relative velocities are similar to those seen in Abell~2199, but again the gradient is not regular enough to signify rotation.
\item Cygnus-A: This poor cooling flow cluster \citep{rey96} is a well studied AGN \citep{jac98,tad94,tad03}. We find a morphology consistent with the images of \citet{jac98}, line ratios telling of an AGN, and a velocity gradient that suggests rotation. The FWHM of the emission lines traces the direction of the jet, as it becomes wider for the lines in that region.
\end{itemize}

Certainly gas exists in BCGs and is excited, and the observations above show that many mechanisms are at play. For instance, signs of emission by hot stars are present in both cooling flow and non-cooling flow systems, and AGN-ionized gas is also present in both cooling flow and non-cooling flow systems. 

\subsection{Scenario}

\citet{wil06}, who conducted a similar study, find little variation of the emission line ratios across the emission nebulae in their high redshift and very luminous sample of 4 cooling flow BCGs, implying a uniform ionization state. They suggest the following single scenario to explain the line emission in all their galaxies: an interaction of smaller cluster galaxies triggers starbursts in cold gas reservoirs,
presumably deposited from the cooling flow. This is supported by their observation
that different ionization states of the gas vary little spatially, implying a single ionization source for the H$\alpha$ emission. Intriguingly, the current observations for the modest cooling flow case of Abell~1204 support this idea (this is the only cluster in this sample where z~$>$~0.1). In this case, the H$\alpha$, [N~{\sc{ii}}] and [S~{\sc{ii}}] emitting gas all share a similar morphology. But, in contradiction to the findings of \citet{wil06}, we do find a difference in line rations with respect to position and H$\alpha$ strength with the lower line luminosity pixels showing more star-forming like activity (consistent with results of Hatch et al. 2007). We calculate a total SFR of  $\simeq$7$\,$M$_{\odot}$$\,$yr$^{-1}$ for this BCG, which is close to the infrared derived SFR of \citet{ode08} and not inconsistent with the mass deposition rate as not all the molecular gas will convert to stars
and as the calculations of MDRs based on the absence of the coolest gas are an order of magnitude {\it below} the rates based on the classical cooling flow scenario - 50$\,$M$_{\odot}$$\,$yr$^{-1}$ \citep{ode08} compared to 675$\,$M$_{\odot}$$\,$yr$^{-1}$ \citep{whi00}. \citet{wil06} also reasoned that the H$\alpha$ and CO gas are related as they share the same kinematics, and it is this CO gas, which is subsequently disturbed by the passing of nearby neighbors or by an AGN, which will emit the H$\alpha$ line. The observations presented here for Abell~1204 show no hard evidence for an interaction with the nearby neighbors, but, it is tempting to speculate
on the likelihood of an interaction with nearby galaxies seen on the acquisition image. It would be interesting to search for molecular gas in this BCG. This is the only cluster in the sample presented in this paper which agrees with the \citet{wil06} hypothesis.

However, the overall ``mixed~bag"  of ionization scenarios seen throughout most of this sample is even more appealing when put into the context of the \citet{hat07} observations. In opposition to the scenario put forth by \citet{wil06}, \citet{hat07} find cases where the line emission properties in cooling flow BCGs suggest motions from strong AGN or starburst driven outflows (Abell~2390, Abell~1068), from galaxy passbys (2A~0335+096), and from rotation (Abell~262). They find that the ionization state is not uniform, and do not conclude that one scenario can account for the emission lines seen in cooling flow BCGs, the results presented here are consistent with such conclusions.

One caveat of our study is the small sample size, nonetheless, no difference in the ionization mechanism has been found that clearly separates the cooling flow and non-cooling flow cluster BCGs. Neither has a consistent picture been developed to explain the origin of the line emission throughout the sample. Although, the emission line characteristics are consistent within each BCG and most systems show a hard ionizing source prominent throughout the central few arcseconds, usually well described by LINER emission line ratios. 

As found by several other authors \citep{wil06,boh02,don00}, this data does not support a simple picture in which X-ray gas cools into molecular clouds subsequently forming stars. This is emphasized by the variation in morphology of the ionized nebula in these systems. The disturbed morphology of the non-AGN ionized gas of Abell~1204 extends towards the direction of several smaller galaxies, suggestive of an interaction with nearby companions. The spectrum of MKW3s is also interesting, with the H$\alpha$ emission line shifted with respected to the underlying spectrum. Although for the majority of cases the line emission is stronger in the central regions of the BCG itself, in Ophiuchus at least, the line emission is localized North of the BCG center, at a distance of $\sim$2$\,$kpc. \citet{don07} find the emission lines in the cooling flow cluster 2A0335+096 are associated with the BCG, as well as a companion. In the case of 2A0335+096, the emission lines are described as dusty, and this is also the case in Ophiuchus and in  NGC~3311 of Abell~1060 where the line emission is constrained by what appears on the acquisition images to be strong dust features. 

There are of course many possible outcomes for any molecular gas that drops out of the cooling flow onto the BCG. In general, the observations from the dataset presented here, of an overwhelming influence of AGN signatures in most of the sample support the scenario currently put forth for the nature of cooling gas in X-ray clusters where AGN feedback is important. That is,  part of the material condensing out of the cooling X-ray gas finds itself in the form of molecular reservoirs at the centers of the BCG, and part rains onto the central black hole. This could trigger an outburst from the AGN which is then hypothesized to reheat the cluster enough to prevent any further cooling. The AGN could simultaneously ionize the molecular gas deposits and influence starbursts. The molecular gas deposits would also be subject to flybys from companions, another mechanism which could trigger a starburst. This complex scenario does explain the observations of line emission resulting from the ionization of hot stars and AGN activity in cooling flows. It could also explain observations of AGN-ionized line emission in non-cooling flow, as the line emission could be triggered after the cooling cluster gas has been reheated by the AGN. The non-cooling flow BCG NGC~3311 of Abell~1060 may also be a part of the nominal $\sim$15\% of line emitting BCGs \citep{edw07b} as it is a non-cooling flow and the emission lines show no evidence for AGN ionization, and thus perhaps the star formation is due to a completely different process. It is however, somewhat more difficult to explain the observations of Ophiuchus, where the line emission is not localized in the BCG. In a future paper, we will combine X-ray and radio observations with our non-detection of line emission from Ophiuchus and Abell~1651 to make a census of the heating and cooling in those systems.

\section{Conclusions}

The observations from integral field spectroscopy have revealed the complex nature of the line emission in these galaxies, for which star formation, AGN, or both are important in several systems, but for which no consistency of the emission line characteristics and the cluster properties is seen.  The line emission is point-like, filamentary, or extended; it exists in regions that are plagued with large patches of dust, and those that are relatively dust free. The AGN ionizes all of the gas within the image in some cases, whereas regions of star formation are discernible in others.  Outflows are seen, as well as gas that has bulk rotations. And the emission could be associated with galaxy interactions in some, but not all cases. Each case with emission lines in a cooling flow shows regions where the emission lines have AGN-like ratios.

The process that is fueling the nominal emission in the non-cooling flows and the other bright galaxies at the center of a cluster is still unknown. However, in cooling flows, the existence of an AGN is  correlated with the presence of emission lines. As seen for example in Abell~1204 (this work), 2A0335+096 \citep{hat07}, and Abell~2052 \citep{bla03}, the ionization mechanism can change with the radial distance from the center, showing lines ionized by a hard source at the center, and those ionized by hot stars further out. 

This work supports the current ideas on the important role of a massive black hole in these galaxies, as much of the optical line emission is dominated by gas ionized by an AGN. The black hole may act as a sink for the cooling gas \citep{piz05}, may trigger a starburst \citep{wil06}, and its energy output may prevent further cooling of the cluster gas \citep{bes05,cro06,mcn07}. This last role is an important factor in our understanding of the building of large galaxies. The observed galaxy luminosity function undergoes a sharp cut-off at the high end. This cut-off can be explained with the inclusion of X-ray gas and AGN feedback, where the amount of cooling (and hence potential star formation) is regulated by heating from the AGN \citep{cro06}.

  \section*{Acknowledgments}
We are grateful to Jorge Iglesias and Sam Rix, who helped with the OASIS observations at WHT. We thank B. McNamara, H. Martel and L. Drissen for very helpful comments and discussion. We also thank Simon Cantin and V\'{e}ronique Petit for advice on the OASIS data reduction and plotting tools, and the Gemini Science team for their support in the observing and data reduction for the Gemini data. We would also like to thank the referee for comments which significantly improved the paper. This work was supported by the Natural Sciences and Engineering Research Council of Canada, Universit\'{e} Laval, and le Fonds qu\'{e}b\'{e}cois de la recherche sur la nature et les technologies through research grants to CR and by the Spanish PNAYA projects  and AYA2007-67965-C03-03 to MM.

\bibliography{le1c}
\clearpage
\appendix
\end{document}